\documentclass[11pt,a4paper]{article}
\pdfoutput=1
\usepackage{jheppub}
\usepackage{afterpage}
\usepackage{xcolor}
\usepackage{epsfig}
\usepackage{cancel}
\usepackage{framed}
\usepackage{multirow}

\usepackage{seqsplit}
\usepackage{enumitem} 

\usepackage{amsmath, amsthm, amsfonts, amssymb, latexsym}
\usepackage{upgreek}
\usepackage[caption=false]{subfig}

\usepackage{url}
\usepackage{ifpdf}

\def\beq{\begin{equation}}
\def\eeq{\end{equation}}

\def\bequ{\begin{equation}}
\def\eequ{\end{equation}}

\title{\Large Two-loop stability of a complex singlet extended Standard Model}

\author[a]{Raul Costa,}
\emailAdd{rauljcosta@ua.pt}
\author[a,b]{Ant\'onio P. Morais,} 
\emailAdd{aapmorais@ua.pt}
\author[b]{Marco O. P. Sampaio,} 
\emailAdd{msampaio@ua.pt}
\author[a,c]{Rui Santos}
\emailAdd{rasantos@fc.ul.pt}

\affiliation[a]{Centro de F\'\i sica Te\'orica e Computacional,
Universidade de Lisboa  \\
1649-003 Lisboa, Portugal} 
\affiliation[b]{Departamento de F\'\i sica da Universidade de Aveiro and I3N \\ 
Campus de Santiago, 3810-183 Aveiro, Portugal} 
\affiliation[c]{
 Instituto Superior de Engenharia de Lisboa - ISEL \\
 1959-007 Lisboa, Portugal
}


\abstract{Motivated by the dark matter and the baryon asymmetry problems, we analyse a complex singlet extension of the Standard Model (SM) with a $\mathbb{Z}_2$ symmetry (which provides a dark matter candidate). 
After a detailed two-loop calculation of the renormalization group equations for the new scalar sector, we study the radiative stability of the model up to a high energy scale (with the constraint that the $126$~GeV Higgs boson found at the LHC is in the spectrum) and find it requires the existence of a new scalar state mixing with the Higgs with a mass larger than $140$~GeV. This bound is not very sensitive to the cut-off scale as long as the latter is larger than $10^{10}\,$GeV.  We then include all experimental and observational constraints/measurements from collider data, dark matter direct detection experiments and from the Planck satellite and in addition force stability at least up to the GUT scale, to find that the lower bound is raised to about $170$~GeV, while the dark matter particle must be heavier than about $50$~GeV.  

}

\begin{document}

\maketitle

\section{Introduction}
\label{sec:Introduction}
The recent discovery of the Higgs boson at CERN's Large Hadron Collider (LHC) by the ATLAS~\cite{ATLAS:2012ae} and CMS~\cite{Chatrchyan:2012tx} collaborations,
and the  measurement of some of its properties (with increasingly greater precision) has proved to be quite demanding for the so-called Beyond the Standard Model (BSM) models.
By the end of the 8 TeV run, it is clear that no large deviations can occur in the Higgs couplings to the remaining Standard Model (SM) particles relative to the ones predicted
for the SM Higgs boson. The LHC run 2 will provide us with even more precise data and will either reveal directly or indirectly the existence of new physics or it will
further restrict the parameter space of BSM models. Furthermore, models with a decoupling limit may just be indistinguishable from the SM for the attained precision. Nevertheless,
some of the BSM models could still be very similar to the SM while providing solutions to some of the outstanding questions of particle physics. Such is the case of the singlet
extension of the scalar sector of the SM, which is the minimal model for dark matter~\cite{Silveira:1985rk, McDonald:1993ex, Burgess:2000yq, Bento:2000ah, Davoudiasl:2004be, 
Kusenko:2006rh, vanderBij:2006ne, He:2008qm, Gonderinger:2009jp, Mambrini:2011ik, He:2011gc, Gonderinger:2012rd, EliasMiro:2012ay, Cline:2013gha, Gabrielli:2013hma, Profumo:2014opa} . The model can simultaneously accommodate electroweak baryogenesis by allowing 
a strong first-order phase transition during the era of EWSB~\cite{Menon:2004wv, Huber:2006wf, Profumo:2007wc,Barger:2011vm, Espinosa:2011ax}, if the singlet is  complex. 

A somewhat related question which is often addressed in BSM models is that of the hierarchy between the Planck scale, $M_{Pl}\simeq 10^{19}$~GeV, and the electroweak symmetry breaking scale (the $Z$ boson mass scale $M_{Z}\simeq 91$~GeV). There are in fact numerous frameworks ranging through supersymmetry~\cite{Martin:1997ns}, models with extra dimensions~\cite{Antoniadis:1990ew,ArkaniHamed:1998rs}, little Higgs models~\cite{ArkaniHamed:2001nc},  just to name a few which address such problem, in which new scalar singlet fields appear in the spectrum. The study of the physical effects of coupling a complex scalar singlet to the SM may then be viewed as a minimal model, for the sector of such frameworks, which explains dark matter and the matter anti-matter asymmetry in the Universe.  

A more complete answer to some of the questions raised above may in fact lie at a very high energy scale. A relatively natural indication for what that scale could be comes from Grand Unified Theories (GUTs) in which, within supersymmetric models, the (running) gauge couplings unify at the GUT scale $M_{GUT}\sim 10^{16}$~GeV~\cite{Dimopoulos:1981yj,Ibanez:1981yh,Ellis:1990wk}. This suggests that the gauge structure of the SM may be a remnant of some larger simple or semi-simple symmetry group (see~\cite{Ramond:1979py} for a complete review). Typically, Grand Unified models fix specific relations among masses and couplings at the GUT scale, therefore, one expects that electroweak scale physics carries some of that information, which may be probed at the LHC~\cite{Miller:2012vn}. Thus, a detailed analysis of the Renormalization Group (RG) is required in order to evolve the couplings of the theory from the high scale down to the low scale. Some recent phenomenological work in light of the recently discovered Higgs boson and dark matter can be found in~\cite{Antusch:2013rxa,Chakrabortty:2013voa,Miller:2013jra,Miller:2014jza,Kadastik:2011aa}.

If the scale for new Physics is indeed as large as the GUT or the Planck scales (that is orders of magnitude beyond any current or planned collider experiment) we may have to work with a minimal theory that remains as the relevant description up to a high energy scale. In fact singlet fields provide a very natural way to couple the SM to hidden sectors if we note that $H^\dagger H$ (where $H$ is the SM Higgs doublet) is one of the (few) singlet operators in the SM which are of dimension less than four (and hence prone to coupling to hidden sector operators in a renormalizable combination). This concept was introduced in~\cite{Patt:2006fw} and is known as the {\em Higgs portal}.
 
Whichever the low energy theory may be, it must be consistent with all known experimental data as well as with theoretical consistency principles. One such principle is stability under the Renormalization Group Evolution (RGE), which will be a focus point in this article. 
This issue is already posed in the SM and, with the precise measurement of the Higgs boson mass $m_{h}\simeq 126$~GeV, state of the art 
calculations indicate that the SM is in a marginally unstable (or metastable) region of parameter space~\cite{Degrassi:2012ry, Alekhin:2012py, Buttazzo:2013uya}. In~\cite{Degrassi:2012ry} it was concluded that, if the electroweak vacuum is indeed metastable, its lifetime could be long enough (around $10^{400}$ years) to highly suppress the probability of decay. In this work we will focus solely on the scenario of strict stability at two-loops and will not
consider the possibility of a meta-stable vacuum.  

In fact we will find that the complex singlet extension we are considering, in addition to providing a dark matter candidate and improving the baryon asymmetry problem, can also improve the stability of the SM. This is related to the presence of a heavier visible scalar state in the spectrum whose mass (we conclude) must be  larger than $140$~GeV, a lower bound which is almost independent of the new physics scale (as long as it is larger than $10^{10}$~GeV).

Finally, although minimal, this complex singlet extension provides a rich collider phenomenology leading to some distinctive signatures that can be tested at the LHC~\cite{Datta:1997fx, Schabinger:2005ei, BahatTreidel:2006kx, Barger:2006sk, Barger:2007im, Barger:2008jx, O'Connell:2006wi,Gupta:2011gd, Drozd:2011aa, Pruna:2013bma, Ahriche:2013vqa, Chen:2014ask}. In this article we combine all bounds from collider experiments, precision electroweak physics, dark matter direct detection experiments and cosmological inference of the relic density, with all theoretical constraints on the model including our RGE evolution analysis. We perform dense parameter space scans to identify the regions of the physical parameters of the model (such as masses and couplings) which are still allowed and will be tested at the next run of the LHC. In particular we identify regions corresponding to scenarios that could provide complete models (explaining dark matter and the baryon asymmetry) up to a high scale (such as the GUT or Planck scale). One important point to stress is that whenever possible we present our results in terms of observables that could in principle be measured at colliders, such as the scalar masses and their couplings to SM particles. Only when it is necessary to trace back features of the results that are determined by theoretical conditions, such as vacuum stability, will we use the parameters of the scalar potential.

The structure of the paper is the following. In Sect.~\ref{sec:TheModel}  we describe the model that we study reviewing its main properties and setting notation. In Sect.~\ref{sec:Veff} we present a detailed analysis of the effective potential where we start by reviewing the procedure to extract the beta functions. The remainder of the section is divided in three parts: i) in Sect.~\ref{sec:BasisIndependent} we provide a proof of some general properties of the effective potential for a scalar theory at any order, ii) in Sect.~\ref{sec:OneTwoLoop} general expressions (which are basis independent) for the beta functions of the scalar sector contributions are derived up to two-loop order for any theory based on the results in~\cite{Martin:2001vx}, and iii)  in Sect.~\ref{sec:relative_error} we define an error measure to assess the differences between the one-loop and the two-loop approximations. In Sect.~\ref{sec:Results} we describe the results of our parameter space scans, first including only theoretical constraints combined with the RGE evolution (Sect.~\ref{sec:RGE_run_alone}) and finally adding the most up to date phenomenological constraints in Sect.~\ref{sec:constraints_pheno} (for which we have developed a new model class in the~\textsc{ScannerS} program~\cite{Coimbra:2013qq,ScannerS}). In the conclusions, Sect.~\ref{sec:conclusion}, we summarize our findings. Many of the technical details, in particular the two-loop beta functions for the complex singlet model, are left to the appendices.

\section{The model}\label{sec:TheModel} 
We will study an extension of the scalar sector of the SM, obtained by adding a complex singlet field $\mathbb{S}=S+i A$, 
which contains a residual $\mathbb{Z}_2$ symmetry $A\rightarrow -A$ after the explicit breaking of a global $U(1)$ symmetry by soft terms (in parenthesis):
\begin{multline}
V=\dfrac{m^2}{2}H^\dagger H+\dfrac{\lambda}{4}(H^\dagger H)^2+\dfrac{\delta_2}{2}H^\dagger H |\mathbb{S}|^2+\dfrac{b_2}{2}|\mathbb{S}|^2+
\dfrac{d_2}{4}|\mathbb{S}|^4+\left(\dfrac{b_1}{4}\mathbb{S}^2+a_1\mathbb{S}+c.c.\right) \, \, . \label{eq:V_general}
\end{multline}
This model is equivalent to adding two real singlet fields to the SM field content with appropriate symmetries imposed as above. 
One should note that for this model (with the exact $\mathbb{Z}_2$ symmetry) the soft breaking parameters must be real, i.e., $a_1\in\mathbb{R}$  and $b_1\in\mathbb{R}$~\cite{Coimbra:2013qq}. Note that this symmetry can be viewed as a CP symmetry defined
by $\mathbb{S} \to  \mathbb{S^*}$. As long as this symmetry remains unbroken we end up with two CP-even and one CP-odd (the dark matter) state under this symmetry. However, when the symmetry is spontaneously broken, we end up with three scalar states with the SM-Higgs CP numbers and no CP-violation in the scalar sector (for a detailed discussion on how to construct basis-invariant quantities that
signal CP violation see \cite{Branco:1999fs, Silva:2004gz}). It is however possible to generate spontaneous CP-violation from
the singlet's phase by adding new particles to the SM. Some of such simple scenarios were discussed in~\cite{Bento:1990wv, Branco:2003rt}.

This model was motivated as a way to provide a dark matter candidate while also contributing to a successful baryon asymmetry generation in the early Universe (see Sect.~\ref{sec:Introduction}). 
However, it also allows for a broken phase where all three neutral scalars mix. We should stress that the potential is exactly the same in the two cases and the only difference between 
the two scenarios is solely a consequence of distinct patterns of symmetry breaking. Furthermore, this model is representative of the most general physical situation one can obtain with 
the addition of a complex singlet\footnote{Except for the
model also derived from~\eqref{eq:V_general} where one obtains two dark matter candidates.}. 
Phenomenological studies at the LHC performed for these two benchmark scenarios are indeed generic in terms of: i) the possible final states to search for experimentally, ii) of mass hierarchies (the scalar masses can both be either larger or smaller than the SM-like Higgs) and iii)of invisible decays.

The main features of the model become clearer by expanding the fields around the two 
types of physical minimae that yield the correct pattern of electroweak symmetry breaking: 
\begin{equation}
H=\dfrac{1}{\sqrt{2}}\left(\begin{array}{c} G^+ \\ v+h+iG^0\end{array}\right) \; \;, \; \; \mathbb{S}=\dfrac{1}{\sqrt{2}}\left[v_S+s+i(v_A+ a)\right]
\end{equation}
where $v=246\;\mathrm{GeV}$ is the SM Higgs vacuum expectation value (VEV), and $v_S,v_A$ correspond, respectively, to the real and imaginary parts of the complex field VEV. 
Organizing the field fluctuations as $h_i=\left\{h,s,a\right\}$ and the mass eigenstates as $H_j$ ($j=1,2,3$), we define the mixing matrix through\footnote{Here we use a convention for the definition of the mixing matrix according to the \textsc{ScannerS} code~\cite{ScannerS} which we will use in our analysis.}
\begin{equation}
H_j=R_{ji}h_i \; 
\end{equation}
where $R_{ij}$ is in general a $3 \times 3$ orthogonal matrix. Imposing a $\mathbb{Z}_2$  symmetry on the $A$ component of the singlet, the minimum conditions lead to two distinct scenarios, namely
\begin{itemize}
\item $ v_A=0$, in which case there is mixing between the doublet fluctuation $h$ and the singlet real fluctuation $s$, while
the singlet imaginary component $A=a$ becomes a dark matter candidate. We call this phase the \textit{dark matter phase}.
\item $v_S\neq 0~{\rm and}~ v_A\neq 0$, that is, both singlet VEVs are non-zero and mixing among all neutral field fluctuations occurs. We call this phase the \textit{broken phase}.
\end{itemize}
In a previous work~\cite{Coimbra:2013qq} we have shown that, interestingly, the theoretical and pre-LHC constraints (including the dark matter ones), together with the LHC results, allow us distinguish
between the two phases in some regions of the parameter space.

One should note that simpler models can be obtained by imposing more symmetry on this same potential. An exact global
$U(1)$ symmetry on the complex singlet implies $a_1=b_1=0$. Depending on the pattern of symmetry breaking
this model can either have one or two dark matter candidates. The same number of dark matter candidates is obtained by
imposing a separate $\mathbb{Z}_2$ symmetry for $S$ and $A$. This implies that the soft breaking parameters are $a_1=0$ 
and $b_1\in \mathbb{R}$. A detailed discussion of the variants obtained from the most general potential for this complex singlet model was presented in~\cite{Coimbra:2013qq}.

Returning to the model under analysis, the mass eigenstates fields $H_i$ always couple to the SM particles through the combination 
\begin{equation}
h=R^{-1}_{hj} H_j = R_{jh} H_j \quad (j=1,2,3)
\end{equation}
because only the Higgs doublet couples to fermions and gauge bosons.
Thus, for any SM coupling $\lambda_{h_{SM}}^{(p)}$ to a given particle $p$, it is easy to conclude that the corresponding coupling in the singlet model for the scalar $H_j$ is given by
\begin{equation}
\lambda_{j}^{(p)}=R_{jh}\lambda_{h_{SM}}^{(p)}\; ,
\end{equation}
\textit{i.e.}, its value relative to the SM coupling is simply a mixing matrix element which is independent of the SM particle to which the coupling corresponds. 
So it is convenient to define a (global) relative coupling to SM particles for each scalar $H_j$, $\kappa_j$, which is normalized to the respective SM couplings 
and is independent (in this model) of the (non-scalar) SM particles: 
\begin{equation}
\kappa_j\equiv \dfrac{\lambda_{H_j}^{(p)}}{\lambda_{h_{SM}}^{(p)}}=R_{jh} \; .
\end{equation}
In the dark matter phase $R_{jh} = (R_{1h} , R_{2h}, 0)$ ($j=3$ corresponds to the dark matter candidate)
while in the broken phase all three $R_{jh}$ are in principle non-zero. In order to make it simpler to identify the couplings and masses let us note  the following. In the dark matter phase there is only one independent reduced coupling $\kappa_{H_{126}}$ which is the SM-like Higgs coupling
to all SM particles. The new non-dark matter scalar has coupling $\kappa_{H_{\rm new}}\equiv\sqrt{1-(\kappa_{H_{126}})^2}$ to SM particles and its mass is denoted by $m_{H_{\rm new}}$.
Regarding the broken phase we use the following notation for the couplings: $\kappa_{H_{126}}$, with the same meaning as before,
$\kappa_{H_{\rm light}}$ and $\kappa_{H_{\rm heavy}}$ for the two extra (non-SM) scalars, where the only restriction is $m_{H_{\rm light}} < m_{H_{\rm heavy}}$
while no order relation is imposed relative to the mass of the SM-like Higgs.
Note that (without loss of generality) we take the modulus of 
$\kappa_j$ when we present our results, since a sign flip can always be absorbed in the scalar eigenstate fluctuation without affecting 
the widths of the scalars or the relevant cross sections -- see Sect.~\ref{sec:collider_bounds}.

\section{The effective potential and the RGEs}\label{sec:Veff} 

The renormalization group equations (RGEs) describing the evolution of the scalar couplings for this theory have been determined at one-loop order in~\cite{Gonderinger:2012rd} from the Coleman-Weinberg potential.
The dark matter phase was then analyzed for some fixed choice of parameters (in our study we will perform a full parameter space scan). In this section we provide a two-loop order calculation of the effective potential for a scalar theory using some generic results known in the literature. We then extract the two-loop RGEs which will allow us (combined with the SM like contributions to be discussed below) to perform a detailed analysis of the vacuum stability constraints combined with the latest constraints from the LHC data. The two-loop analysis is important because it provides more reliable results and it can be used to assess the error of the one-loop approximation (for which we develop a quantitative measure). 

The radiative corrections to the couplings, $\{L\}$, of a given quantum field theory are described by their evolution equations. The latter are functions of the energy scale of the process, $\mu$, the renormalization scale. Such evolution equations, the RGEs, are obtained by requiring renormalization scale invariance of all simply connected $n$-point functions  which are generated by the effective action $\Gamma[\phi_i]$. In this section we focus on the scalar sector to illustrate the procedure to obtain the RGEs. This is because the new interactions in this model are only introduced through the scalar potential, so the main novel ingredients of the RGEs are in the scalar sector. All other contributions are similar, in form, to the ones calculated in the SM. Our analysis will allow us to obtain all of the (purely) scalar contributions which we have checked in special limits against the SM~\cite{Arason:1991ic, Ford:1992pn}, the two Higgs doublet model (2HDM)~\cite{Ferreira:2009jb} and $U(N)$ symmetric complex singlet models~\cite{Tamarit:2014dua}. In appendix~\ref{app:2loopBetas} we present the full RGEs with all fermion and gauge boson contributions at two loops which were checked by implementing the model (with the linear term removed) in the \textsc{Sarah} package~\cite{Staub:2013tta}.\footnote{The \textsc{Sarah} package does not deal with the linear term of this model. Our analysis based on the effective potential provides the RGE of such term.}

 We denote the set of (real) scalar fields used to expand the potential by  $\Phi_i$. The scale invariance conditions of the $n$-point functions are known as the Callan-Symanzyk equations. Let us define $t\equiv \log \mu$. It can be shown~\cite{LeBellac:1991cq} that, for a translation invariant vacuum, the scale invariance of all $n$-point functions is equivalent to the scale invariance of the effective potential defined as the effective action per unit volume ($V^{(4)}$ is the 4-volume under which the effective action is integrated)
\begin{equation}
V_{\rm eff}\equiv \dfrac{\Gamma[\Phi_i]}{V^{(4)}}\; .
\end{equation} 
Then, the scale invariance condition reads
\begin{equation}\label{eq:scale_invarianceVeff}
\dfrac{dV_{\rm eff}}{dt}=\left[\dfrac{\partial}{\partial t}+\sum_L\beta_L\dfrac{\partial}{\partial L}-\sum_i\gamma_i\Phi_i\dfrac{\partial}{\partial \Phi_i}\right]V_{\rm eff}=0
\end{equation}
where, for a given coupling $L$ and field value component $\Phi_i$, we define the beta function and anomalous dimension, respectively
\begin{equation}
\beta_L\equiv \dfrac{d L}{d t}\qquad , \qquad  \gamma_i\equiv - \dfrac{d \log \Phi_i}{d t}\;.
\end{equation}
This condition must hold for any field value configuration. If we expand the scale invariance condition, Eq.~\eqref{eq:scale_invarianceVeff}, as a Taylor series in the fields, then each coefficient in the expansion must vanish on its own. The vanishing of each coefficient corresponds precisely to imposing the scale invariance of all $n$-point functions. 

The effective potential and the RGEs have been derived in a number of forms in the literature for generic field theories. Detailed derivations based on the direct calculation of $n$-point functions have been performed to obtain generic expressions~\cite{Machacek:1983tz,Machacek:1983fi,Machacek:1984zw,Luo:2002ti}. However, often, such derivations shadow part of the underlying simplicity of the procedure, especially in the purely scalar case. A somewhat simpler procedure consists of using the effective potential (which in fact encodes all $n$-point functions). The effective potential calculation for generic field theories at two-loop order was reviewed by Martin in~\cite{Martin:2001vx}. Typically the diagrammatic calculation proceeds in the eigen-basis where all field fluctuations are the mass eigen-states of the theory. The final form of the effective potential, in general, is presented in a form which depends on the mixing matrices at the vacuum state. However, the RGEs are independent of the vacuum state. Focusing on a pure scalar theory, here we start by providing a general proof showing that it is always possible to write the effective potential (at any order) in terms of the original couplings of the theory independently of the vacuum choice, as expected. We then use the results summarized in~\cite{Martin:2001vx} for the two-loop scalar contributions to the effective potential in order to write an explicit (basis independent) form of the effective potential. Finally we apply the scale invariance constraint to obtain general expressions for the RGEs and observe the generic cancellation of logarithmic terms which is an important consistency check. 

\subsection{Basis independent form} \label{sec:BasisIndependent}
In what follows we will use the potential of the theory in three different forms. First we use the parametrization of~\cite{Martin:2001vx} for a generic scalar potential in an arbitrary basis which we call the $L$-basis: 
\begin{eqnarray}
V=L+L^{i}\Phi_{i}+\dfrac{1}{2!}L^{ij}\Phi_{i}\Phi_{j}+\dfrac{1}{3!}L^{ijk}\Phi_{i}\Phi_{j}\Phi_{k}+\dfrac{1}{4!}L^{ijkl}\Phi_{i}\Phi_{j}\Phi_{k}\Phi_{l}\;,
 \label{eq:V-Ls}
\end{eqnarray}
where the couplings constants $\{L,L^i,L^{ij},L^{ijk},L^{ijkl}\}$ are completely symmetric under interchange of the indices which run over the $N$ real scalar degrees of freedom, $\Phi_i$. The second form of the potential is obtained by expanding the fields around the classical (vacuum) configuration through the shift
\begin{eqnarray}
\Phi_{i}(x)=v_{i}+\phi_{i}(x)\;.
 \label{eq:shift}
\end{eqnarray}
The vacuum expectation values are denoted as $v_{i}$, whereas the quantum fluctuations around the classical field configuration as $\phi_i$. The second form of the potential is then
\begin{eqnarray}
V=V^{(0)}+\dfrac{1}{2!}\Lambda^{ij}\phi_{i}\phi_{j}+\dfrac{1}{3!}\Lambda^{ijk}\phi_{i}\phi_{j}\phi_{k}+\dfrac{1}{4!}\Lambda^{ijkl}\phi_{i}\phi_{j}\phi_{k}\phi_{l} \;.
 \label{eq:V-Lambdas}
\end{eqnarray}
The relation between the $L$-couplings in the first basis and the $\Lambda$-couplings in this basis is provided in appendix~\ref{app:DetailsBasisIndep}, Eq.~\eqref{eq:Lambdas}. Note that, in this basis (which we name $\Lambda$-basis), the minimum conditions impose that the coefficients of the linear terms vanish. 

The third basis (named $\lambda$-basis) is used to make contact with the general results described in~\cite{Martin:2001vx}. It is obtained through a rotation, $M_{j}^{\phantom{a}i} \subset SO(N)$, on the space of quantum field fluctuations such that it diagonalizes the quadratic part of the potential to obtain mass eigenstates:
\begin{eqnarray}
\phi_{j}=M_{j}^{\phantom{a}i}R_{i} \;.
\label{eq:Ri}
\end{eqnarray}
In this basis, the scalar potential is given by
\begin{eqnarray}
 V=V^{(0)}+\dfrac{1}{2!}(m^i)^{2}R_{i}^{2}+\dfrac{1}{3!}\lambda^{ijk}R_{i}R_{j}R_{k}+\dfrac{1}{4!}\lambda^{ijkl}R_{i}R_{j}R_{k}R_{l} \;.
\label{eq:VR}
\end{eqnarray}
The relation between the $\Lambda$-couplings of the second form and the $\lambda$-couplings of this form is again provided in appendix~\ref{app:DetailsBasisIndep}, Eq.~\eqref{eq:lambdas}. It is easy to see that the relation between the two parametrizations is obtained simply by applying a rotation matrix to each of the indices of the coupling.

In the perturbative regime, the RGEs are obtained by imposing the scale invariance condition on the loop expansion of the effective potential $V_{\rm eff}$. The loop expansion can be thought of as a power series in the Planck constant. Defining the perturbative parameter, $\varepsilon \equiv \hbar/(16 \pi^2)$, the general loop expansion is represented by
\begin{eqnarray}
V_{{\rm eff}}=\sum_{n=0}^{+\infty}\varepsilon^{n}V^{(n)}(L(t),v_{i}(t),t) \;.
\label{eq:VeffLoop}
\end{eqnarray}
The perturbative formulation of quantum field theory is naturally performed in terms of field fluctuations which are mass eigenstates (as it is the case in the $\lambda$-basis, Eq.~\eqref{eq:VR}). While this procedure simplifies the diagrammatic calculations, the final form for the effective potential is explicitly dependent on the mixing matrices $M_{j}^{\phantom{a}i}$. We now provide a proof which shows that it is always possible to write the $n$-loop effective potential without referring to a particular basis, hence without any explicit dependence on the mixing matrices.

The general loop expansion of the effective potential is a sum of one-particle irreducible vacuum diagrams (i.e. with no external lines). This statement immediately imposes a constraint relating the number of propagator lines $P_D$ in the diagram (the subscript $D$ labels the diagram), and the number of vertices, $V_{D,k}$, with $k$ external lines. Noting that each propagator line leaving a vertex must enter another vertex we have that 
\begin{eqnarray}
2P_{D}=\sum_{k}kV_{D,k} \;,
\label{eq:2PD}
\end{eqnarray}
i.e. if we count the number of propagators for each vertex and sum them all, we obtain twice the number of propagator lines in the diagram.  Therefore, in the physical $\lambda$-eigenbasis, we conclude that the general $n$-loop contribution to the effective potential must be a scalar  with the following form
\begin{eqnarray}
V^{(n)}=\sum_{D=1}^{N_{D}^{(n)}}\left(\lambda_{V_1}\ldots\lambda_{V_{D}}\right)^{j_{1},\ldots,j_{P_{D}},j_{1},\ldots,j_{P_{D}}}I_{D}^{(n)}(m_{j_{1}}^{2},\ldots,m_{j_{p_{D}}}^{2},\mu^{2}) \;,
\label{eq:Vn_lambdas}
\end{eqnarray}
where $I_{D}^{(n)}$ represents the loop integral for the diagram and $\lambda_{V_1}\ldots\lambda_{V_{D}}$ represents the product of all vertex couplings in the diagram. Note that the exact assignment of indices to each vertex is not essential to the argument. The only important points are that: i) to each propagator line in the diagram corresponds a mass $m_j^2$, and ii) there are twice the number of free indices in the vertices compared with the number of masses (or equivalently propagator lines) -- see Eqs.~\eqref{eq:2PD} and~\eqref{eq:Vn_lambdas}. Finally, observe that we have included a sum over $D$, that is, over the number $N_D^{(n)}$ of $n$-loop vacuum diagrams. Here we continue to use the Einstein convention for the sum over scalar field (latin) indices\footnote{There is a slight ambiguity in the repeated up-type indices in Eq.~\eqref{eq:Vn_lambdas} which are not summed over. This is a remnant of working in the diagonal basis which disappears once we rotate back to a generic basis.}.
 
To rotate back to the $\Lambda$-basis (which does not depend on the mixing matrices) we need to observe two facts:
\begin{enumerate}
\item The loop functions $I_{D}^{(n)}$ are typically analytic in the masses. Thus one can Taylor expand them around $m_j^2=\mu^2$ to end up with a sum of monomials with products of $m_j^2-\mu^2$ (each multiplying the vertex product pre-factor).  
\item Each mass squared factor is contracted {\em exactly} with two vertex indices which are set to be equal.
\end{enumerate}
Thus, if we use the property that the masses, $m_j^2$, are obtained by applying two rotation matrices to $\Lambda^{ij}$, we obtain exactly the necessary number of rotation matrices to rotate back all $\lambda$-vertices to $\Lambda$-vertices. In appendix~\ref{app:DetailsBasisIndep} we detail some of the intermediate steps leading to the general result
\begin{eqnarray}
V^{(n)}&=&\sum_{D=1}^{N_{D}^{(n)}}\left(\Lambda_{1}\ldots\Lambda_{V_{D}}\right)^{m_{1},\ldots,m_{P_{D}},m_{P_{D}+1},\ldots,m_{2P_{D}}}\left[\mathbf{I}_{D}^{(n)}\left(\Lambda^{ij},\mu^{2}\right)\right]_{m_{1},\ldots,m_{2P_{D}}} \nonumber\\
&\equiv&\mathbf{\Lambda}^{(n,D)}\cdot\mathbf{I}_{D}^{(n)}\left(\mathbf{\Lambda_{(2)}},\mu^{2}\right)\; , \label{eq:AllOrderVn}
\end{eqnarray}
where the $\mathbf{I}_{D}^{(n)}$ are the matrix versions of the loop integrals (see appendix~\ref{app:DetailsBasisIndep}), and $\mathbf{\Lambda_{(2)}}$ is the matrix form of $\Lambda^{ij}$. In the last line we have introduced a condensed notation where the sum over diagrams is operated by the Einstein convention (through the repetition of $D$) and the contraction of the field indices in the loop function $\mathbf{I}_{D}^{(n)}$ with the field indices in the vertex product $\mathbf{\Lambda}^{(n,D)}$ is represented by the dot $\cdot$ operator. 

This simple argument shows that it is always possible to write the effective potential, at any loop order, in an invariant way, in the same spirit of the trace form in which the one-loop Coleman-Weinberg potential is usually presented. In the next section we make this discussion concrete by writing the two-loop effective potential in this form.

\subsection{Two-loop effective potential and RGEs}
\label{sec:OneTwoLoop}
In this section we present the basis independent form of the two-loop effective potential, renormalized in the minimal-subtraction $\overline{\rm{MS}}$ scheme. At two-loop order the expansion, Eq.~\eqref{eq:VeffLoop}, becomes
\begin{eqnarray}
V_{\rm{eff}}= V^{(0)}+ \varepsilon V^{(1)}+ \varepsilon^2 V^{(2)} + \mathcal{O}(\varepsilon^3)\; . \label{eq:2-loopVeff}
\end{eqnarray}
Here $V^{(0)}$ is the tree-level scalar potential evaluated at the minimum (which in the generic \mbox{$L$-basis} is given by Eq.~\eqref{eq:Lambdas}) and $V^{(1)}$ is the Coleman-Weinberg potential. The latter is often written in the literature in the form~\eqref{eq:AllOrderVn}. As for the two-loop term, $V^{(2)}$, the one-particle irreducible diagrams that contribute involve only cubic and quartic vertices. Therefore, in a pure scalar theory, the two-loop correction for the effective potential is the sum of a cubic $V^{(2)}_{sss}$ with a quartic $V^{(2)}_{ssss}$ contribution, i.e.
\begin{eqnarray}
 V^{(2)} =  V^{(2)}_{sss} + V^{(2)}_{ssss} \; , \label{eq:2-loopV2}
\end{eqnarray}
which were calculated in~\cite{Martin:2001vx} in the $\lambda$-basis. In appendix~\ref{app:lambdaVeff} we provide a summary of the one and two-loop contributions. Using the procedure leading to Eq.~\eqref{eq:AllOrderVn}, we rewrite such contributions in the \mbox{$L$-basis}, which, in a matrix condensed notation, take the simple form
\begin{eqnarray}
V^{(1)}&=& \frac{1}{2}\left[\mathbf{\Lambda_{(2)}}^{2}\right]^{ij}\left[\frac{1}{2}\log(\mathbf{\Lambda_{(2)}})-t-\frac{3}{4}\right]_{ij}\; , \label{eq:VCWL}\\[2mm]
V_{sss}^{(2)}&=&\dfrac{t}{2}\Lambda_{\phantom{a}mn}^{i}\Lambda^{jmn}\left[\mathbf{\Lambda_{(2)}}\left(t+2-\log\mathbf{\Lambda_{(2)}}\right)\right]_{ij} + t_{\rm{independent}} , \label{eq:vsssL}\\[2mm]
V_{ssss}^{(2)}&=&\dfrac{t}{2}\Lambda^{ijkl}\Lambda_{kl}\left[\mathbf{\Lambda_{(2)}}\left(t+1-\log\mathbf{\Lambda_{(2)}}\right)\right]_{ij}+ t_{\rm{independent}} \; , \label{eq:vssssL}
\end{eqnarray}
where we have omitted the $t$-independent part of the two-loop contributions (since it only enters the derivation of the RGEs at three-loop order) and we have used the matrix $\log$ (as defined by the Taylor series over matrices).

We are now in the position to derive the generic two-loop scalar beta functions. In appendix~\ref{app:2loopBetas} we summarize the procedure to obtain them. Similarly to the effective potential they are expanded in powers of $\varepsilon$ so we denote the $n$-loop contributions by 
\begin{equation}
\left\{\beta^{(n)},\;\beta^{(n)i},\;\beta^{(n)ij},\:\beta^{(n)ijk},\;\beta^{(n)ijkl},\;\gamma^{(n)i}\right\}\; ,
\end{equation}
respectively for the beta functions of the vacuum energy ($L$); the linear ($L^i$), quadratic ($L^{ij}$), cubic ($L^{ijk}$) and quartic couplings ($L^{ijkl}$); and the anomalous dimension of the field $\Phi_i$. The final results are more conveniently written using the condensed notation
\begin{equation}\label{eq:general_beta:scalar}
\beta^{(n)i_1,\ldots,i_p}=L^{i_1,\ldots,i_p}\sum_k\gamma^{(n)i_k}+\delta^{(n)(i_1,\ldots,i_p)}\; ,
\end{equation}
where $(i_1,\ldots,i_p)$ denotes symmetrization of the indices in the $\delta^{(n)}$. The one and two loop $\delta^{(n)}$ are then
\begin{eqnarray}
\delta^{(1)}	&=&	\dfrac{1}{2}L^{ab}L_{ab} \nonumber\\
\delta^{(1)i}	&=&	L^{ab}L_{ab}^{\phantom{ab}i} \nonumber\\
\delta^{(1)ij}	&=&	L^{ab}L_{ab}^{\phantom{ab}ij}+L^{abi}L_{ab}^{\phantom{mn}j} \nonumber\\
\delta^{(1)ijk}	&=&	L_{ab}^{\phantom{ab}ij}L^{abk}+2L_{ab}^{\phantom{ab}ik}L^{abj} \nonumber\\
\delta^{(1)ijkl}	&=&	L^{abij}L_{ab}^{\phantom{ij}kl}+2L^{abik}L_{ab}^{\phantom{ij}jl} \label{eq:deltas_n1}
\end{eqnarray}
and
\begin{eqnarray}
\delta^{(2)}	&=&	\dfrac{1}{2}L_{ab}\left(2L^{ab}\gamma^{(1)a}-L^{cda}L_{cd}^{\phantom{cd}b}\right) \nonumber\\
\delta^{(2)i}	&=&	2L_{ab}L^{abi}\gamma^{(1)a}-L_{ab}L^{cdai}L_{cd}^{\phantom{cd}b}-\dfrac{1}{2}L_{ab}^{\phantom{ab}i}L^{cda}L_{cd}^{\phantom{cd}b}\nonumber \\
\delta^{(2)ij}	&=&	2\left(L_{ab}L^{abij}+L_{ab}^{\phantom{ab}i}L^{abj}\right)\gamma^{(1)a}-L_{ab}L^{cdai}L_{cd}^{\phantom{cd}bj}-2L_{ab}^{\phantom{ab}i}L_{cd}^{\phantom{ab}aj}L^{cdb}-\dfrac{1}{2}L_{ab}^{\phantom{ab}ij}L^{cda}L_{cd}^{\phantom{cd}b} \nonumber\\
\delta^{(2)ijk}	&=&	2\left(L_{ab}^{\phantom{ab}i}L^{abjk}+L_{ab}^{\phantom{ab}j}L^{abik}+L_{ab}^{\phantom{ab}k}L^{abij}\right)\gamma^{(1)a}-3L_{cd}^{\phantom{ab}ak}\left(L_{ab}^{\phantom{ab}i}L^{cdbj}+L_{ab}^{\phantom{ab}ij}L^{cdb}\right) \nonumber \\
\delta^{(2)ijkl}	&=&	2\left(L_{ab}^{\phantom{ab}ij}L^{abkl}+L_{ab}^{\phantom{ab}ik}L^{abjl}+L_{ab}^{\phantom{ab}il}L^{abjk}\right)\gamma^{(1)a}-6L_{ab}^{\phantom{ab}ij}L^{cdak}L_{cd}^{\phantom{cd}bl} \; . \label{eq:deltas_n2}
\end{eqnarray}
Using these results we have obtained all two-loop scalar contributions to the RGEs of the complex singlet model, Eq.~\eqref{eq:V_general}.
In appendix~\ref{app:RGExSM} we present the full two-loop RGEs for this model also with the contributions from the other SM particles fully included.

\subsection{A measure of the one to two-loop truncation error}\label{sec:relative_error}
Since the RGEs are computed through a truncation of a perturbative series, it is useful to define a measure of the error of the approximation by comparing the one-loop with the two-loop approximation. To define such measure let us first define a (functional) norm of a given  real function $f(t)$ defined on an interval $t\in[t_0,t_0+\Delta T]$
by
\begin{equation}
N[f]\equiv \sqrt{\int_{t_0}^{t_0+\Delta T}\dfrac{dt}{\Delta T}f(t)^2}\; .
\end{equation}
Then we define the relative distance between a curve $f(t)$ and a curve $g(t)$ with respect to a scale function $s(t)$ by 
\begin{equation}
\delta[f,g;s]\equiv \dfrac{N[f-g]}{N[s]}\; ,
\end{equation}
which for smooth functions is positive definite and non-singular (if $s(t)$ is not identically zero everywhere).
Let us consider now a given coupling $L(t)$ with mass dimension $d$. Consider the one-loop approximation $L^{(1)}(t)$ and the two-loop approximation $L^{(2)}(t)$. We define its relative error as the relative distance between the curves $L^{(1)}(t)$ and $L^{(2)}(t)$ with an appropriate scale as follows 
\begin{equation}
\delta_{12}^L\equiv\begin{cases}  
\left(\dfrac{N[L^{(1)}-L^{(2)}]}{N[L^{(2)}]}\right)^{\frac{1}{d}}&,\;  \frac{N[L^{(2)}]^{\frac{1}{d}}}{M_Z}> \epsilon \vspace{2mm}\\ 
\dfrac{\left(N[L^{(1)}-L^{(2)}]\right)^{\frac{1}{d}}}{M_Z}&,\;  \frac{N[L^{(2)}]^{\frac{1}{d}}}{M_Z}< \epsilon \end{cases} \; ,
\end{equation} 
where $M_Z$ is the $Z$ boson mass (i.e. the typical electroweak mass scale at which the couplings are set) and $\epsilon=10^{-2}$ is a small constant. The latter is introduced to safeguard against numerical errors in couplings which are very close to zero so the differences would be due to round off errors rather than  a real difference between the one and two-loop approximations. In the case $d=0$ we do not have to divide by any mass scale so we use
\begin{equation}
\delta_{12}^L\equiv\begin{cases}  
\dfrac{N[L^{(1)}-L^{(2)}]}{N[L^{(2)}]}&,\;  N[L^{(2)}]> \epsilon \vspace{2mm}\\ 
N[L^{(1)}-L^{(2)}]&,\;  N[L^{(2)}]< \epsilon \end{cases} \; ,
\end{equation} 
Finally, we define a global quality factor to evaluate the error of the one-loop approximation, $\Delta_{12}$, to be the largest of all computed relative distances
\begin{equation}
\Delta_{12}=\max_L\delta_{12}^L \; .
\end{equation} 
This error measure should be interpreted with care. Firstly because it corresponds to a difference between the one and two loop approximations, so it assesses the error of the one-loop approximation and the importance of the two-loop approximation\footnote{To assess the error of the two-loop approximation a three-loop calculation would be required. Nevertheless one expects that the two-loop approximation is a substantial improvement of the one-loop if the perturbative expansion is to hold.}. Secondly (see discussion in Sect.~\ref{sec:RGE_run_alone}) the global error will indicate that the two-loop approximation is important, but the individual errors, $\delta_{12}^L$, will be small for the couplings which determine the interesting features of the results. Finally we should note that we use the full two-loop results in our analysis.


\section{Results of the parameter space scans}\label{sec:Results}
In this section we present a detailed analysis of the allowed parameter space of the complex singlet model. We start by presenting in the next section (Section~\ref{sec:RGE_run_alone}) 
a theoretical study of the effect of the RGE evolution of the couplings,
where we consider as initial conditions at the $Z$ mass scale: 
\begin{itemize}
\item that the minimum is global and provides the right pattern of electroweak symmetry breaking;
\item that the vacuum is stable (the potential is bounded from below);
\begin{equation}\label{eq:boundedness_below}
\lambda>0 \; \; \; \wedge \; \; \; d_2>0 \; \; \; \wedge \; \; \; \delta_2>-\sqrt{\lambda d_2} \; ;
\end{equation}
\item that perturbative\footnote{A model that breaks perturbative unitarity is not necessarily wrong. However, to deal with such possibility is largely beyond the scope of this work.} unitarity holds
\begin{multline}
|\lambda|\leq 16\pi \;\wedge \; |d_2|\leq 16\pi \; \wedge \; |\delta_2|\leq 16 \pi \; \\\wedge \; \left|\frac{3}{2}\lambda+d_2\pm \sqrt{\left(\frac{3}{2}\lambda+d_2\right)^2+d_2^2}\right|\leq 16\pi \; . \label{eq:unitaritycond}
\end{multline}
\end{itemize}
Two distinct studies are performed, one for the dark matter phase and one for the broken phase.
We will then move on, in Section~\ref{sec:constraints_pheno}, to a complete phenomenological analysis where we will add, to these electroweak scale conditions, all available experimental constraints from dark matter experiments and from collider experiments (to be described in detail later). 

\subsection{Low scale input}

In this study, we will be interested in assessing the importance of the two-loop running compared with one-loop running when drawing conclusions on parameter bounds from global scans. In fact our strategy is not to focus on very constrained or particularly tuned scenarios (which may lead to conclusions which are not generic) but to visualise millions of scenarios in parameter space projections to determine conditions on the physical parameters that are imposed by the combination of all the phenomenological and theoretical constraints the model is subject to. 

For internal consistency, in the two loop running, one must provide one loop input relations among the parameters of the theory and the physical  parameters (such as masses and couplings). In particular, it is well known that top quark contributions are dominant relative to other SM like contributions in the one-loop relations between the scalar masses/couplings relations, so we have made the approximation of neglecting light fermion and gauge boson contributions. In appendix~\ref{app:OneLoopInput} we analyse the effect of correcting the initial data, used in the RGE running, with such one-loop relations. There we show that such corrections to the initial data have a small effect in the shape of the regions we obtain, so they do not change our conclusions from the perspective of a global scan (see Sect.~\ref{sec:RGE_run_alone} and appendix~\ref{app:OneLoopInput}). 

We set all our input at the $Z$-boson mass scale ($\mu=M_Z$). In particular, we have extracted the SM top Yukawa coupling value at the $Z$-scale from the GAPP code by J. Erler~\cite{GAPP} which performs fits to electroweak precision data. We have set the code to the current best fit point to all the latest electroweak data.  

Unless stated otherwise, the scans are performed according to the ranges defined in table~\ref{scan_table}. The SM Higgs mass is varied between
$124.7$ and $127.1$~GeV while the SM VEV is fixed at $246$~GeV. The ranges for the VEVs both in the broken phase ($v_A,v_S\neq 0$) and in the dark matter
phase ($v_S\neq 0$ and $v_A=0$) were chosen to be in the interval between $0$ and $1000$~GeV. Regarding the scalar masses, for the theoretical study of the RGE effects, we used all new particle masses in the range $[0,1000]$~GeV. As for the phenomenological study, the new visible scalar masses vary in the interval between $12$ and $1000$~GeV and the mass of the dark matter candidate varies in the interval between $6$ and $1000$~GeV\footnote{In the phenomenological analysis we have a lower bound on the masses due to the lack of data from colliders (mainly LEP) and from dark matter searches (LUX).}.
Finally, in the dark matter phase, the $a_1$ coupling is an input parameter and its range has been set in the $[-10^8,0]$ $\mathrm{GeV}^3$ interval. Note that the fact that $a_1<0$ in the dark matter phase is a consequence of the choice of vacuum combined with the (conventional) choice of positive VEVs for the scan.

\begin{table}
\begin{center}
\begin{tabular}{|r|c|c|c|c|}
\hline
\multicolumn{1}{|c|}{\multirow{2}{*}{Scan parameter}} & \multicolumn{2}{c|}{Dark matter phase}        & \multicolumn{2}{c|}{Broken  phase}            \\ \cline{2-5} 
\multicolumn{1}{|c|}{}                                & Min                   & Max                   & Min                   & Max                   \\ \hline
$m_h$ (GeV)                                           & 124.7                 & 127.1                 & 124.7                 & 127.1                 \\ \hline
$m_{s_1}$ (Gev)\;\;\;                                       & \multicolumn{1}{l|}{} & \multicolumn{1}{l|}{} & \multicolumn{1}{l|}{} & \multicolumn{1}{l|}{} \\
\emph{-Theoretical}                               & 0                     & 1000                   & 0                     & 1000                   \\
\emph{-Phenomenological}                          & 12                    & 1000                   & 12                    & 1000                   \\ \hline
$m_{s_2}$ (Gev)\;\;\;                                                                             & \multicolumn{1}{l|}{} & \multicolumn{1}{l|}{} & \multicolumn{1}{l|}{} & \multicolumn{1}{l|}{} \\
\emph{-Theoretical}                               & 0                     & 1000                   & 0                     & 1000                   \\
\emph{-Phenomenological}                          & 6                    & 1000                   & 12                    & 1000                   \\ \hline
$v_h$ (GeV)                                           & 246                   & 246                   & 246                   & 246                   \\ \hline
$v_S$ (GeV)                                           & 0                     & 1000                   & 0                     & 1000                   \\ \hline
$v_A$ (GeV)                                           & 0                     & 0                   & 0                     & 1000                   \\ \hline
$a_1$ ($\mathrm{GeV}^3$)                              & $\mathrm{-10^8}$      & 0                     & \multicolumn{2}{c|}{n/a}                      \\ \hline
\end{tabular}
\end{center}
\caption{Range of parameters in the scans with and without phenomenological constraints for the dark matter and broken phase. In the dark matter phase, $m_{s_1}$ is the new visible scalar's mass and $m_{s_2}$ is the mass of the dark matter candidate whereas in the broken phase both are visible. The parameter $a_1$ is an input parameter
only in the dark matter phase.}
\label{scan_table}
\end{table}
As previously explained in~\cite{Coimbra:2013qq}, we use the VEVs as input parameters for numerical convenience in the scan. This explores the linearity of the vacuum condition in the couplings. Then, using the VEVs, the masses
and the angles from the mixing matrix as independent parameters, the vacuum conditions become a linear system for the dependent couplings which can be solved efficiently.

\subsection{RGE running with no phenomenological constraints}\label{sec:RGE_run_alone}
In this section we study the effect of the renormalization group running of couplings on the allowed parameter space of the theory. We first perform a dedicated scan  over the free parameters of the theory at the low scale where we apply: i) the tree level perturbative unitarity test (which is inbuilt, for any model, in the~\textsc{ScannerS} code -- see also~\cite{Coimbra:2013qq}), ii) the requirement that the electroweak minimum is the global one, and iii) that the potential is bounded from below. For each point which is accepted under these constraints we perform an RGE running of all couplings assuming their values are set at the electroweak scale, the $Z$ boson mass, i.e. $\mu=M_Z$. We let the evolution proceed up to the Planck scale, $M_{\rm Pl}\sim 10^{19}\,\mathrm{GeV}$, (if neither of the conditions discussed below is violated) and keep information on the cut-off scale (denoted henceforth the {\em stopping scale}) where one of the following occurs for the running couplings:
\begin{itemize}
\item Violation of the boundedness from below condition, Eq.~\eqref{eq:boundedness_below}. The quartic couplings reach values such that the potential of the theory acquires runaway directions for large field values.
\item Perturbative unitarity violation, i.e. any of the conditions in Eq.~\eqref{eq:unitaritycond} fails. Typically this is also related with the appearance of a Landau pole in one of the couplings. In practice we have checked that, for all points in our scan, the poles only appear in the quartic couplings so this condition automatically prevents them. 
\end{itemize}
The purpose of this separate study is to understand what are the cuts imposed by radiative effects at higher scales before introducing the phenomenological constraints.

We found that, for all points in our scan, the only couplings for which perturbative unitarity was violated before the Planck scale were $\lambda$ and $d_2$. As for boundedness from below, Eq.~\eqref{eq:boundedness_below}, all conditions except $d_2>0$ were violated in several points of the scan. In fact, we found that the final $d_2$, for all points in our scan, was always greater than the initial one. 
\subsubsection{Dark matter phase}
\begin{figure}
\begin{center}
\includegraphics[width=0.488\linewidth,clip=true,trim=35 40 20 50]{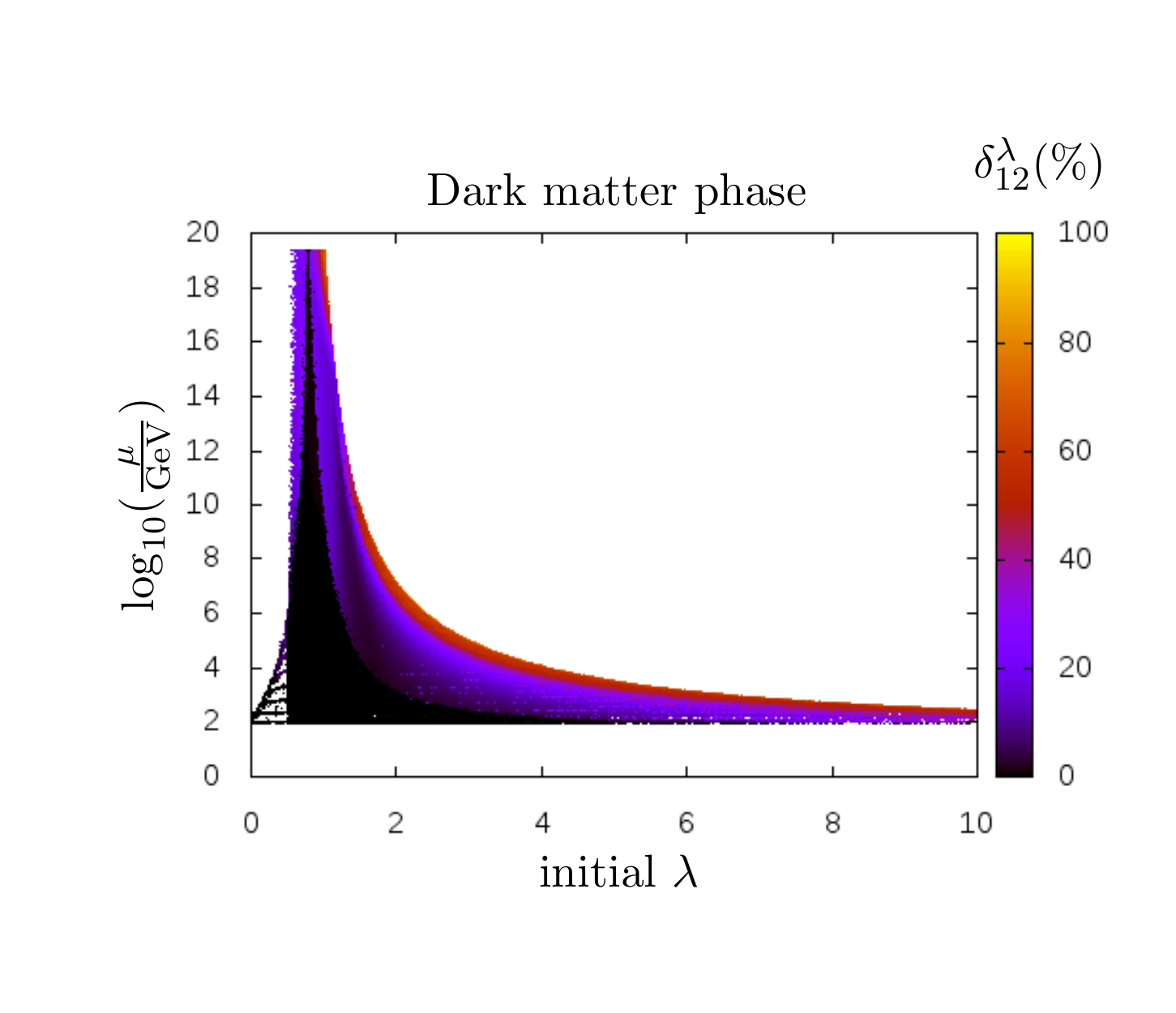}\hspace{0.02\linewidth}\includegraphics[width=0.488\linewidth,clip=true,trim=35 40 20 50]{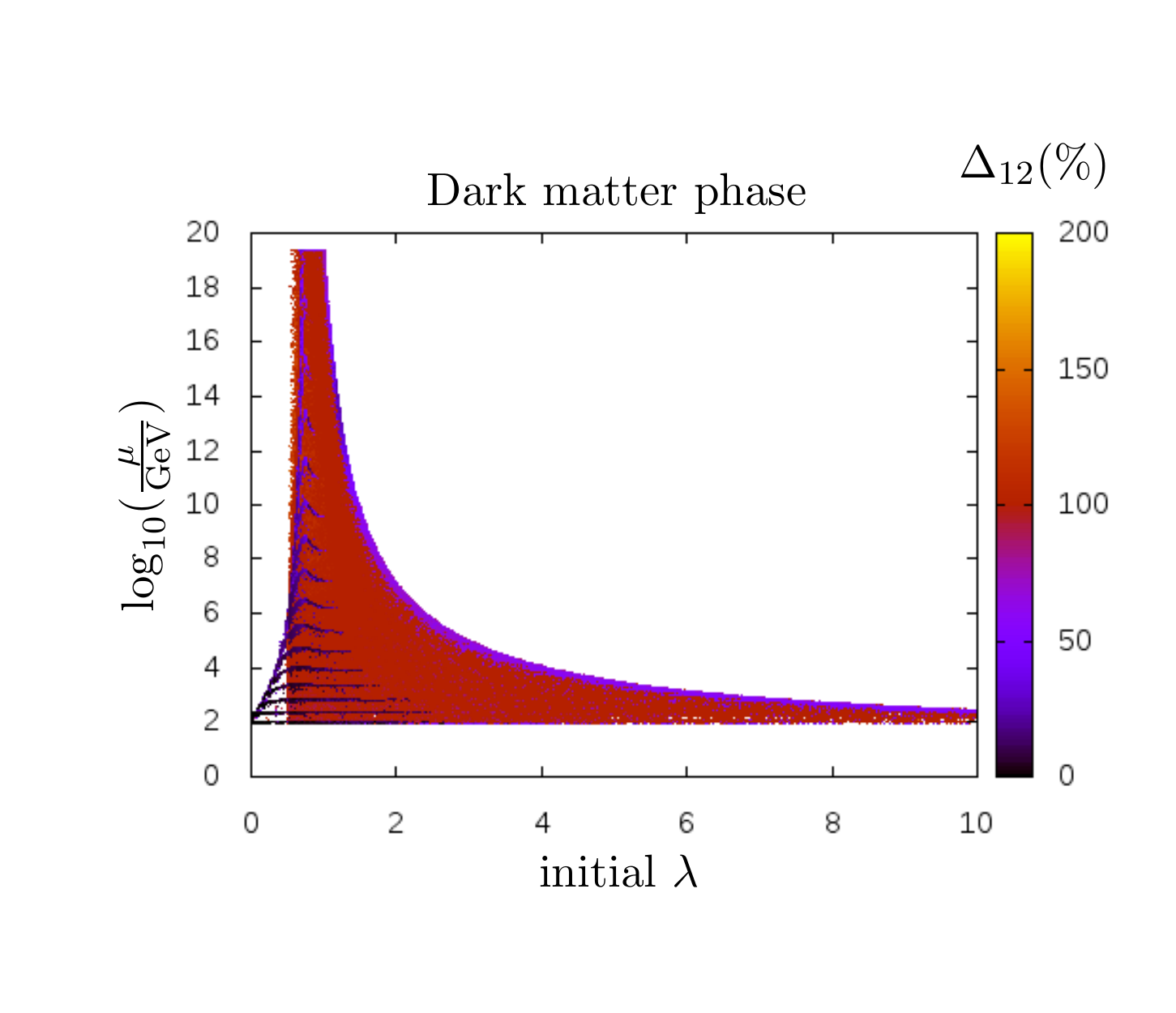}
\end{center}
\caption{{\em  Relative one-loop to two-loop error}: On the left panel we show a projection of the parameter space where, for each point, we represent the scale at which the evolution stopped as a function of the initial condition for $\lambda$ at the electroweak scale. The color gradation corresponds to the one to two-loop relative error parameter for $\lambda$, i.e.  $\delta^\lambda_{12}$ as defined in section \ref{sec:relative_error}. The right panel shows the same projection, with the global error parameter $\Delta_{12}$ in the color scale. In both panels, points with lowest error parameters are overlaid on top of points with higher error parameters. 
}
\label{1Loop2LoopError}
\end{figure}
In figure~\ref{1Loop2LoopError} we start by highlighting the allowed parameter space for the stopping scale as a function of the initial $\lambda$ coupling. The behavior of the stopping scale as a function of $\lambda$ will turn out to be crucial in explaining our results. We also include the measure of the one-loop error in the color scale for the $\lambda$ coupling ($\delta_{12}^\lambda$, on the left) and for the global error ($\Delta_{12}$, right). The figure contains a projection of points where the stopping scale (vertical axis) is compared with the initial condition for $\lambda$ at the electroweak scale. We see that there is only a narrow range of $\lambda$ couplings, between $\lambda\simeq 0.5$ and $\lambda\simeq 1$,  for which the Planck scale is reached with none of the stability/perturbative unitarity conditions being violated.

 Regarding the error measures displayed in the color scale, the left panel shows that there are points everywhere with a small one-loop truncation error for $\lambda$  (i.e. $\delta_{12}^\lambda$). Only points close to the boundaries, which correspond to points where perturbative unitarity is violated (signaling that some coupling is evolving to a Landau pole) have truncation errors close to $50\%$. Note however that our results contain full two-loop order corrections, whereas this error measure is the difference between the one-loop and the two-loop calculation. Thus we expect our result to have a smaller error. 
 Regarding the global one-loop error ($\Delta_{12}$ in the right panel) it shows that there are always some couplings with order one error which supports the importance of using the two-loop approximation, especially if one wants to study quantities involving the new scalar couplings which are absent in the SM. Finally, one should note that we have checked that the quartic couplings and the top Yukawa coupling, which are the ones that determine if the stability/perturbative unitarity conditions are violated, generically have one-loop errors smaller than $10\%$. So we expect our results to be robust against higher loop corrections.

Moving on to the discussion of physical quantities which are in principle directly measurable, such as masses and couplings (or equivalently, in this model, mixings), we first note an important characteristic of our study: we present, simultaneously, two possible scenarios within the dark matter phase study. One where the \textit{new} visible mixing scalar has a mass smaller than the known SM-like Higgs (we refer to this as the \textit{light scenario}) and the other one where the \textit{new}  scalar is heavier than the SM-like Higgs (which we call the \textit{heavy scenario}).

\begin{figure}
\begin{center}
\includegraphics[width=0.49\linewidth,clip=true,trim=40 35 25 45]{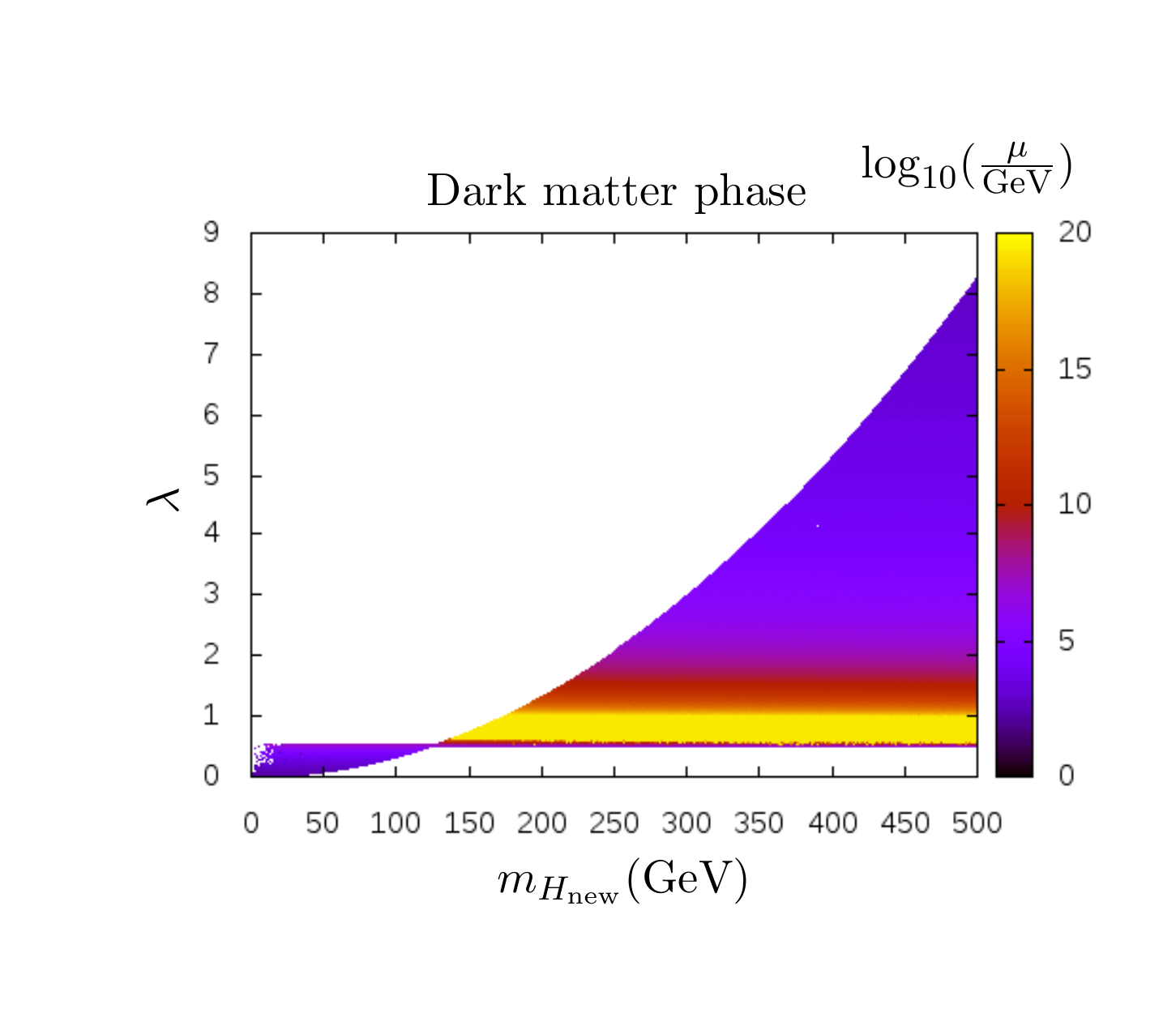}\hspace{0.01\linewidth}\includegraphics[width=0.49\linewidth,clip=true,trim=40 35 25 45]{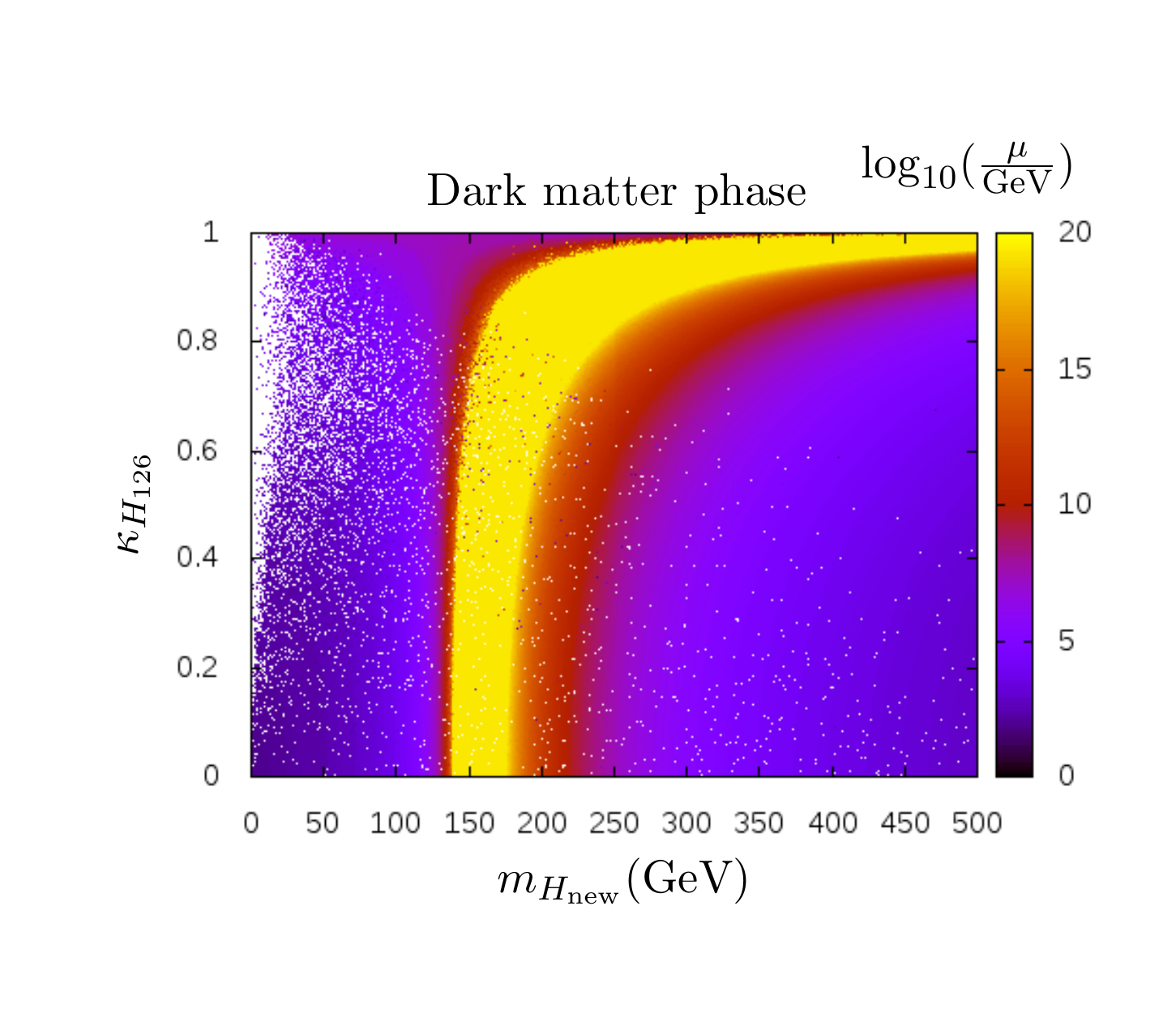}
\end{center}
\caption{{\em Dark matter phase}:  On the left panel we present a projection on the $(m_{H_{\rm new}},\, \lambda)$ plane where the color gradation corresponds to the stopping scale for each point. The right panel shows another projection with the SM-like Higgs coupling,  $\kappa_{H_{126}}$, on the vertical axis. In both panels, points with higher stopping scales are overlaid on top of points with lower stopping scales.}
\label{onlyRGE_mvisible_lambda}
\end{figure}
In figure~\ref{onlyRGE_mvisible_lambda} we present a projection on the $(m_{H_{\rm new}},\, \lambda)$  plane (left) and another projection, in the right panel,  
where $\lambda$ is replaced by the reduced SM-like Higgs coupling ($\kappa_{H_{126}}$) on the vertical axis. The color gradation corresponds to the stopping
scale for each point (i.e. the maximum scale up to which the theory remains stable for the given point). In the dark matter phase $m_{H_{\rm new}}$ stands for the mass of the non-SM-like non-DM particle (i.e. which mixes with the SM Higgs). As discussed in figure~\ref{1Loop2LoopError} only values
of $\lambda$ between $\simeq 0.5$ and $\simeq 1$ remain valid up to the Planck scale. This range is almost independent of the mass despite the clear correlation in the boundary region of the left panel scan. In the right panel we see that imposing stability conditions up to the Planck scale imposes a relatively sharp lower bound on 
the mass of the new scalar state at around $m_{H_{\rm new}}\simeq 140$~GeV. Furthermore, this cut-off is quite sharp in the sense that even if one requires 
a stopping scale as low as $\sim 10^{7}$~GeV we still get a bound of about $130$~GeV
\footnote{Note that we haven't yet combined this with the phenomenological constraints which tend to push this bound to higher values since $\kappa_{H_{126}}$ is typically closer to 1.}.
Furthermore the right panel shows that, in this model, not only we need a heavier scalar to stabilize the SM but also that it has to mix with the SM Higgs. This is clear if we recall, that for the blue points at lower masses, there are cases where the DM particle is much heavier. Thus the DM candidate does not play an important role here. Also note that for generic values of fixed $\kappa_{H_{126}}$ (away from one), there is in fact also an upper bound, i.e. the stability region (yellow) is in an interval of masses between $\sim 140$~GeV and $\sim 250$~GeV, whereas as we move towards $\kappa_{H_{126}}\rightarrow 1$ (i.e. for weaker mixing between the two non-DM states), both ends of the interval move to larger values. This is consistent with the claim that the dark matter candidate is not contributing to the stabilization of the theory, since in the limit $\kappa_{H_{126}}\rightarrow 1$ all new scalar states are effectively dark. From now on we refer to the yellow band of points in figure~\ref{onlyRGE_mvisible_lambda} as the ``{\em stability band}''. We will see in the broken phase that this band is also important to explain some boundaries of the stability band of each of the two mixing scalars.

The origin of this lower bound on the mass of the new heavy scalar is in fact related to the local minimum conditions combined with fixing the $\simeq 126$~GeV Higgs mass within $3\sigma$ of the measured central value. Considering the dark matter phase as an example, one can check that the (linear) minimum conditions provide two independent constraints. In addition we have another condition that fixes the mass of one of the mixing states to be $m_{H_{126}}$, the mass of the observed Higgs. Using all such conditions one finds that~\footnote{Forcing the potential to be in a global minimum implies \mbox{$v_S \, v\, \delta_2 = \pm 2 \, \kappa_{H_{126}} \, \kappa_{H_{\rm new}} (m_{H_{\rm new}}^2-m_{H_{126}}^2)$}, and therefore $\lambda$ is always real.}
\begin{equation}
\lambda =\frac{m_{H_{\rm new}}^2+m_{H_{126}}^2}{v^2}\pm \sqrt{\left[\dfrac{m_{H_{\rm new}}^2-m_{H_{126}}^2}{v^2}\right]^2-\left( \frac{v_S}{v}\delta_2\right)^2}\; .
\end{equation}
In the limiting case of no mixing ($v_S\rightarrow 0$) we obtain $\lambda=2m_{H_{\rm new}}^2/v^2$ or $\lambda=2m_{H_{126}}^2/v^2$, which are precisely the two boundary lines that we see in figure~\ref{onlyRGE_mvisible_lambda}, left panel~\footnote{One should note at this point that we have chosen to display only the mass range up to $500$~GeV because there is nothing qualitatively different that we found for larger masses, i.e. all the interesting Physics can be observed in this range}. Furthermore, an expression can also be obtained for the mixing matrix element, or equivalently
\begin{equation}
\kappa_{H_{126}}^2=\dfrac{1}{2}\left[1\pm\frac{\sqrt{\left(m_{H_{\rm new}}^2-m_{H_{126}}^2\right)^2-\left(v\,v_S\delta_2\right)^2}}{m_{H_{\rm new}}^2-m_{H_{126}}^2}\right]\; .
\end{equation}
One can identify the upper boundary of the stability band of figure~\ref{onlyRGE_mvisible_lambda}, right panel, with a solution with the minus sign. Then noting that in the stability region $m_{H_{\rm new}}^2-m_{H_{126}}^2>0$, we conclude  that $\kappa_{H_{126}}^2=1$ is only possible when $m_{H_{\rm new}}^2\rightarrow+\infty$.\footnote{We have also verified in our scans that $v_S\delta_2\neq 0$ in the stability band.}

Summarizing, the main conclusions we can draw from figure~\ref{onlyRGE_mvisible_lambda} are
\begin{itemize}
\item Stability up to the Planck scale imposes a lower bound on the mass of the new scalar state, $m_{H_{\rm new}} \gtrsim 140$~GeV. The closer we move to the SM-like limit, the wider the allowed range  in the stability band, but at the same time the lower bound on the mass moves to larger values. 
\item The light scenario only survives up to a scale of a few TeV, in which case new Physics would be needed at relatively low energy scales.
\end{itemize}

\begin{figure}
\begin{center}
\includegraphics[width=0.49\linewidth,clip=true,trim=40 40 20 50]{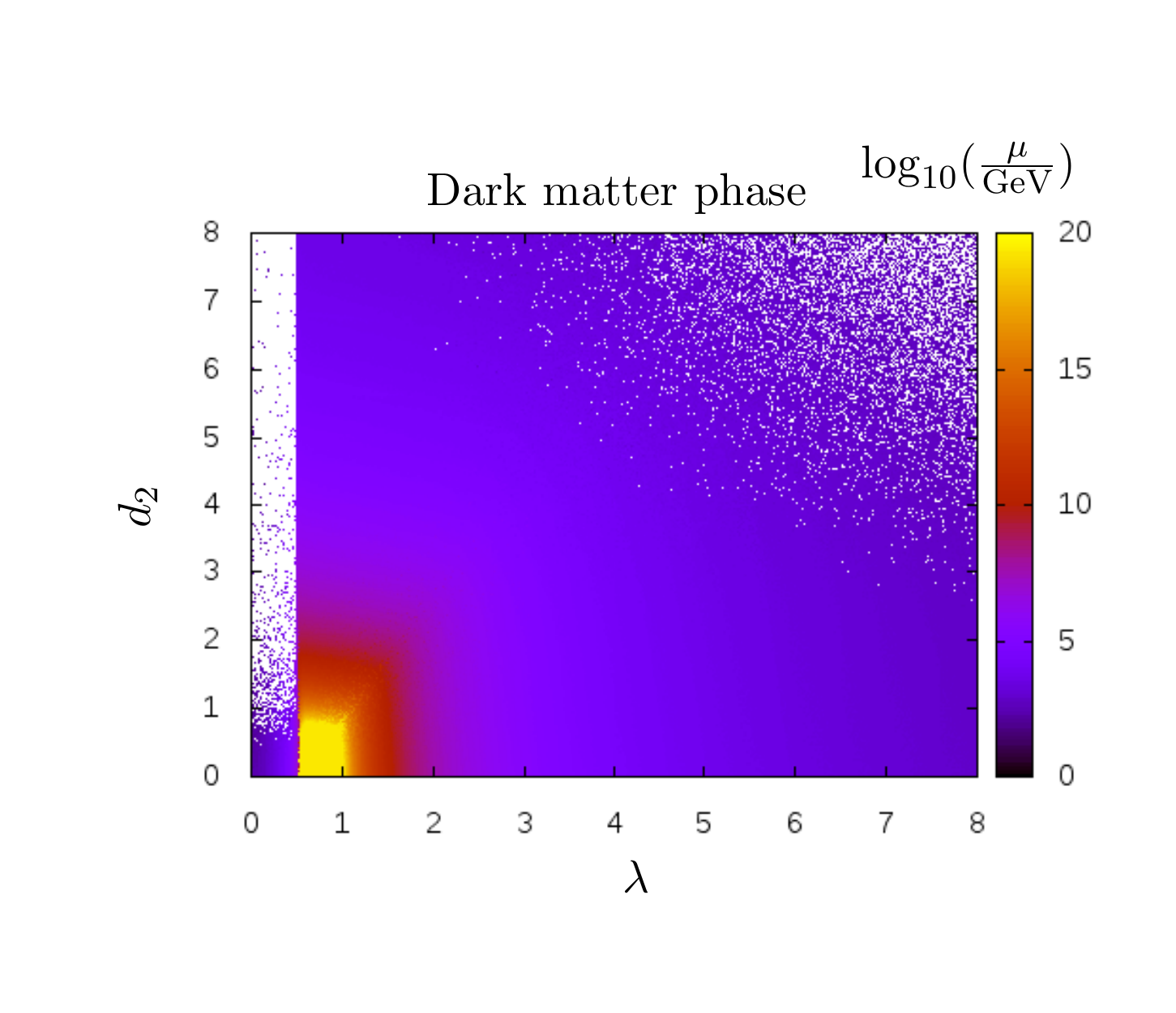}\hspace{0.01\linewidth}\includegraphics[width=0.49\linewidth,clip=true,trim=40 40 20 50]{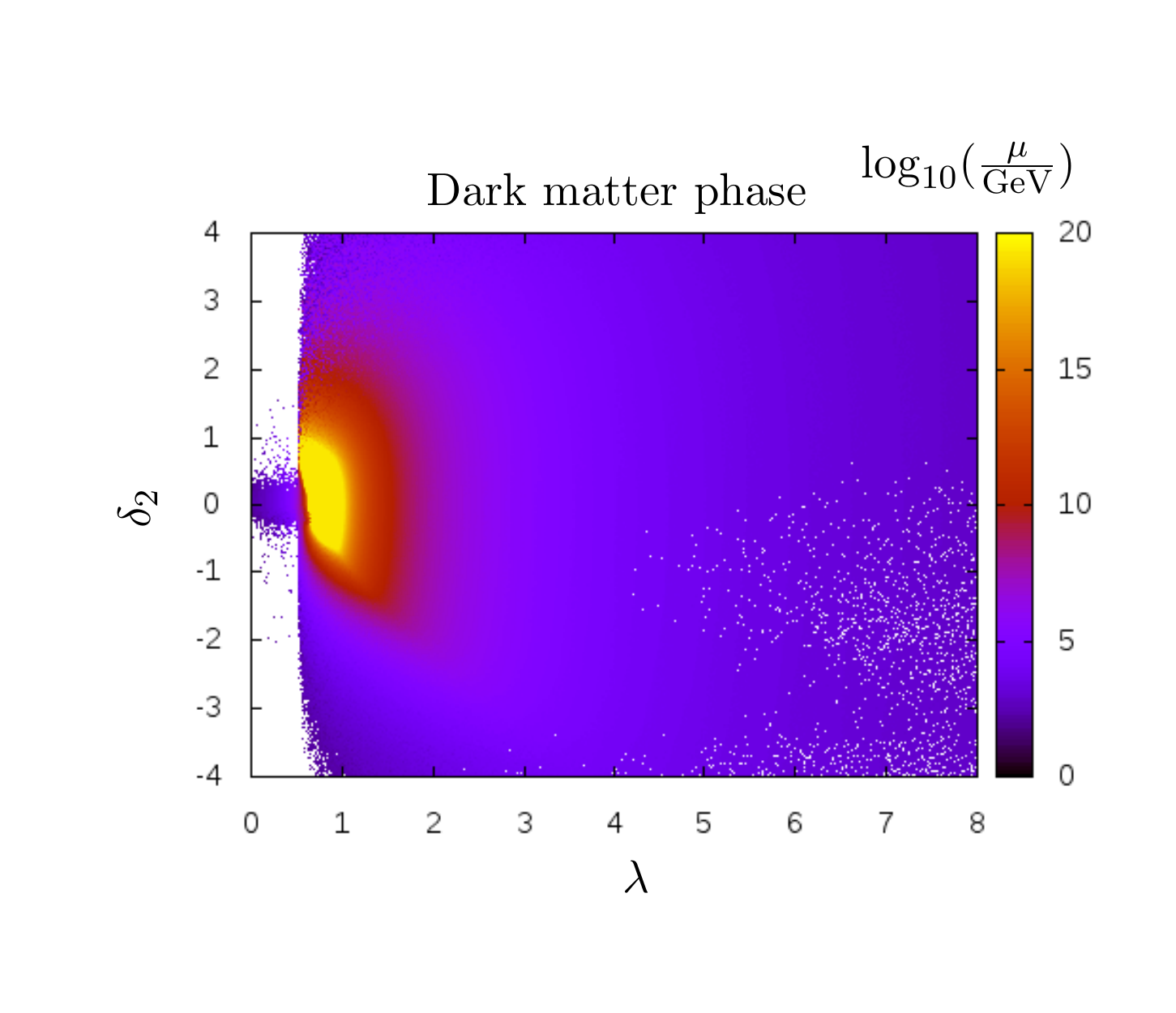}
\end{center}
\caption{{\em Dark matter phase}: the left panel shows a projection on the $(\lambda, d_2)$ plane where again the color gradation corresponds to the stopping scale for each point. The right panel shows another projection on the $(\lambda, \delta_2)$ plane. The color gradation corresponds to the stopping
scale for each point. In both panels, points with higher stopping scales are overlaid on top of points with lower stopping scales.}
\label{onlyRGE_DM_others}
\end{figure}
In figure~\ref{onlyRGE_DM_others} we present a projection on the $(\lambda, d_2)$ plane (left panel) and another projection on the $(\lambda, \delta_2)$ plane (right panel). The plots clearly show that stability up to the Planck scale requires not only $\lambda$ below 1 but also that both $|\delta_2| \lesssim 1$ and $d_2 \lesssim 1$. As we move away from this order $1$ bounds we quickly reach a region where stability holds only up to a few TeV.

\subsubsection{Broken phase}

We now move on to discuss the broken phase where the RGE running has similar effects on the parameters. In fact, given that the RGEs are independent of the type of minimum, the only difference between the two scenarios (for the purpose of the evolution) is that the allowed regions for the initial data obtained at the low scale are different (i.e. the type of minimum is different for each case). 

Before discussing the results we should clarify the notation and the interval of variation for the mass of each particle. All particles have the same quantum numbers and therefore they all mix. One of the scalars is the SM-like Higgs boson detected at the LHC, with a mass of $\simeq 126$~GeV. The remaining two scalar masses are denoted as $m_{H_{\rm light}}$ and $m_{H_{\rm heavy}}$ such that $m_{H_{\rm light}} < m_{H_{\rm heavy}}$. These scalars masses can be both lighter, both heavier, or one lighter and one heavier than the  $\simeq 126$~GeV one.  

\begin{figure}
\begin{center}
\includegraphics[width=0.49\linewidth,clip=true,trim=40 30 20 50]{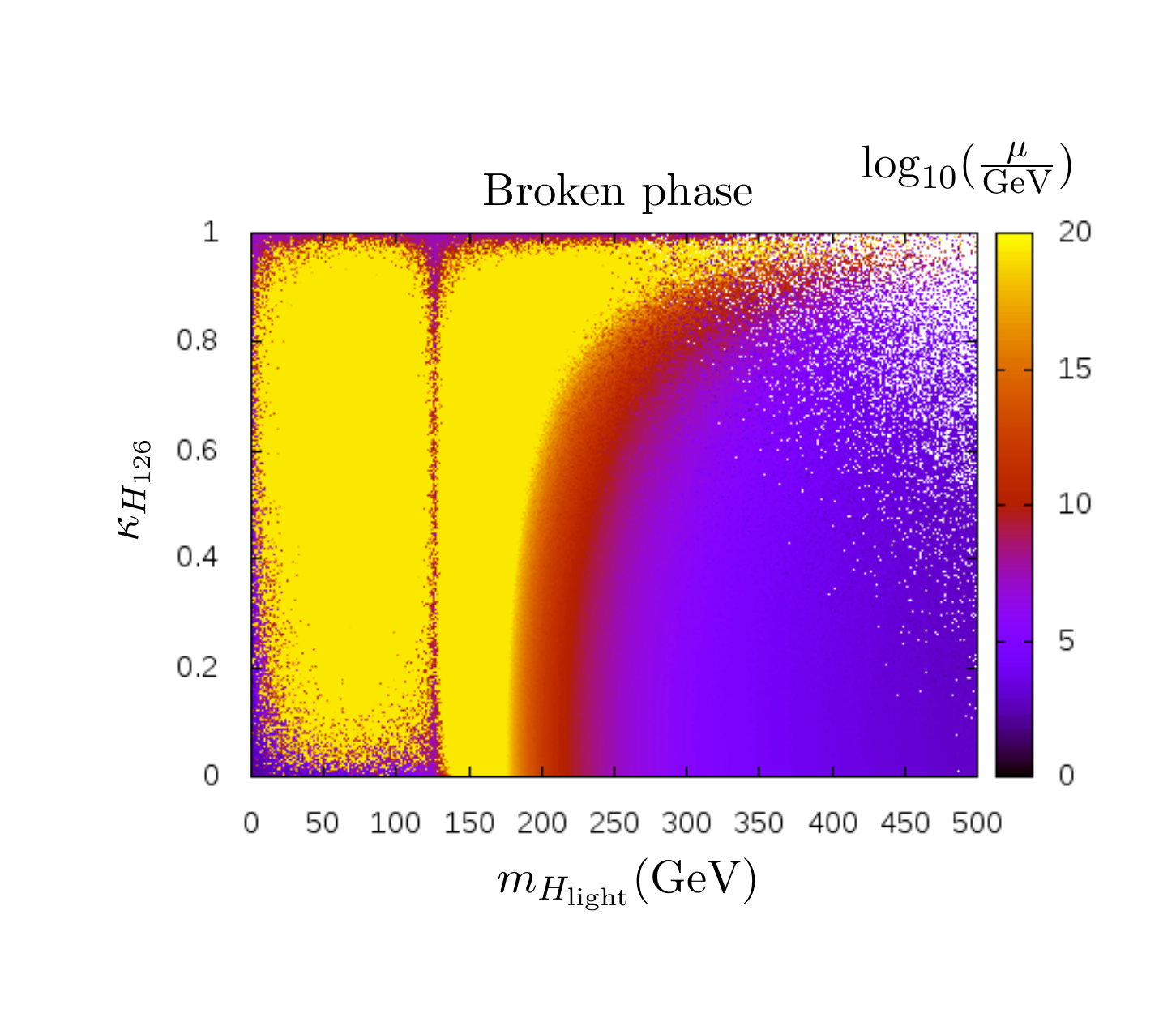}\hspace{0.01\linewidth}\includegraphics[width=0.49\linewidth,clip=true,trim=40 30 20 50]{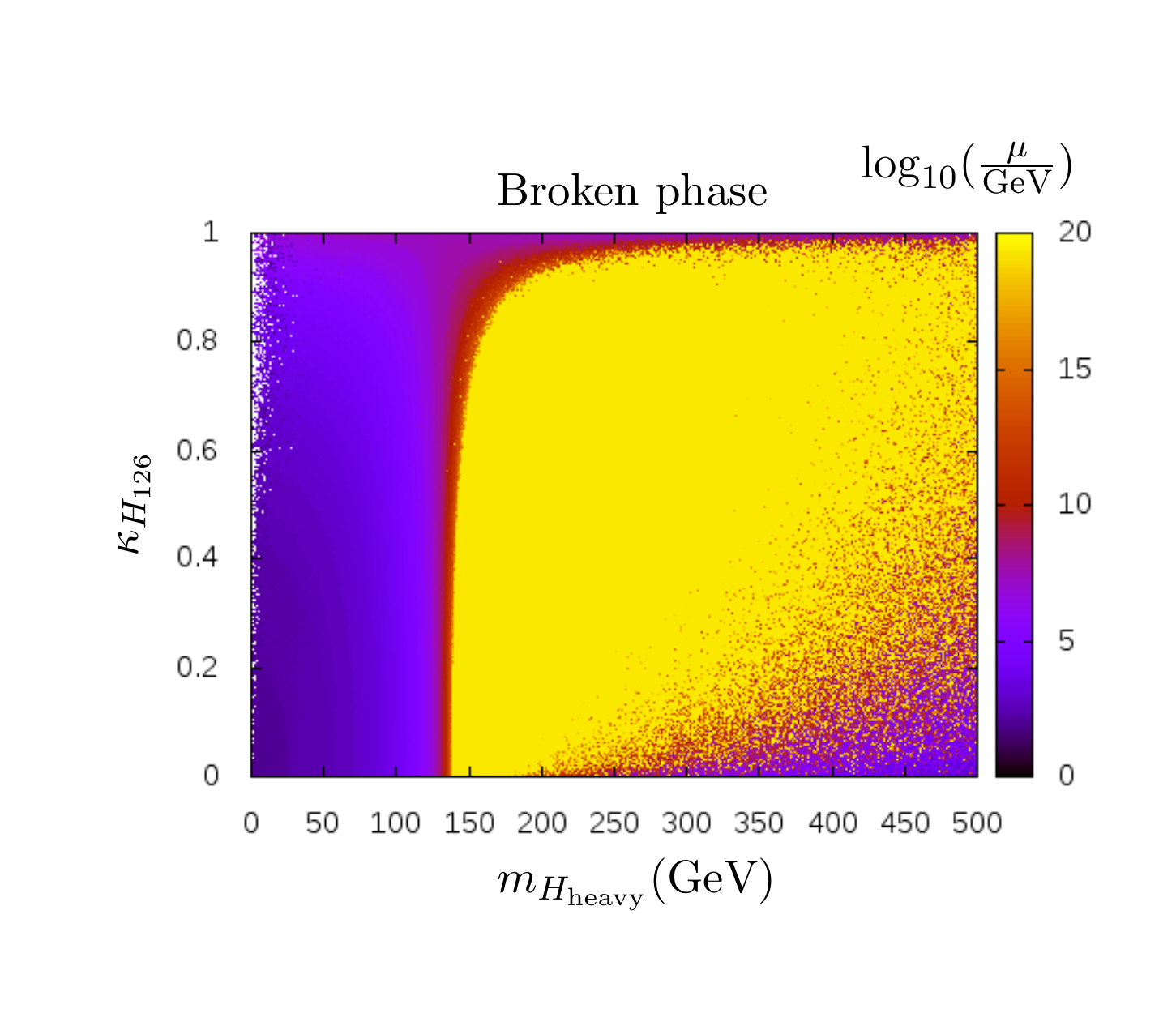}
\end{center}
\caption{{\em Broken phase}:  On the left panel we present a projection along the $(m_{H_{\rm light}} ,\, \kappa_{H_{126}})$ plane 
while on the right panel we show a projection along the $(m_{H_{\rm heavy}} ,\, \kappa_{H_{126}})$. The color gradation corresponds to the stopping
scale for each point. In both panels, points with higher stopping scales are overlaid on top of points with lower stopping scales.
}
\label{onlyRGE_mheavy_mlight}
\end{figure}
In figure~\ref{onlyRGE_mheavy_mlight}  we present a projection on the $(m_{H_{\rm light}},\, \kappa_{H_{126}})$ plane (left) and another one on the $(m_{H_{\rm heavy}},\, \kappa_{H_{126}}) $ (right). Again, the color gradation corresponds to the stopping
scale for each point. Similarly to the dark matter phase, we can find a lower bound, but only on the heavier mixing scalar, at a mass of about $m_{H_{\rm heavy}} \simeq 140$ GeV. Furthermore we observe that, in fact, the left boundary for the yellow region in the right panel is exactly the same as the left boundary of the stability band of figure~\ref{onlyRGE_mvisible_lambda}. On the other hand, for the lighter scalar, the yellow region's boundary on the right corresponds precisely to the right boundary of the stability band of figure~\ref{onlyRGE_mvisible_lambda}. 
Thus we conclude that also in this scenario, we need at least one scalar with mass larger than $\sim 140$~GeV to stabilize the theory up to the Planck scale. Observe that a scenario where the mass of $H_{\rm light}$ is smaller than the SM-like Higgs mass, can be stable up to Planck scale (as long as $H_{\rm heavy}$ is in the stability band), which could be a phenomenologically interesting scenario at the next run of the LHC.

One can also show that in the limit where one of the new scalars ($i=H_{\rm light}$ or $H_{\rm heavy}$) is very weakly coupled  to the remaining SM particles ($R_{ih} \to 0$) we recover figure~\ref{onlyRGE_mvisible_lambda} for the other new scalar as expected. This property is quite interesting for it means that  there is a continuous limit connecting to the DM phase. The closer $R_{ih}$ is to zero the more indistinguishable the two phases become. 
Whether we may distinguish between an exact dark matter phase or a quasi-dark limit of the broken scenario will depend on how close to the vicinity of the dark phase limit the model is and it would require a detailed analysis\footnote{In a previous work~\cite{Coimbra:2013qq} we have discussed some scenarios where the two phases could be distinguished experimentally.}. 

It is interesting to note that this limit is not possible in some simple extensions of the SM such as the 2HDM. In fact, the 2HDM with an exact $\mathbb{Z}_2$
symmetry (one doublet is odd and the other is even under $\mathbb{Z}_2$) also allows for a dark matter phase (known as the inert model~\cite{Deshpande:1977rw, Barbieri:2006dq}) and a spontaneously broken phase. However in that case the continuous limit from the broken phase to the dark phase (i.e. by taking the limit of vanishing VEV) is not allowed if perturbative unitarity~\cite{Gorczyca:2011he} constraints are imposed. Therefore only the inert version contains a dark matter candidate.
 
\subsection{Phenomenological constraints}\label{sec:constraints_pheno}

In this section we present the results from our phenomenological scans. We have already described in detail the theoretical constraints. We will now describe what we have included
as bounds coming from various experimental sources. Part of our procedure has been described in detail in a previous work~\cite{Coimbra:2013qq} which we will often refer to for details.

\subsubsection{Electroweak precision observables}

We start with the electroweak precision observables, using the $S,T,U$ variables~\cite{Maksymyk:1993zm,Peskin:1991sw}. In the singlet extension of the SM the new contributions to the radiative corrections of the $W$ and $Z$ bosons self energies (respectively $\Pi_{WW}(q^2)$ and $\Pi_{ZZ}(q^2)$), appear only through the states which mix with the SM Higgs doublet fluctuation.
The general expressions are available for example in~\cite{Barger:2007im}. 
The relative shift between the oblique observables calculated in the BSM model and the reference SM , $\Delta \mathcal{O}_i\equiv \mathcal{O}_i-\mathcal{O}^{SM}_i\rightarrow \left(\Delta S,\Delta T,\Delta U\right)$, are required to be consistent with the electroweak fit within a 95\% C.L. ellipsoid of the best fit point $\Delta\mathcal{O}_i^{(0)}$, i.e.
\begin{equation}\label{eq:oblique_test}
\Delta \chi^2\equiv\sum_{ij}\left(\Delta \mathcal{O}_i-\Delta \mathcal{O}^{(0)}_i\right){\left[(\sigma^2)^{-1}\right]}_{ij}\left(\Delta \mathcal{O}_j-\Delta \mathcal{O}^{(0)}_j\right)<7.815\; .
\end{equation}
The covariance matrix is defined using the correlation matrix, $\rho_{ij}$, and the standard deviation of each parameter, $\sigma_i$, 
through the expression $\left[\sigma^2\right]_{ij}\equiv \sigma_i\rho_{ij}\sigma_j$. In order to test these observables we use the latest SM global fit from the Gfitter collaboration~\cite{Baak:2012kk} with a reference Higgs mass $m_{h,\rm ref}= 126\, {\rm GeV}$ and top mass $m_{t,\rm ref}=173\, {\rm GeV}$. The values of the best fit point (i.e. the shift with respect to the reference model at the origin) and the correlation matrix are respectively
\begin{equation}
\begin{array}{l}
\Delta S^{(0)} = 0.03 \pm 0.10\\ 	
\Delta T^{(0)} = 0.05 \pm 0.12 \\	
\Delta U^{(0)} = 0.03 \pm 0.10
\end{array}\;,\qquad 
\rho_{ij}=\left(\begin{array}{ccc} 1&0.891 &-0.540\\0.891&1 &-0.803 \\-0.540&-0.803 &1\end{array}\right) \; .
\end{equation}

\subsubsection{Collider bounds}
\label{sec:collider_bounds}
The ranges we have chosen for the scalar masses in our scans, table~\ref{scan_table}, are such that the spectrum contains a Higgs boson of mass $\simeq 126$~GeV to explain the observed signals at the LHC. Furthermore the LHC (and previous colliders such as LEP and the Tevatron), also provide strong experimental limits on new scalars. To apply all available experimental constraints on the Higgs couplings and on new scalars, we make use of \textsc{ScannerS}' external interfaces with other codes. $95\%$ C.L. exclusion limits were applied using \textsc{HiggsBounds}~\cite{Bechtle:2013wla}, and~\textsc{HiggsSignals}~\cite{Bechtle:2013xfa} was used to check for consistency  with the observed Higgs boson at the LHC (i.e. to obtain the probability for the fit of the model point to all known signal data). 

\textsc{HiggsBounds/Signals} needs as input all branching ratios (BR) and decay widths of the new scalar particles to all possible final states, as well as cross-section ratios for all possible production modes (i.e. normalized to SM cross-section). This information is then used to compute experimental quantities such as the signal rates
\begin{equation}
\mu_i =\dfrac{\sigma_{\rm New}(H_i){\rm Br}_{\rm New}\left(H_i\rightarrow X_{\rm SM}\right)}{\sigma_{\rm SM}(h_{SM}){\rm Br}_{\rm SM} \left(h_{SM}\rightarrow X_{\rm SM}\right)} \, \,  \; .
\label{mu}
\end{equation} 
Here $\sigma_{\rm New}(H_i)$ and $\sigma_{\rm SM}(h_{SM})$ are, respectively,  the Higgs production cross sections for $H_i$ and for a SM Higgs with mass $m_{H_i}$;
$ {\rm Br}_{\rm New}\left(H_i\rightarrow X_{\rm SM}\right)$ is the $H_i$ BR to SM particles  while ${\rm Br}_{\rm SM} \left(h_{SM}\rightarrow X_{\rm SM}\right)$ is the SM Higgs BR (again evaluated at the mass $m_{H_i}$).

As previously discussed in Sect.~\ref{sec:TheModel}, each scalar couples to SM particles exactly as the SM Higgs with a suppressing factor given by a mixing matrix element, $R_{ih}$. Furthermore, because there are new scalars involved, the BRs have to be re-weighted to account for new decay channels to new scalar particles whenever they are kinematically allowed. In this case the signal rates then become
\begin{equation}\label{eq:mu_xSM}
\mu_i = R_{ih}^2 \dfrac{R_{ih}^2\Gamma(h_{SM}\rightarrow  X_{\rm SM})}{R_{ih}^2 \Gamma(h_{SM}\rightarrow  X_{\rm SM})+\sum\Gamma(H_i\rightarrow \mathrm{new\,scalars})}  \, .
\end{equation}
The latter reduces to $\mu_i = R_{ih}^2$, i.e. to the cross-section ratio, whenever the given scalar is not allowed to decay to other new scalars. Note that because the experimental results are given in terms of rates there is no need to calculate the Higgs production cross sections. 

The experimental tables in \textsc{HiggsBounds} only contain (so far) searches for one scalar decaying to two identical scalars. In our model this  proceeds via $H_i\rightarrow H_j H_j$, corresponding to the partial width
\begin{equation}
\Gamma\left(H_i\rightarrow H_j H_j\right) = \dfrac{g^2_{ijj}}{32\pi m_{i}}\sqrt{1-\dfrac{4m_{j}^2}{m_{i}^2}}\; ,
\end{equation}
where $g_{ijj}$, is the coupling between scalars $i,j,j$ and $m_{j}$ is the mass of the scalar state $H_j$ (see also~\cite{Coimbra:2013qq}). 

Finally we must note that we have used \textsc{HiggsBounds-4.1.3} which includes data both for  visible and invisible decay searches at colliders, and \textsc{HiggsSignals-1.2.0} which contains all the LHC Higgs measurements to date.

\subsubsection{Dark Matter constraints}

\begin{itemize}
\item {\bf Relic density:} For the dark matter phase, we used the \textsc{MicrOmegas}~\cite{Belanger:2014hqa} software package to calculate the relic density $\Omega_A h^2$ for the dark matter particle, $A$. We then reject points for which $\Omega_A h^2$ is larger than the upper $3\sigma$ band ($\Omega_ch^2+3\sigma$) of the combination of measurements from the WMAP and Planck satellites~\cite{Ade:2013zuv,Hinshaw:2012aka} 
\begin{eqnarray}
\Omega_c h^2=0.1199 \pm 0.0027 \; .
 \label{eq:relic_density}
\end{eqnarray}

\item {\bf Direct detection - nucleon scattering cross-section:} Another constraint to be applied to the DM phase comes from limits obtained in experiments attempting to detect directly the dark candidate. Such experiments place bounds on the spin-independent scattering cross section of weakly interacting massive particles (WIMPS) on nucleons. Using the procedure described in~\cite{Coimbra:2013qq}, we compute the scattering cross section for the dark scalar with \textsc{MicrOmegas} which is then re-weighted by the factor\footnote{This is such that it is taken into account that the dark candidate cannot explain all observed relic density if smaller than the Planck/WMAP measurements.} $\Omega_A/\Omega_c$. Finally, the point is rejected if this prediction is larger than the upper bound set by the \texttt{LUX2013} collaboration~\cite{Akerib:2013tjd}.
\end{itemize}

\subsection{Discussion}
With all the previous theoretical and experimental constraints taken into account we now move on to the discussion of the parameter space that is still allowed for each of the two phases of the model.
\subsubsection{Dark Matter phase}
In this section we analyze the results for the dark matter phase. We start by applying the collider bounds at $1$, $2$ and $3 \sigma$. 
\begin{figure}
\begin{center}
\includegraphics[width=0.49\linewidth,clip=true,trim=40 40 30 30]{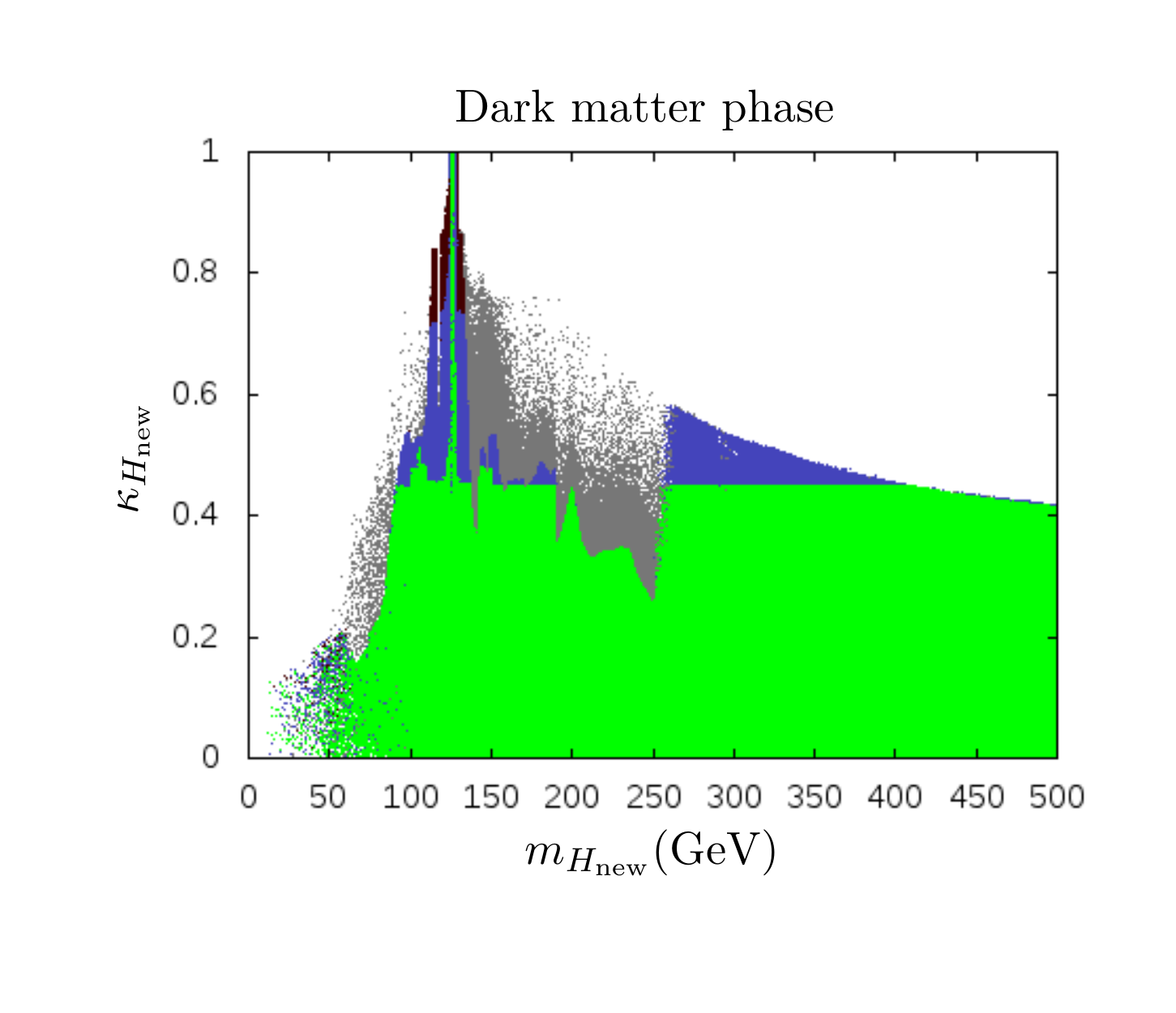}\hspace{0.01\linewidth}\includegraphics[width=0.49\linewidth,clip=true,trim=40 40 30 30]{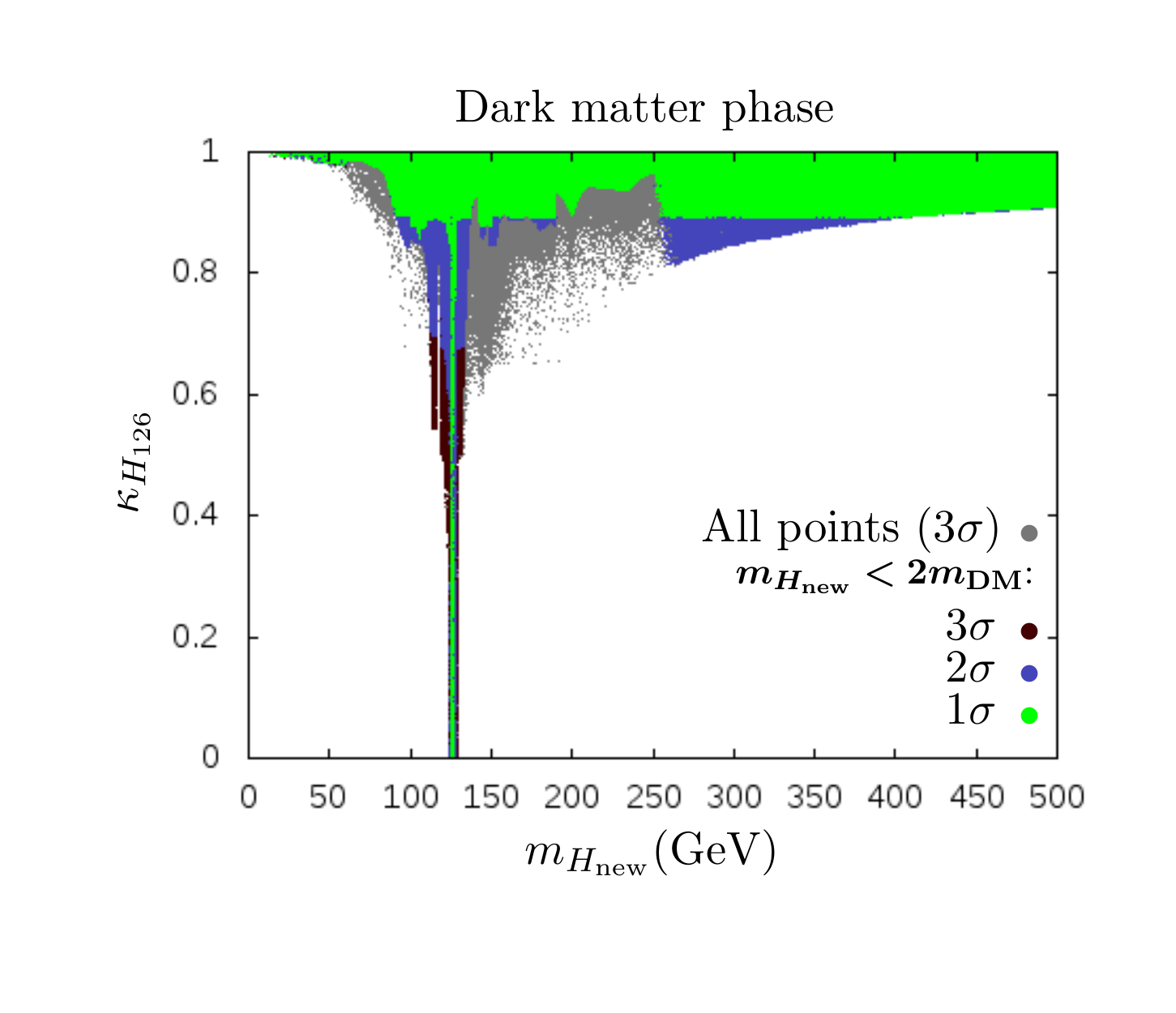}
\end{center}
\caption{{\em Dark matter phase}: We show projections of the new visible scalar mass ($m_{H_{\rm new}}$) versus its coupling ($\kappa_{H_{\rm new}}$) on the left and versus the observed Higgs coupling ($\kappa_{H_{\rm 126}}$) on the right. 
The bottom layer (grey) is the full scan of points allowed within $3\sigma$. The remaining layers (overlaid on top in the order, $3\sigma,2\sigma,1\sigma$) contain the cut that $H_{\rm new}$ is not allowed to decay to a pair of dark matter particles.}
\label{m_Mix_DM_Pvalue}
\end{figure}
In figure~\ref{m_Mix_DM_Pvalue}, left, we present a projection on the plane of the new visible scalar mass versus its coupling, and on the right the same but against the observed Higgs coupling. As expected, close to the observed Higgs mass ($\simeq 126$ GeV) all values of the Higgs couplings are allowed. This is because in this scenario the coupling is shared by the two states such that their squares add up to one corresponding to the SM Higgs coupling. This scenario was discussed as the twin peak Higgs in the context of the singlet model in~\cite{Heikinheimo:2013cua, Ahriche:2014cpa}. 
One would expect that there would be more allowed points at $2$ and $3 \sigma$. However, the electroweak precision constraints shrink the allowed region to $\kappa_{H_{126}} \gtrsim 0.8$ for masses $\gtrsim 250$~GeV.

\begin{figure}
\begin{center}\includegraphics[width=0.49\linewidth,clip=true,trim=40 30 10 30]{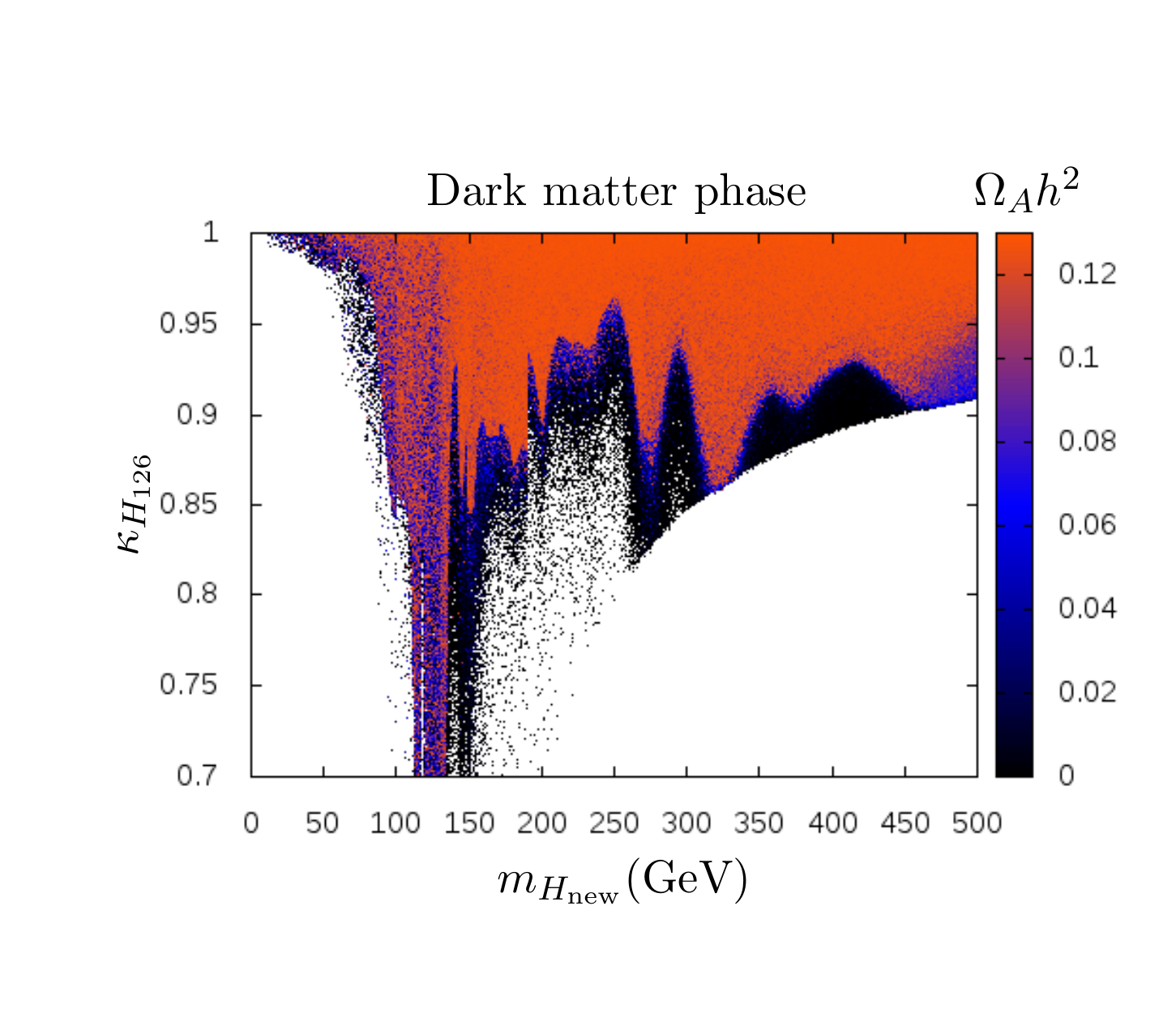}\hspace{0.01\linewidth}\includegraphics[width=0.49\linewidth,clip=true,trim=40 30 10 30]{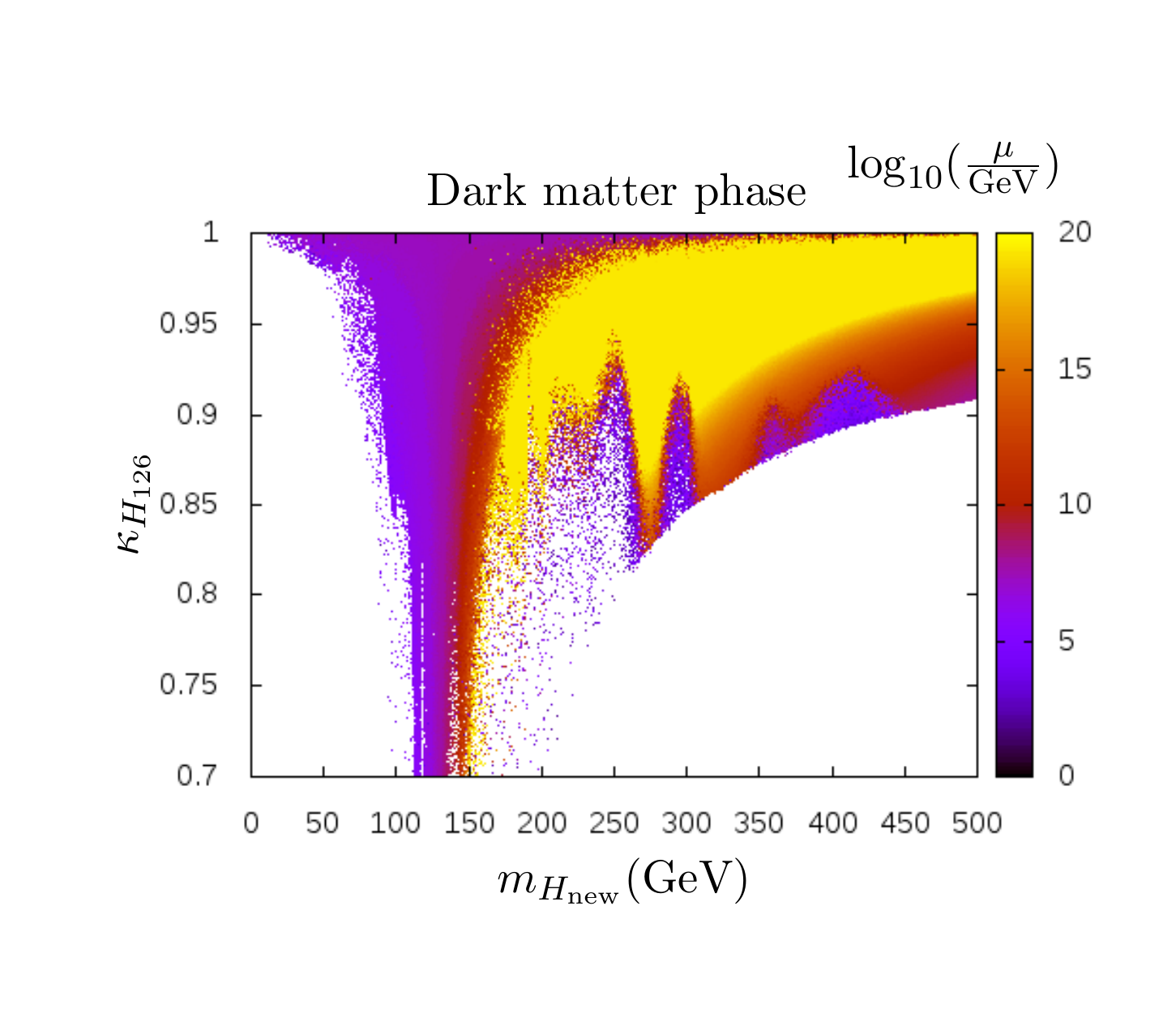}
\end{center}
\caption{{\em Dark matter phase}: Observed Higgs coupling, $\kappa_{H_{126}}$ , as a function of the new visible scalar mass. On the left panel the color gradation corresponds to the dark matter relic density, whereas on the right panel it represents the scale at which the evolution has stopped. Points with higher values in the color scale are overlaid on top of points with lower values.}
\label{DM_Omega}
\end{figure}
In figure~\ref{DM_Omega} we show the accepted scenarios on the \mbox{$(m_{H_{\rm new}},\kappa_{H_{126}})$} plane. In the left panel the color gradation indicates the relic density of the new invisible particle. We note that plenty of points saturate the experimental measurement $\Omega_c h^2$ within the $3 \sigma$ error band. Comparing it with figure~\ref{m_Mix_DM_Pvalue}, we see that those points are densely populated in the region where the new visible scalar and the observed Higgs boson signal are consistent within $1 \sigma$ with the LHC data. Furthermore,  if we insist in RG stability up to scales of the order of the GUT scale ($M_{GUT} \sim 10^{16}~\rm{GeV}$), we find that a significant subset of these model points overlap with the yellow band in the right panel of figure~\ref{DM_Omega}. 

\begin{figure}
\begin{center}
\includegraphics[width=0.49\linewidth,clip=true,trim=30 30 10 30]{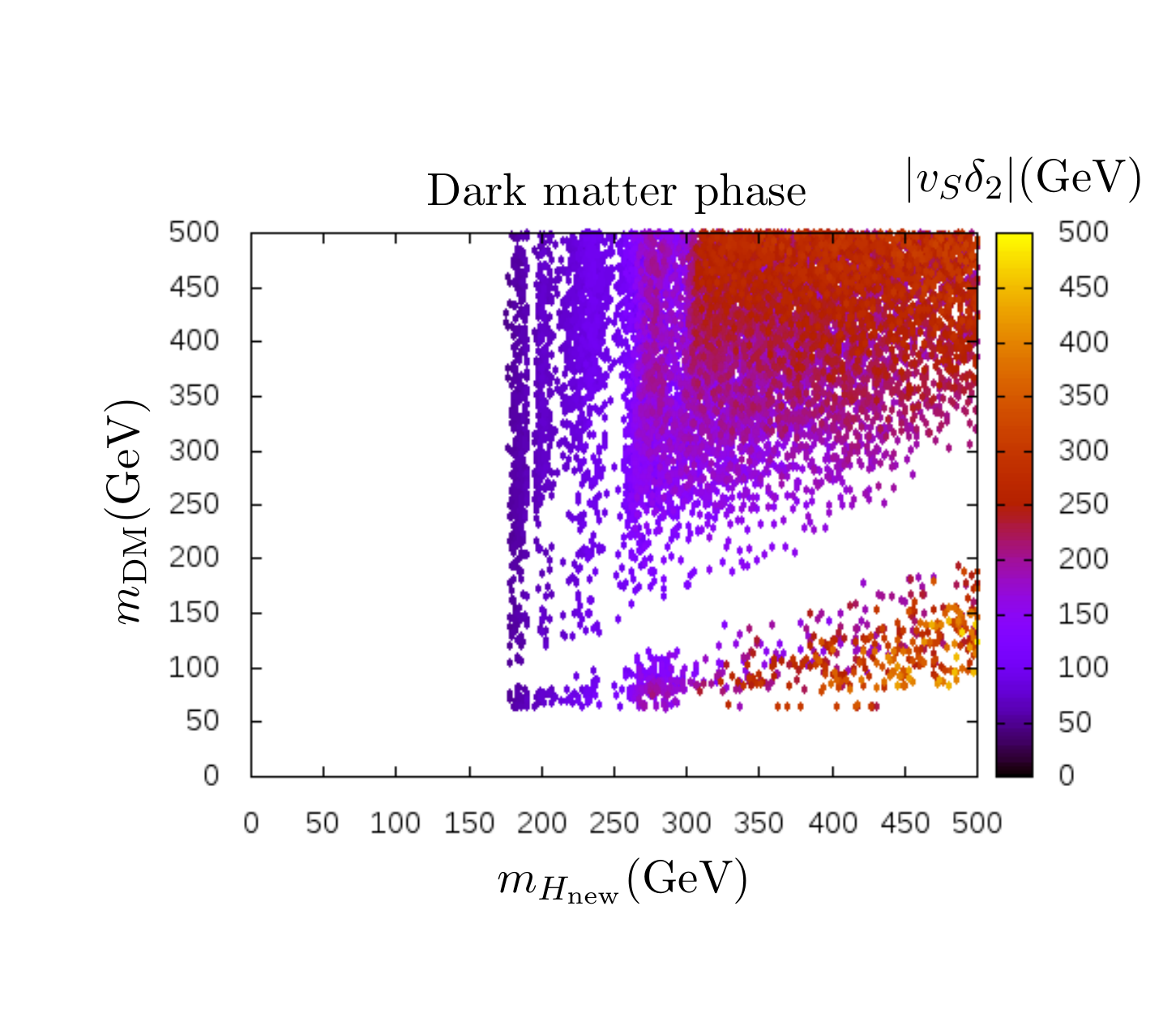}\hspace{0.01\linewidth}\includegraphics[width=0.49\linewidth,clip=true,trim=30 30 10 30]{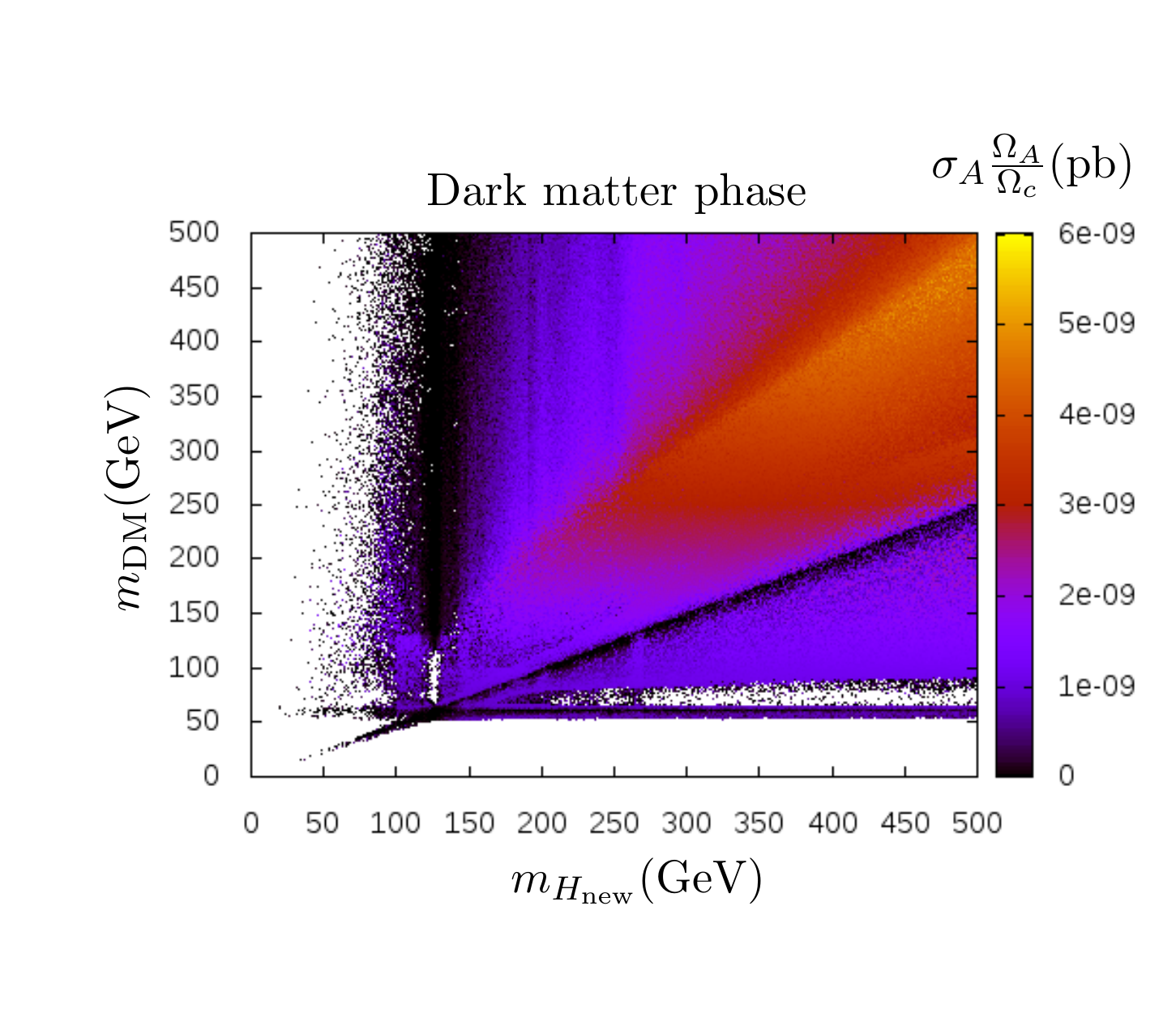}
\end{center}
\caption{{\em Saturated Model}: The two panels show a projection of the dark matter mass ($m_{\rm DM}$) versus the new scalar mass ($m_{H_{\rm new}}$). On the left, the color gradation corresponds to $|v_S\delta_2|$ (which is a measure of the degree of mixing). Points were accepted with stopping scale larger than the GUT scale and which explain fully the relic density, $\Omega_c$, within $3\sigma$. On the right, we represent in the color scale the spin-independent direct detection cross section (with no cuts) which is used in the text to explain some feature of the left panel.}
\label{saturated_model_mixings}
\end{figure}

This shows that if the mass of the new visible scalar is larger than about  
$\sim 170~\rm{GeV}$, there is an island of complete models (i.e. that explain fully  $\Omega_c$ - see figure~\ref{saturated_model_mixings}) which are also RG stable and consistent with experimental data (for large $m_{H_{\rm new}}$ such island shrinks to a line towards decoupling). If one of such model points is realized in nature, the coupling of the new visible scalar, to SM particles, can then be as large as $\kappa_{H_{\rm new}} \sim 0.4$, which may be observable at the $13/14~\rm{TeV}$ LHC runs\footnote{A detailed analysis of the experimental reach to such scenarios is however necessary to confirm if they can be excluded at the LHC.}.

In figure~\ref{saturated_model_mixings}, left panel,  we analyze the allowed parameter space for the complete models on the $(m_{H_{\rm New}},m_{\rm DM})$ plane. These are defined as the points which: have been accepted in the scan within $3\sigma$ and with all limits imposed; that are stable at least up to the GUT scale; and for which the relic density of the model saturates the Planck/WMAP measurement within $3 \sigma$. One should note that the results do not change much if the high scale is changed by a few orders of magnitude. Thus, for models where the UV completion appears at an intermediate symmetry breaking scale (such as the seesaw~\cite{Lindner:2001hr} scale $10^{11} - 10^{16}$ GeV) the results are qualitatively the same.

The main features of the result are as follows. There is a lower bound around $m_{H_{\rm new}}\simeq 170$~GeV which results from the combination of all imposed constraints (see figure~\ref{saturated_model_mixings}, left). There is also a lower bound on the dark matter particle mass just below $m_{\rm DM}\simeq \frac{1}{2}m_{H_{126}}$~GeV, and an excluded wedge around $m_{\rm DM}=\frac{1}{2}m_{H_{\rm new}}$. These correspond to regions where the annihilation channels $A A \rightarrow H_i$ (to visible Higgses) are very efficient in reducing the relic density so it becomes difficult to saturate the measured $\Omega_c$. These two lines can also be observed in the right panel where the (re-weighted) direct detection cross-section is represented in the color scale and where completeness was not imposed for $\Omega_c$. The two black lines correspond precisely to the annihilation channels above, since the weight factor ($\Omega_A/\Omega_c$) drops abruptly. In this panel, the excluded region below $m_{\rm DM}\simeq 50$ and away from $m_{\rm DM}=\frac{1}{2}m_{H_{\rm new}}$ is due to the strongest exclusion power from the LUX data for masses  in this range. We again have opted for showing the plots with the mass
range only up to $500$ GeV to better capture the details of the low mass region and also because the extended mass range is an obvious continuation of regions shown in figure~\ref{saturated_model_mixings}.

\subsubsection{Broken phase}

In this scenario we need three mixing matrix elements ($R_{1h}$, $R_{2h}$ and $R_{3h}$) to describe all scalar couplings to the SM particles. Since $R_{1h}^2+R_{2h}^2+R_{3h}^2=1$, only two of them are independent. As previously explained, our scan boxes are such that one of the scalars has a mass within the experimental band for the SM-like Higgs while the remaining two can either be heavier, lighter or degenerate with the SM-like one. Therefore we will be analyzing several scenarios simultaneously: i)  scenarios with up to three degenerate scalar states, ii) with both new scalars heavier or both lighter than the SM-like Higgs and finally, iii) with one new scalar heavier than the SM-like Higgs and the other one lighter.
Observe that while in the dark matter phase $\kappa_{H_{126}}$ fixes all non-DM scalar couplings to the SM particles, in this phase we need two of the reduced couplings, $\kappa_j$, to determine the third one. 

\begin{figure}
\begin{center}
\includegraphics[width=0.49\linewidth,clip=true,trim=40 30 30 30]{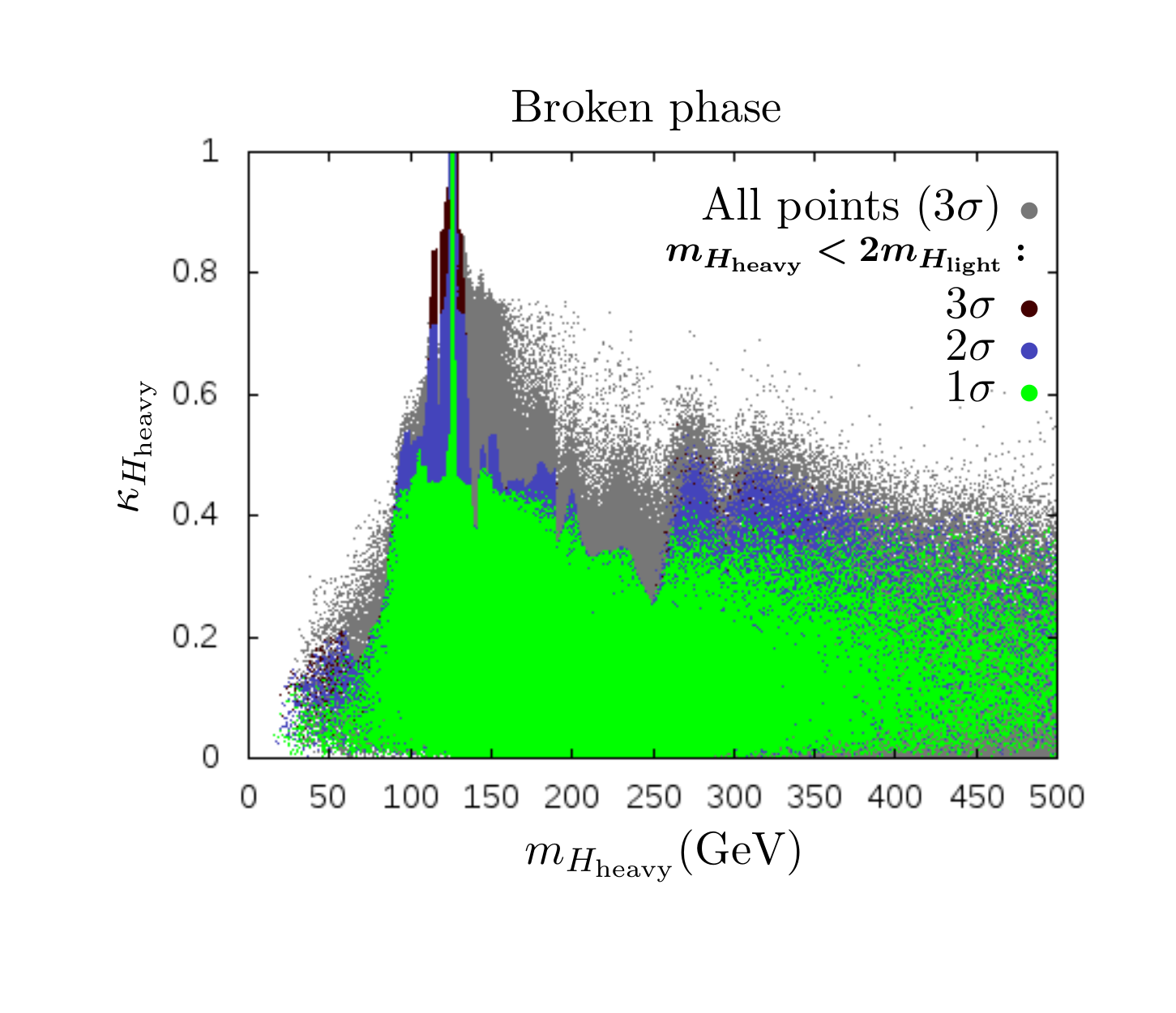}\hspace{0.01\linewidth}\includegraphics[width=0.49\linewidth,clip=true,trim=40 30 30 30]{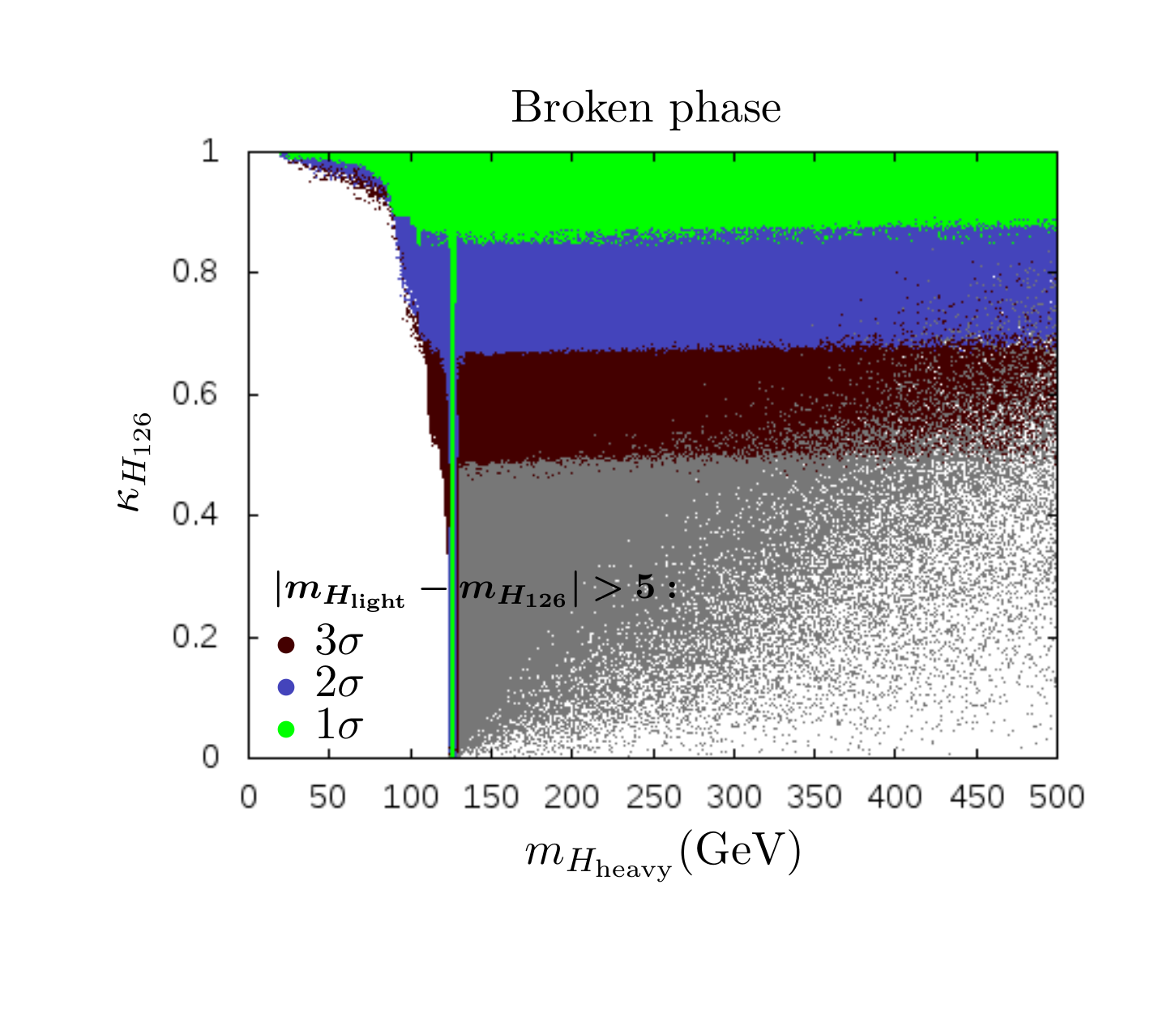}
\includegraphics[width=0.49\linewidth,clip=true,trim=40 40 30 30]{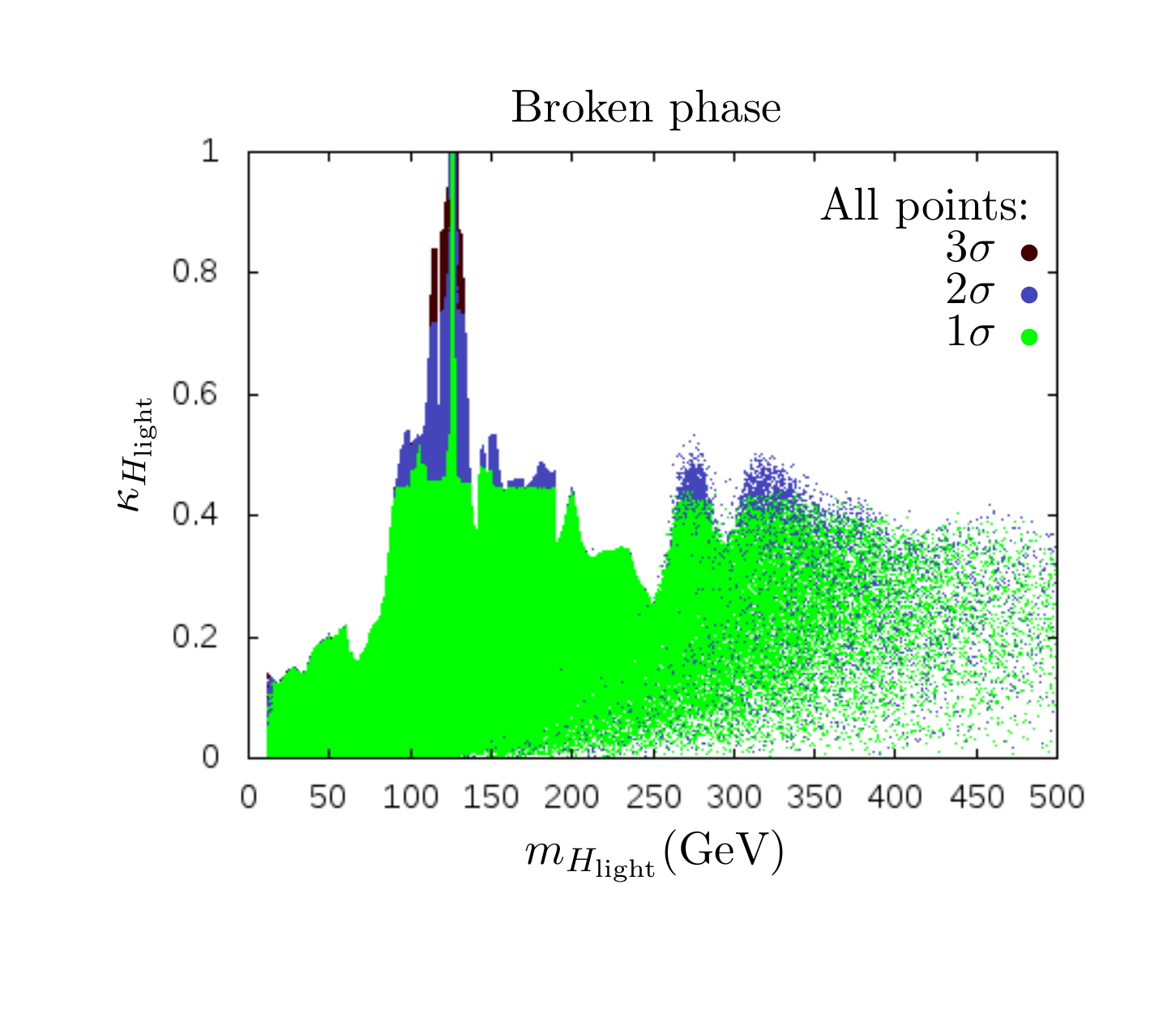}\hspace{0.01\linewidth}\includegraphics[width=0.49\linewidth,clip=true,trim=40 40 30 30]{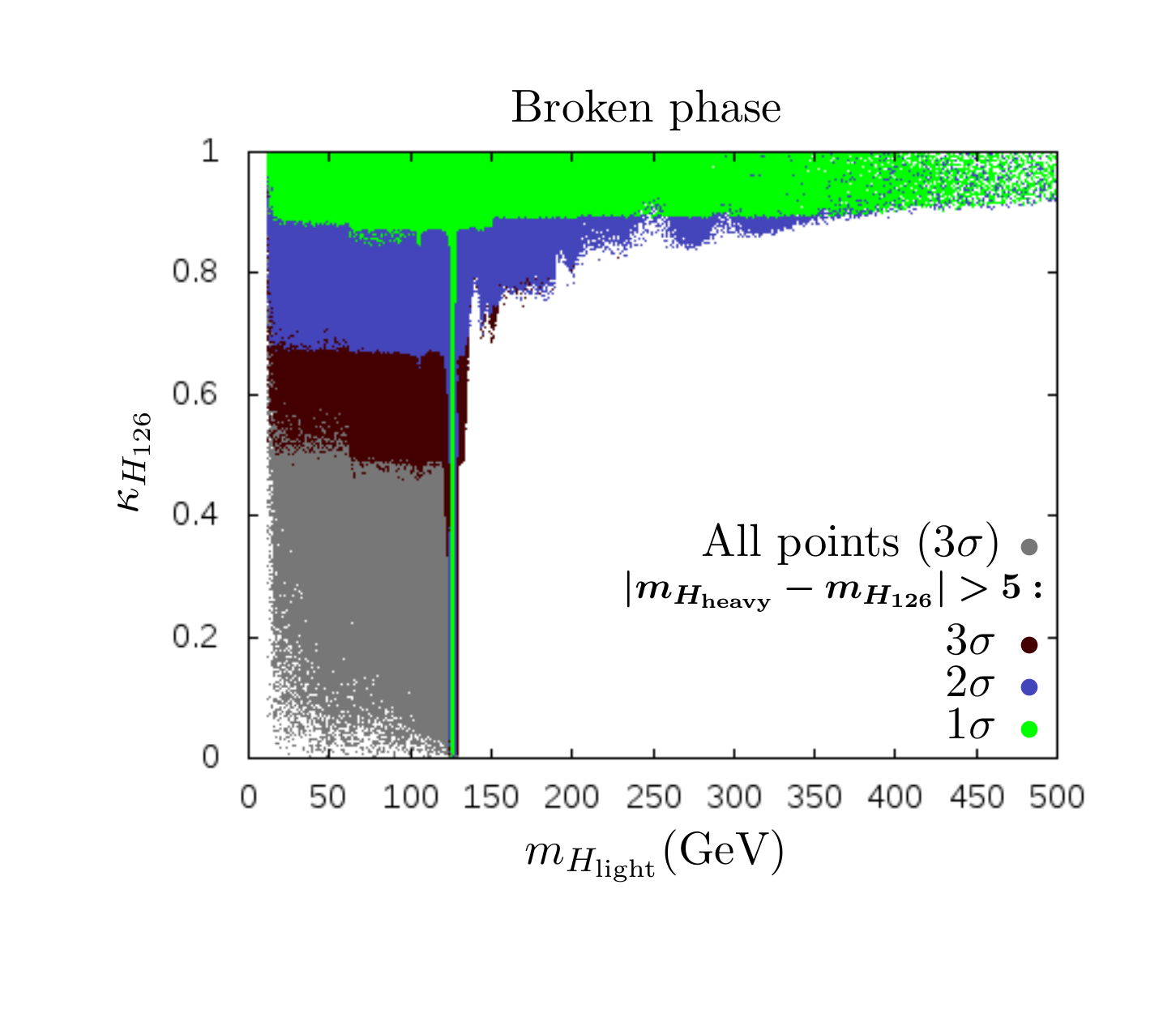}
\end{center}
\caption{{\em Broken phase}: In the top row of panels we represent the mass of $H_{\rm heavy}$ versus its own coupling to SM particles (left) and versus the SM-like Higgs coupling (right). The bottom panels are similar but for $H_{\rm light}$ in the horizontal axis. The bottom layer of points in all panels (grey) corresponds to the full set of points which are consistent with all experimental bounds and the LHC measurements within $3\sigma$. On top we overlay points which are within $3\sigma,2\sigma$ and $1\sigma$ with the following cuts for each panel: {\em Top left} - $H_{\rm heavy}$ cannot decay to $H_{\rm light}$; {\em Top right} - $H_{\rm light}$ is away from degeneracy with the SM-like Higgs by $5$~GeV; and {\em Bottom right} - $H_{\rm heavy}$ is away from degeneracy with the SM-like Higgs by $5$~GeV.
}
\label{mixall_pvalue}
\end{figure}
In figure~\ref{mixall_pvalue} we present the allowed parameter space after including all phenomenological bounds. The top left panel
shows the projection of the heavy scalar mass versus its coupling to the SM particles, while in the bottom left the heavy scalar mass is replaced by the light one. The right panels are the same except that in the vertical axis we represent the coupling of the SM-like Higgs mass. The grey points in the {\em bottom right} and both {\em top} panels (bottom layer) correspond to the full set of solutions that survive the LHC Higgs measurements at $3 \sigma$. 
In the {\em top left} panel, the additional constraint that the heavy scalar is not allowed to decay to the light one is represented in three layers of points that survive at $3 \sigma$ (brown), $2 \sigma$ (blue) and $1 \sigma$ (green) -- overlaid in that order from the bottom to the top. In the right panels, the same representation is used for the colored points, except that the cuts are: {\em top right}, the light Higgs is $5~\rm{GeV}$ away from degeneracy with the $\simeq 126$~GeV SM-like one; {\em bottom right}, the heavy Higgs is $5~\rm{GeV}$ away from degeneracy with the SM-like one. 

\begin{figure}[t!]
\begin{center}
\includegraphics[width=0.49\linewidth,clip=true,trim=40 30 25 30]{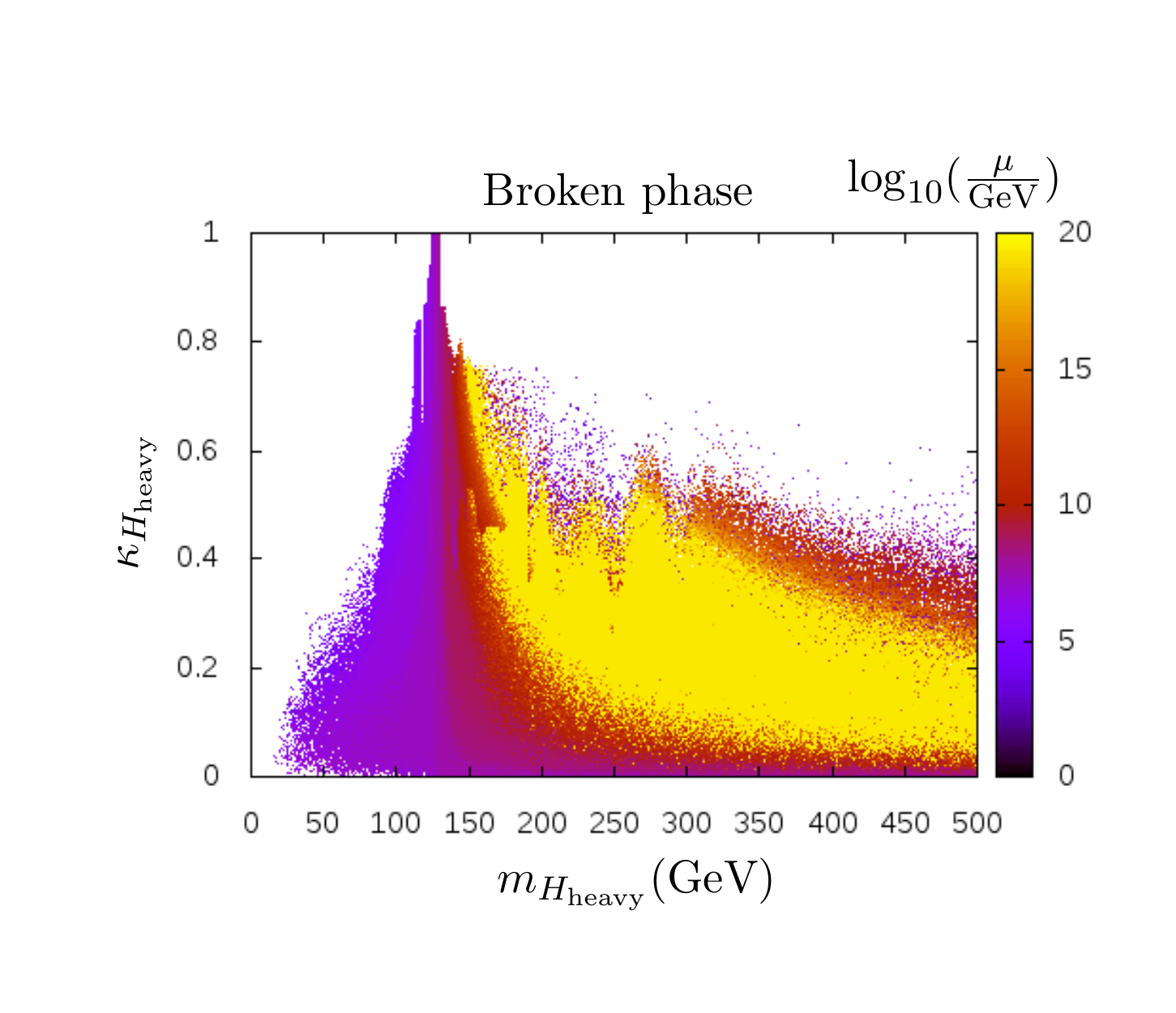}\hspace{0.01\linewidth}\includegraphics[width=0.49\linewidth,clip=true,trim=40 30 25 30]{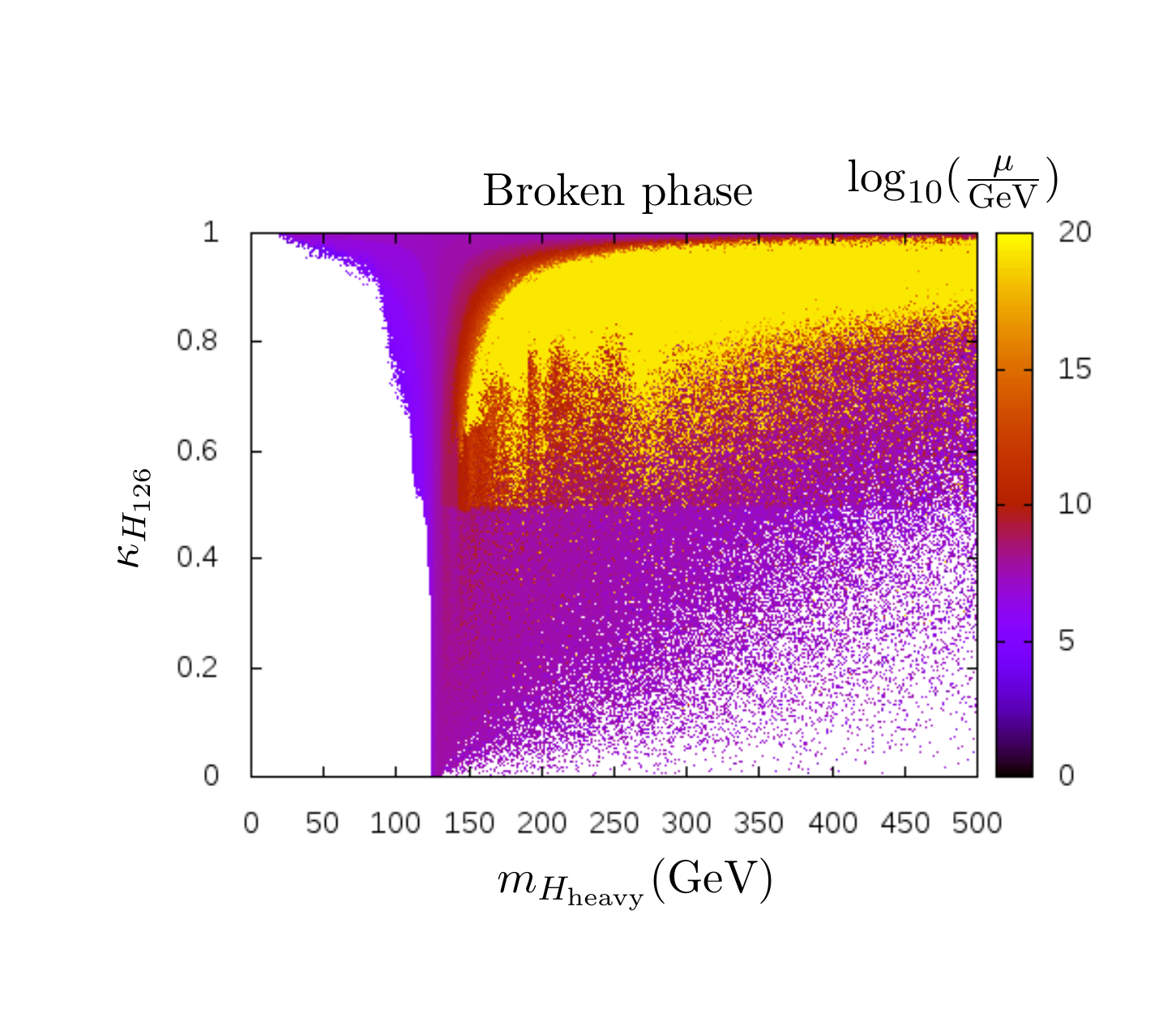}
\includegraphics[width=0.49\linewidth,clip=true,trim=40 30 25 30]{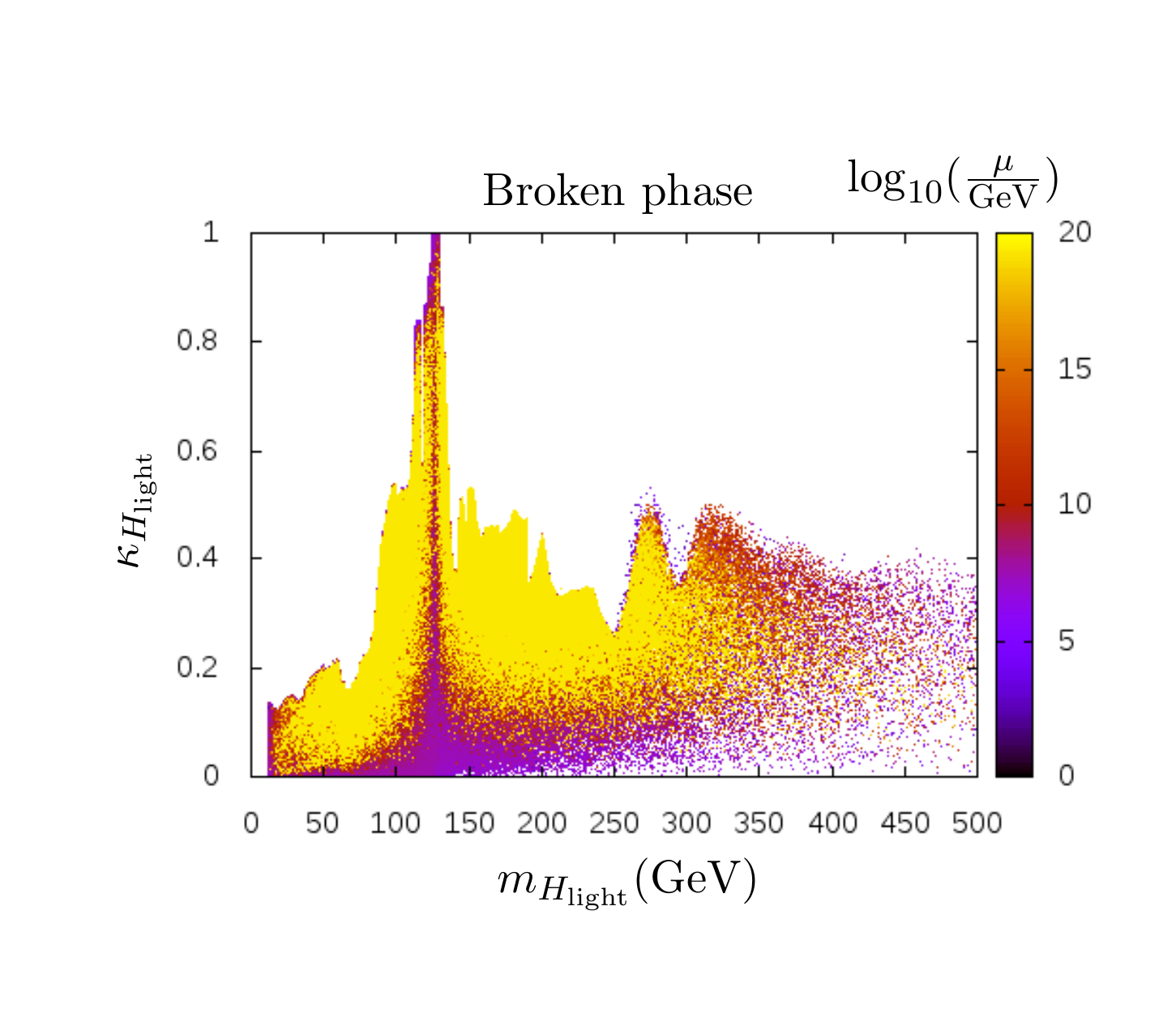}\hspace{0.01\linewidth}\includegraphics[width=0.49\linewidth,clip=true,trim=40 30 25 30]{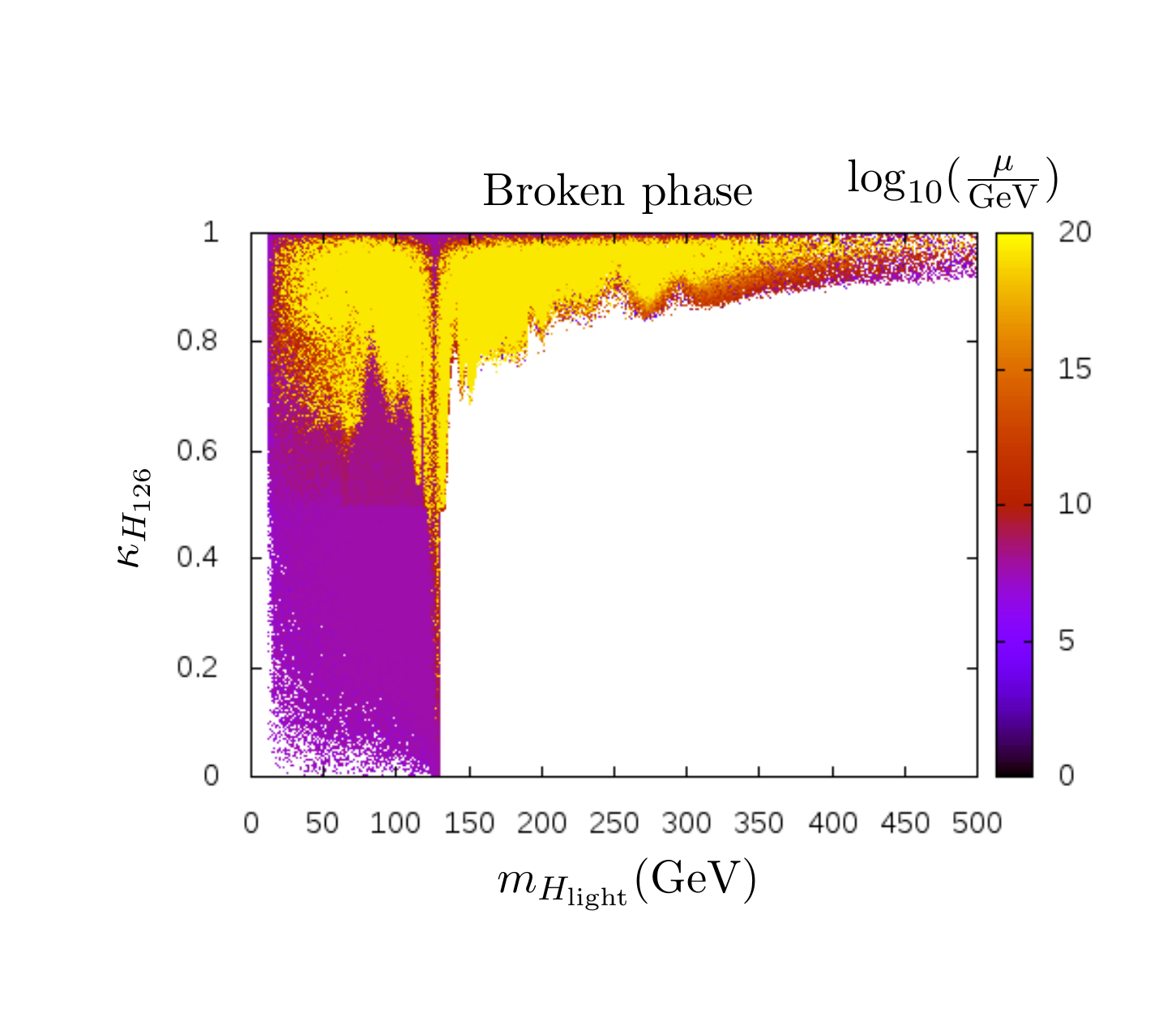}
\end{center}
\caption{{\em Broken phase}:  In the top row of panels we represent the mass of $H_{\rm heavy}$ versus its own coupling to SM particles (left) and versus the SM-like Higgs coupling (right). The bottom panels are similar but for $m_{H_{\rm light}}$ in the horizontal axis. All points correspond to the full set which is consistent with all experimental bounds and the LHC measurements within $3\sigma$. For all plots the color gradation corresponds to the scale at which the evolution stopped. Points are overlaid in order of the stopping scale with higher stopping scales on top of points with lower stopping scale.
\label{mixall_scale}
}
\end{figure}
One of the most striking features of this model is that after the 8 TeV run, the reduced SM-like Higgs coupling to SM particles, $\kappa_{H_{126}}$, 
appears to be basically unconstrained at $3 \sigma$. However, one must note that this is a remnant of the various degenerate scenarios as follows.  The visible grey points on the top right panel correspond to points on the green peak of the bottom right one. This means that such grey points correspond to a scenario where two of the masses are degenerate. Similarly the grey points on the bottom right panel correspond to points on green peak of the top right plot. The scenario where the three scalars are degenerate in mass is captured by points that pile at the peak around $\simeq 126$ GeV  in all panels (the SM-like coupling is shared by the three scalars). We would then have a \textit{triplets}  peak scenario instead of a twin peak one. 

When the scalar masses are away from degeneracy we have an almost constant bound on  $\kappa_{H_{126}}$ (except in mass regions where limits from the non-observation of scalars are stronger). This is a consequence of the property that this coupling is universal for all SM particles. Because $R_{1h}^2+R_{2h}^2+R_{3h}^2=1$ or in terms of $kappa$s,  $\kappa_{H_{126}}^2+\kappa_{H_{\rm light}}^2+\kappa_{H_{\rm heavy}}^2=1$, the other two couplings depend on the value of  $\kappa_{H_{126}}$ but we can see in the right 
plots that some freedom is still allowed.

Finally, the bounds are clearly stronger if the heavy state is not allowed to decay to two light ones. This is why we have grey points above the $3 \sigma$ level on the top left but not on the bottom left. 

In figure~\ref{mixall_scale} we present the allowed parameter space within $3\sigma$ and no other restrictions.
On the left panels we show the projection of the new scalar mass ($m_{H_{\rm heavy}}$ on the top, and $m_{H_{\rm light}}$ on the bottom) versus their couplings to SM particles, whereas on the right panels, the vertical axis contains the observed reduced SM-like Higgs coupling to SM particles. For all plots the color gradation indicates the scale at which the evolution stopped and points with a higher stopping scale are overlaying points that stopped at a lower scale. As previously discussed at length, it is clear that one needs a heavy scalar which mixes with the SM-Higgs to stabilize the theory up to the Planck scale. Also, when  $\kappa_{H_{126}}$ is exactly $1$, stability up to the Planck scale no longer holds. 

It is interesting to note that, regarding the twin peak scenarios, the one where the lighter Higgs is almost degenerate with the SM-like one is, not only allowed, but stable up to the Planck scale. On the contrary, and because one needs a scalar heavier than about 140 GeV for stability, the scenario where the heavier one is almost degenerate with the SM-like Higgs is not stable up to high energy scales.

\section{Conclusions}
\label{sec:conclusion}

In this work we have performed the first stability study of a complex singlet extension of the SM using the full two-loop renormalization group equations. We have first provided a general proof showing that the effective potential of a pure scalar theory can always be written in a form which does not depend explicitly on the vacuum choice, at any order in perturbation theory. Using these results we wrote the two-loop effective potential in terms of the couplings of the model, and derived the scalar contributions to the RGEs from its scale invariance. These were then used in a numerical study of the effects of the RGE evolution. 

Following the RGE study, we have analyzed the effect of all phenomenological constraints at the electroweak scale namely from the LEP, Tevatron and LHC experiments, electroweak precision bounds and direct and indirect constraints on dark matter. For this type of model this exhausts the experimental constraints on the parameter space because there is neither CP-violation in the theory nor charged Higgs scalars (that would be subject to B-physics constraints). Finally we have combined the RGE study with the phenomenological study to discuss the interplay between the two sources of constraints.

The model we have analyzed contains two distinct phases, one where the $\mathbb{Z}_2$ symmetry is unbroken, which we named {\em dark matter phase} and predicts a dark matter candidate alongside with a new visible scalar, and one where the symmetry is spontaneously broken, denoted as {\em broken phase}, predicting two new scalar states. We have shown that there is a continuous limit connecting the broken phase to the dark matter phase, a feature which is not allowed in models such as the inert version of the 2HDM due to perturbative unitarity constraints. The broken phase also contains a triplets peak scenario, that could only be probed via the measurement of the scalar self interactions, whereas twin peak scenarios can occur in both phases. 

Our main findings were presented (whenever possible) in terms of measurable quantities such as the physical scalar masses and their couplings to other SM particles. This has allowed us to show, clearly, the following important results:
\begin{itemize}
\item In the dark matter phase there is a stability band, that is, there is a range of masses of the new scalar where the theory is stable
up to the Planck scale. The lower limit of this new scalar mass is about 140~GeV but it depends on the SM-like Higgs coupling to the SM particles. To be more precise, stability needs a non-zero mixing between the two scalars. After combining the phenomenological constrains within $3\sigma$, forcing stability at least up to the GUT scale and and making that the relic density of the model saturates the Planck/WMAP measurement within $3 \sigma$ the lower bound for the new scalar mass is raised to about $\simeq 170$~GeV.
\item We have also shown that there are vast regions of the parameter space in agreement with all experimental data, simultaneously
saturating the experimental bounds for dark matter and stable to very high energy scales such as, at least, the GUT scale. The mass of the dark matter candidates is restricted to $\gtrsim$ 50~GeV.
\item The most striking feature of the broken phase is its phenomenological potential. It is clear that there is still plenty of parameter space left
to be scrutinized at the LHC, but several different mass hierarchy scenarios are possible leading to interesting final states, such as having two new scalars lighter
than 126 GeV.
\item  The broken phase also has a stability band. However, it is interesting to note that stability up to the Planck scale is possible even
with a new scalar lighter than 126 GeV, provided the heavy state is heavier than approximately 140 GeV. All this type of scenarios can be probed at the 
next run of the LHC.
\end{itemize}

As a final note, it would be interesting to explore if the central result of this paper is robust against changes in the structure of this minimal version of the complex singlet model, that is, if the shape of the stability band and its lower cut-off of $140$~GeV does not change if more general scalar interaction terms are allowed.

\section*{Acknowledgements}
We thank Carlos Tamarit for a thorough comparison of the two-loop RGE equations. We also thank Florian Staub  for correspondence  regarding the \textsc{Sarah} code, David Miller for  interesting discussions and Jo\~ao P. Silva for discussions about the CP properties of the model. The work in this article 
is supported in part by the Portuguese
\textit{Funda\c{c}\~{a}o para a Ci\^{e}ncia e a Tecnologia} (FCT)
under contract PTDC/FIS/117951/2010. R.S. is also partially supported by PEst-OE/FIS/UI0618/2011. A.M. and M.S. are funded by FCT through the grants SFRH/BPD/97126/2013 and SFRH/BPD/ 69971/2010 respectively.

\appendix

\section{Details of the basis independent calculations}
\label{app:DetailsBasisIndep}
The field dependent expressions for the couplings in the $\Lambda$-basis in terms of the vacuum expectation values $v_i$ and the couplings in the $L$-basis are  
\begin{eqnarray}
V^{(0)}	&=& L+L^{i}v_{i}+\dfrac{1}{2!}L^{ij}v_{i}v_{j}+\dfrac{1}{3!}L^{ijk}v_{i}v_{j}v_{k}+\dfrac{1}{4!}L^{ijkl}v_{i}v_{j}v_{k}v_{l} \;, \nonumber\\
\Lambda^{ij}	&=&	L^{ij}+L^{ijk}v_{k}+\dfrac{1}{2}L^{ijkl}v_{k}v_{l} \;, \nonumber\\[1mm]
\Lambda^{ijk}	&=&	L^{ijk}+L^{ijkl}v_{l} \;, \nonumber\\[1mm]
\Lambda^{ijk}	&=&	L^{ijkl} \;.  \label{eq:Lambdas}
\end{eqnarray}
The couplings in the rotated $\lambda$-basis (where the field fluctuations are the physical eigenstates) are related to the $\Lambda$-basis by
\begin{eqnarray}
m_i^2&=&M_{mi}M_{ni}\Lambda^{mn} \nonumber\\
\lambda^{ijk}&=&M_{m}^{\phantom{a}i}M_{n}^{\phantom{a}j}M_{p}^{\phantom{a}k}\Lambda^{mnp} \nonumber\\
\lambda^{ijkl}&=&M_{m}^{\phantom{a}i}M_{n}^{\phantom{a}j}M_{p}^{\phantom{a}k}M_{q}^{\phantom{a}l}\Lambda^{mnpq} \;.
\label{eq:lambdas}
\end{eqnarray}
Note that the field space indices are lowered and raised by the flat Euclidean space metric $\delta_{ij}$. We also use the Einstein convention that repeated indices which are one up and one down are summed over. 

The main steps to prove the general result, Eq.~\eqref{eq:AllOrderVn}, are as follows. Using the property that the loop functions must be totally symmetric in the interchange of the masses, and assuming the Taylor expansion of the loop functions  $I_{D}^{(n)}$ converges with non-zero radius around $\mu^{2}$, we have that 
\begin{equation}
I_{D}^{(n)}(m_{j_{1}}^{2},\ldots,m_{j_{p_{D}}}^{2})=\sum_{q_{1},\ldots,q_{P_{D}}}a_{q_{1},\ldots,q_{P_{D}}}^{(n)}\left(m_{j_{1}}^{2}-\mu^{2}\right)^{q_{1}}\ldots\left(m_{j_{p_{D}}}^{2}-\mu^{2}\right)^{q_{P_{D}}}\; .
\end{equation}
Inserting in Eq.~\eqref{eq:Vn_lambdas} we obtain
\begin{multline}
V^{(n)}=\sum_{D=1}^{N_{D}^{(n)}}\sum_{q_{1},\ldots,q_{P_{D}}}a_{q_{1},\ldots,q_{P_{D}}}^{(n)}\left(\lambda_{1}\ldots\lambda_{V_{D}}\right)^{j_{1},\ldots,j_{P_{D}},j_{1},\ldots,j_{P_{D}}}\times \\ \times\left(m_{j_{1}}^{2}-\mu^{2}\right)^{q_{1}}\ldots\left(m_{j_{p_{D}}}^{2}-\mu^{2}\right)^{q_{P_{D}}}
\end{multline}
 Now we use the fact that, when rotating back each $\lambda_{A}$ coupling in the vertex product, each up index is rotated by a mixing matrix so that 
\begin{eqnarray}
V^{(n)} & = & \sum_{D=1}^{N_{D}^{(n)}}\sum_{q_{1},\ldots,q_{P_{D}}}a_{q_{1},\ldots,q_{P_{D}}}^{(n)}\left(\Lambda_{1}\ldots\Lambda_{V_{D}}\right)^{m_{1},\ldots,m_{P_{D}},m_{P_{D}+1},\ldots,m_{2P_{D}}}\times\\
 &  & \times M_{m_{1}}^{\phantom{a}j_{1}}\ldots M_{m_{P_{D}}}^{\phantom{a}j_{P_{D}}}M_{m_{P_{D}+1}}^{\phantom{a}j_{1}}\ldots M_{m_{2P_{D}}}^{\phantom{a}j_{P_{D}}}\left(m_{j_{1}}^{2}-\mu^{2}\right)^{q_{1}}\ldots\left(m_{j_{p_{D}}}^{2}-\mu^{2}\right)^{q_{P_{D}}}\; .\nonumber
\end{eqnarray}
Using the orthogonality condition obeyed by the mixing matrices we can finally transfer the rotation matrices to the mass factors. Denoting the two-by-two matrix formed with the components $\Lambda^{ij}$ by $\Lambda_{(2)}$, then we obtain
\begin{eqnarray}
 V^{(n)}& = & \sum_{D=1}^{N_{D}^{(n)}}\sum_{q_{1},\ldots,q_{P_{D}}}a_{q_{1},\ldots,q_{P_{D}}}^{(n)}\left(\Lambda_{1}\ldots\Lambda_{V_{D}}\right)^{m_{1},\ldots,m_{P_{D}},m_{P_{D}+1},\ldots,m_{2P_{D}}}\times \nonumber\\ 
&&\times\left[\left(\Lambda_{(2)}-\mu^{2}\right)^{q_{1}}\right]_{m_{1}m_{P_{D}+1}}\ldots\left[\left(\Lambda_{(2)}-\mu^{2}\right)^{q_{P_{D}}}\right]_{m_{P_{D}}m_{2P_{D}}}\label{eq:TaylorDefIn}\\
 & \equiv & \sum_{D=1}^{N_{D}^{(n)}}\left(\Lambda_{1}\ldots\Lambda_{V_{D}}\right)^{m_{1},\ldots,m_{P_{D}},m_{P_{D}+1},\ldots,m_{2P_{D}}}\left[\mathbf{I}_{D}^{(n)}\left(\Lambda_{(2)},\mu^{2}\right)\right]_{m_{1},\ldots,m_{2P_{D}}} \; .
\end{eqnarray}
Eq.~\eqref{eq:TaylorDefIn} defines the matrix version of the loop functions $\mathbf{I}_{D}^{(n)}$. 

\subsection{Effective potential at two loops in the $\lambda$-basis}
\label{app:lambdaVeff}
In this section we summarize the scalar contributions to the effective potential in the basis of physical scalar states. 
The Coleman-Weinberg potential for a generic QFT is given by the supertrace
\begin{eqnarray}
V^{(1)} = \dfrac{1}{4} \sum_n \left( -1 \right)^{2 s_n} \left( 2 s_n +1 \right) \left( m^2_n \right)^2 \left( \log\left( m_n^2 \right) -2 t - c_n \right)\label{eq:CWapp} \; ,
\end{eqnarray}
where $s_n$ is the spin of some $\psi_n$ field and $m_n$ its physical mass ($\lambda$-basis). In the $\overline{\rm{MS}}$ scheme $c_n = (3/2, 3/2, 5/6)$ respectively for a scalar, a fermion or a vector field. In particular, for a scalar theory
\begin{eqnarray}
V^{(1)}&=&\sum_{i}\frac{1}{4}\left(m_{i}^{2}\right)^{2}\left(-2t-\frac{3}{2}+\log(m_{i}^{2})\right)\; . \label{eq:VCWlambda}
\end{eqnarray}
As for the two-loop cubic and quartic contributions they are~\cite{Martin:2001vx}
\begin{eqnarray}
V_{sss}^{(2)}&=&\dfrac{1}{6}\sum_{i,j,k}\left(\lambda^{ijk}\right)^{2}\left\{\left(m_{i}^{2}+m_{j}^{2}+m_{k}^{2}\right)t^{2}+\left[2\left(m_{i}^{2}+m_{j}^{2}+m_{k}^{2}\right)-m_{i}^{2}\log m_{i}^{2} -m_{j}^{2}\log m_{j}^{2} \right. \right.\nonumber\\ 
&& \left. \left. -m_{k}^{2}\log m_{k}^{2}\right]t+P^{(2)}(m_{i}^{2},m_{j}^{2},m_{k}^{2})\right\}\; , \label{eq:vsss}\\[2mm]
V_{ssss}^{(2)}&=&\dfrac{1}{2}\sum_{i,j}\lambda^{iijj}m_{i}^{2}m_{j}^{2}\left[t\left(t+1-\frac{1}{2}\left(\log m_{i}^{2}+\log m_{j}^{2}\right)\right)+\frac{1}{4}\left(1-\log m_{i}^{2}-\log m_{j}^{2}\right. \right.\nonumber\\ 
&& \left. \left. +\log m_{i}^{2}\log m_{j}^{2}\right)\right] \; , \label{eq:vssss}
\end{eqnarray}
where the cubic scale independent term is
\begin{eqnarray}
 P^{(2)}(m_{i}^{2},m_{j}^{2},m_{k}^{2}) = -\dfrac{5}{2} \left( m_{i}^{2} + m_{j}^{2} + m_{k}^{2} \right) - \dfrac{1}{2}\xi\left(  m_{i}^{2}, m_{j}^{2}, m_{k}^{2} \right) \label{eq:Poly} \; ,
\end{eqnarray}
with 
\begin{eqnarray}
\xi\left(  m_{i}^{2},m_{j}^{2},m_{k}^{2} \right) &=& \rho \left[ 2 \log\left(\dfrac{m_{k}^{2}+ m_{i}^{2} - m_{j}^{2} - \rho}{2 m_{k}^{2}} \right) \log\left(\dfrac{m_{k}^{2}+ m_{j}^{2} - m_{i}^{2} - \rho}{2 m_{k}^{2}} \right)  \right. \nonumber\\
&& \left. - \log\left( \dfrac{m_{i}^{2}}{m_{k}^{2}} \right) \log\left( \dfrac{m_{j}^{2}}{m_{k}^{2}} \right) - 2 {\rm Li}_2\left(\dfrac{m_{k}^{2}+ m_{i}^{2} - m_{j}^{2} - \rho}{2 m_{k}^{2}} \right)  \right. \nonumber\\
&& \left.  - 2 {\rm Li}_2\left(\dfrac{m_{k}^{2}+ m_{j}^{2} - m_{i}^{2} - \rho}{2 m_{k}^{2}} \right) + \dfrac{\pi^2}{3} \right] \label{eq:Poly2}\; ,
\end{eqnarray}
and
\begin{eqnarray}
\rho =  \left(  m_{i}^{4} + m_{j}^{4} + m_{k}^{4} - 2 m_{i}^{2}m_{j}^{2} - 2 m_{i}^{2}m_{k}^{2} - 2 m_{j}^{2}m_{k}^{2} \right)^{1/2}\label{eq:rho}\; .
\end{eqnarray}
 and ${\rm Li}_2(x)$ is the dilogarithm function.

Now it is explicit that $V^{(1)},V^{(2)}_{sss}$ and $V^{(2)}_{ssss}$ are in form~\eqref{eq:Vn_lambdas} so one can apply the transformations leading to the general form~\eqref{eq:AllOrderVn} to obtain the results in Eqs.~\eqref{eq:VCWL}, \eqref{eq:vsssL} and \eqref{eq:vssssL}.

\subsection{Scale invariance conditions and logarithm cancellations}
\label{app:2loopBetas}
The scale invariance of the effective potential is expressed in the \textit{Callan-Symanzyk} conditions, Eq.~\eqref{eq:scale_invarianceVeff}. Similarly to the effective potential, all beta functions and anomalous dimensions can be expanded perturbatively in powers of $\varepsilon$:
\begin{equation}
 \beta_{L}(t) = \sum_{n = 0}^{+ \infty} \varepsilon^{n+1} \beta^{(n+1)}_{L}(t) \; , \;
 \gamma_{i}(t) = \sum_{n = 0}^{+ \infty} \varepsilon^{n+1} \gamma^{(n+1)}_{i}(t) \; .
\end{equation}
 For the purpose of argument, here we label any coupling by $L$. If we insert these expansions in the scale invariance condition, Eq.~\eqref{eq:scale_invarianceVeff}, and equate order by order in powers of $\varepsilon$, we get a tower of equations
\begin{equation}\label{eq:DVt}
\mathcal{D}^{(m)}V^{(0)}=-\dfrac{\partial V^{(n)}}{\partial t}-\sum_{m=1}^{n-1}\mathcal{D}^{(m)}V^{(n-m)} \; ,
\end{equation}
where we have defined
\begin{equation}\label{eq:Dm}
\mathcal{D}^{(m)}\equiv\sum_{L}\beta_{L}^{(m)}\dfrac{\partial}{\partial L}-\sum_{i}\gamma_{i}^{(m)}v_{i}\dfrac{\partial}{\partial v_{i}} \; .
\end{equation}
The left hand side of Eq.~\eqref{eq:DVt} contains the $n$th order beta functions and anomalous dimensions and the right hand side contains lower order beta functions and anomalous dimensions. Thus, this provides an iterative procedure by which the $n$th order evolution functions (left hand side), are extracted from the $n$th order effective potential and evolution functions of order $m<n$ (right hand side of Eq.~\eqref{eq:DVt}).

Let us now apply~\eqref{eq:DVt} to a purely scalar theory . First we note that
\begin{eqnarray}
\mathcal{D}^{(m)}V^{(0)}&=&\mathcal{D}^{(m)}\left(L+L^{i}v_{i}+\dfrac{1}{2!}L^{ij}v_{i}v_{j}+\dfrac{1}{3!}L^{ijk}v_{i}v_{j}v_{k}+\dfrac{1}{4!}L^{ijkl}v_{i}v_{j}v_{k}v_{l}\right) \label{eq:left_hand_side}\\
&=& \beta^{(n)}+\left[\beta^{(n)i}-L^i\gamma^{(n)i}\right]v_{i}+\dfrac{1}{2!}\left[\beta^{(n)ij}-L^{ij}\gamma^{(n)i}-L^{ij}\gamma^{(n)j}\right]v_{i}v_{j}+\nonumber\\
&& \dfrac{1}{3!}\left[\beta^{(n)ijk}-L^{ijk}\gamma^{(n)i}-L^{ijk}\gamma^{(n)j}-L^{ijk}\gamma^{(n)k}\right]v_{i}v_{j}v_{k}+ \nonumber\\
&&\dfrac{1}{4!}\left[\beta^{(n)ijkl}-L^{ijkl}\gamma^{(n)i}-L^{ijkl}\gamma^{(n)j}-L^{ijkl}\gamma^{(n)k}-L^{ijkl}\gamma^{(n)l}\right]v_{i}v_{j}v_{k}v_{l}\; , \nonumber
\end{eqnarray}
so indeed the left hand side contains the $n$th order evolution functions. As for the right hand side, we assume it has a similar polynomial form in the VEVs, i.e. 
\begin{equation}
-\dfrac{\partial V^{(n)}}{\partial t}-\sum_{m=1}^{n-1}\mathcal{D}^{(m)}V^{(n-m)}=\delta^{(n)}+\delta^{(n)i}v_i+\delta^{(n)ij}v_iv_j+\delta^{(n)ijk}v_iv_jv_k+\delta^{(n)ijkl}v_iv_jv_kv_l\,. \label{eq:right_hand_side}
\end{equation}
However, specializing to one and two-loops we know that the effective potential is not polynomial in the VEVs -- it contains various $\log$ and ${\rm Li}_2$ functions. Such terms must always cancel out in the Callan-Symanzyk equations, so they provide an internal consistency check of the calculation. Inserting~\eqref{eq:left_hand_side} and~\eqref{eq:right_hand_side} in~\eqref{eq:DVt}  and equating, we finally get the general result, Eq.~\eqref{eq:general_beta:scalar}. 

Finally, we outline the steps leading to the general results Eqs.~\eqref{eq:deltas_n1} and~\eqref{eq:deltas_n2}. At one-loop order, using Eq.~\eqref{eq:VCWL} we have that 
\begin{eqnarray}
-\dfrac{\partial V^{(1)}}{\partial t}&=&\dfrac{1}{2}\Lambda^{ij}\Lambda_{ij}\nonumber\\
&=&\dfrac{1}{2}\left(L^{ij}+L^{ijk}v_{k}+\dfrac{1}{2}L^{ijkl}v_{k}v_{l}\right)\left(L_{ij}+L_{ij}^{\phantom{ab}m}v_{m}+\dfrac{1}{2}L_{ij}^{\phantom{ij}mn}v_{m}v_{n}\right) \; .
\end{eqnarray}
Expanding, and collecting the various coefficients of the powers of the VEVs we obtain the $\delta^{(1)i_1,\ldots,i_p}$, Eq.~\eqref{eq:deltas_n1}. At two-loops
\begin{eqnarray}
-\dfrac{\partial V^{(2)}}{\partial t}-\mathcal{D}^{(1)}V^{(1)}&=&\dfrac{1}{2}\Lambda_{ij}\left(\mathcal{D}^{(1)}\Lambda^{ij}-\Lambda^{ijmn}\Lambda_{mn}-2\Lambda_{\phantom{a}mn}^{i}\Lambda^{jmn}\right) \\
&&-\left[\Lambda_{(2)}\left(\dfrac{1}{2}\log\Lambda_{(2)}-t\right)\right]_{ij}\left( \mathcal{D}^{(1)}\Lambda^{ij}-\Lambda_{\phantom{a}mn}^{i}\Lambda^{jmn}-\Lambda^{ijkl}\Lambda_{kl}\right)\nonumber\; ,
\end{eqnarray}
where we have separated the terms which are not polynomial in the VEVs and are scale dependent, on the second line. Expanding the $\Lambda$-couplings in terms of the $L$-couplings we obtain
\begin{eqnarray}
-\dfrac{\partial V^{(2)}}{\partial t}-\mathcal{D}^{(1)}V^{(1)}&=&\delta^{(2)}+\delta^{(2)i}v_i+\delta^{(2)ij}v_iv_j+\delta^{(2)ijk}v_iv_jv_k+\delta^{(2)ijkl}v_iv_jv_kv_l \nonumber \\
&&+\left[\Lambda_{(2)}\left(\dfrac{1}{2}\log\Lambda_{(2)}-t\right)\right]_{ij}\Lambda^{ij}\gamma^{(1)i} \; .
\end{eqnarray}
In the $\log$ terms on the second line all first order beta functions coming from  $\mathcal{D}^{(1)}\Lambda^{ij}$ have canceled the terms $-\Lambda_{\phantom{a}mn}^{i}\Lambda^{jmn}-\Lambda^{ijkl}\Lambda_{kl}$ as expected. All that remains is a contribution which is proportional to the anomalous dimensions. The one-loop anomalous dimensions are, however, strictly zero before we include non-scalar contributions. Note however that the cancellation of this term with fermion contributions to the effective potential can be observed in the SM contributions (which we include in the final RGEs for the complex singlet model).

\section{Two-loop RGEs for the complex singlet model} \label{app:RGExSM}

We present here a summary of the one and two-loop RGEs for the complex singlet model Eq.~\eqref{eq:V_general}, which were calculated as described in Sec.~\ref{sec:Veff}. The structure of both the gauge and fermion sectors of the complex singlet model are identical to the SM. Therefore, the two loop beta functions of the gauge and Yukawa\footnote{We do not consider in this paper contributions from the tau and bottom Yukawa couplings, and the first two generations Yukawa couplings, due to their smallness.} couplings are
\begin{eqnarray}
\beta_{g_i}^{(1)} &=& b_i g^3_i\;,\qquad \beta_{g_i}^{(2)} = g^3_i \left[ \sum_j b_{j i} g^2_j + C_i  y^2_t\right] \; \\
\beta_{y_t}^{(1)} &=& y_t \left[\dfrac{9}{2}y_t^2 - \dfrac{17}{20}g_1^2 - \dfrac{9}{4} g_2^2 - 8 g_3^2 \right] \nonumber\\
\beta_{y_t}^{(2)} &=& y_t\left[  \dfrac{3}{8} \lambda^2 + \dfrac{\delta^2_2}{8} - 3 y_t^2 \lambda - 12 y_t^4 + \dfrac{393}{80} g_1^2 y_t^2 
 + \dfrac{225}{16} g_2^2 y_t^2 + 36 g_3^2 y_t^2 + \dfrac{1187}{600} g^4_1 \right.\nonumber\\ 
&& \left.  \phantom{11}- \dfrac{23}{4} g_2^4  - 108 g_3^4 - \dfrac{9}{20} g_1^2 g_2^2  + \dfrac{19}{15} g_1^2 g_3^2 + 9 g_2^2 g_3^2 \right] \; ,
\end{eqnarray}
with
\begin{equation}
b_i = \left( \dfrac{41}{10} , -\dfrac{19}{6}, -7 \right) \;,\;\; 
C_i = \left( -\dfrac{17}{10} , -\dfrac{3}{2}, -2 \right) \; , \;\;
b_{j i} = \begin{pmatrix}
 \dfrac{199}{50}&  \dfrac{9}{10} &  \dfrac{11}{10}\\[2.5mm]
 \dfrac{27}{10} &  \dfrac{35}{6} &  \dfrac{9}{2}\\[2.5mm]
\dfrac{44}{5} & 12 & -26
\end{pmatrix}\; .
\end{equation}
For the scalar sector one must consider the SM contributions and those arising from the complex singlet field. For the quartic and bilinear couplings we have, respectively,
\begin{eqnarray}
\beta_{\lambda}^{(1)}&=&  \dfrac{27}{50} g^4_1 + \dfrac{9}{5} g^2_1 g^2_2 + \dfrac{9}{2} g^4_2 - 24 y_t^4 + 6 \lambda^2 + \delta_2^2 + 4 \lambda \gamma^{(1)}_h \\[1mm]
\beta_{\lambda}^{(2)}&=& - \dfrac{3411}{500}g_1^6 - \dfrac{1677}{100} g_1^4 g_2^2 - \dfrac{289}{20}g_1^2 g_2^4 + \dfrac{305}{4} g_2^6-y_t^2 \left( \dfrac{171}{25} g_1^4 - \dfrac{126}{5} g_1^2 g_2^2 + 9 g_2^4 \right) \nonumber\\ 
&& - y_t^4 \left( \dfrac{32}{5} g_1^2 +128 g_3^2 \right)+ 120 y_t^6+ \lambda\left(\dfrac{297}{100}g_1^4  +  \dfrac{9}{2}g_1^2 g_2^2 + \dfrac{99}{4} g_2^4 + 24y_t^4 \right) \nonumber\\ 
&&  + \lambda^2\left(\dfrac{54}{5}g_1^2 + 54 g_2^2  -72 y_t^2 \right)- 21 \lambda^3 - 2 \delta_2^3 - 3 \lambda \delta_2^2  + 4\lambda \gamma^{(2)}_h + 12 \lambda^2\gamma^{(1)}_h  \; , \nonumber \\[2mm]
\beta_{m^2}^{(1)} &=& 3 m^2 \lambda + b_2 \delta_2 + 2 m^2 \gamma^{(1)}_h \\[1mm]
\beta_{m^2}^{(2)} &=& m^2 \left[ \dfrac{189}{200} g_1^4 + \dfrac{9}{20} g_1^2 g_2^2 + \dfrac{63}{8} g_2^4  +\lambda\left( \dfrac{63}{10} g_1^2 + \dfrac{63}{2} g_2^2   - 36 y_t^2 \right)\right]\nonumber \\
&&- \dfrac{9}{2} m^2\lambda^2 -  \dfrac{1}{2} m^2 \delta_2^2 - b_2 \delta^2_2  + 2 m^2 \gamma^{(2)}_h + 6 \lambda m^2 \gamma^{(1)}_h  \; ,\nonumber
\end{eqnarray}
where, in limit $\delta_2 \rightarrow 0$ and $b_2 \rightarrow 0$, we recover the usual Standard Model equations consistent with~\cite{Arason:1991ic, Ford:1992pn}, up to (conventional) normalizations of the couplings.
The \mbox{$\beta$-functions} for the remaining parameters are:
\begin{eqnarray}
\beta_{\delta_2}^{(1)} &=& \delta_2\left[ 2 d_2 + 2\delta_2 + 3 \lambda + 2 \gamma^{(1)}_h + 2 \gamma^{(1)}_S \right] \label{eq:delta2} \\[1mm]
\beta_{\delta_2}^{(2)} &=&\delta_2\left[\dfrac{189}{200}g_1^4 +\dfrac{9}{20} g_1^2 g_2^2 + \dfrac{63}{8}g_2^4 + \delta_2\left( \dfrac{6}{5} g_1^2 + 6 g_2^2  - 12 y_t^2 \right)+ \lambda\left( \dfrac{63}{10} g_1^2+ \dfrac{63}{2} g_2^2 - 36 y^2_t\right)   \right. \nonumber\\
&& \phantom{\dfrac{9}{2}}\left.-3 d_2^2 - 6 d_2 \delta_2 - \dfrac{7}{2} \delta_2^2 - 9 \delta_2 \lambda - \dfrac{9}{2} \lambda^2+ \right.\nonumber\\
&&\left.\phantom{\dfrac{9}{2}}+ 2 \left(\gamma^{(2)}_h +  \gamma^{(2)}_S+\left( \delta_2 + 3 \lambda \right) \gamma^{(1)}_h   + \left(2 d_2 + \delta_2 \right) \gamma^{(1)}_S\right) \right] \; ,\nonumber\\[2mm]
\beta_{b_2}^{(1)} &=&  2 b_2 d_2 + 2 m^2 \delta_2 + 2 b_2 \gamma^{(1)}_S \label{eq:b2} \\[1mm]
\beta_{b_2}^{(2)} &=& m^2 \delta_2\left[\dfrac{21}{5} g_1^2 + 21 g_2^2 - 24 y^2_t \right]-3 b_2 d_2^2 - b_2 \delta_2^2 - 2 m^2 \delta_2^2 \nonumber\\
&&+ 4 m^2 \delta_2 \gamma^{(1)}_h + 4 b_2 d_2 \gamma^{(1)}_S + 2 b_2 \gamma^{(2)}_S\; , \nonumber\\[2mm]
\beta_{d_2}^{(1)} &=& 5 d_2^2 + 2 \delta_2^2 + 4 d_2 \gamma^{(1)}_S\label{eq:d2}\\
\beta_{d_2}^{(2)} &=& \delta_2^2\left(\dfrac{21}{5} g_1^2 + 21 g_2^2 - 24 y^2_t \right)-16 d_2^3 - 6 d_2 \delta_2^2 - 4 \delta_2^3  
 + 4 \delta_2^2 \gamma^{(1)}_h + 10 d_2^2 \gamma^{(1)}_S + 4 d_2 \gamma^{(2)}_S\; , \nonumber\\[2mm]
\beta_{b_1}^{(1)} &=& b_1 d_2 + 2 b_1 \gamma^{(1)}_S \;,\;\;\beta_{b_1}^{(2)} =-2 b_1 d_2^2 - b_1 \delta_2^2 + 2 b_1 d_2 \gamma^{(1)}_S + 2 b_1 \gamma^{(2)}_S\; , \label{eq:b1}\\[2mm]
\beta_{a_1}^{(1)} &=& a_1 \gamma^{(1)}_S \;,\;\beta_{a_1}^{(2)} = a_1 \gamma^{(2)}_S \; . \label{eq:a1}
\end{eqnarray}
The anomalous dimensions of the SM Higgs, $S$ and $A$ fields are obtained from the general formalism in~\cite{Machacek:1983tz} and can be checked in the SM limit~\cite{Arason:1991ic,Ford:1992pn}:
\begin{eqnarray}
\gamma_{h}^{(1)} &=& 3 y^2_t - \dfrac{9}{20} g^2_1 - \dfrac{9}{4} g^2_2 \label{eq:gammah} \\[1mm]
\gamma_{h}^{(2)} &=&  \dfrac{1293}{800}g_1^4 + \dfrac{27}{80}g_1^2 g_2^2 - \dfrac{271}{32} g^4_2 + \dfrac{17}{8} g_1^2 y_t^2  + \dfrac{45}{8} g_2^2 y_t^2 + 20 g_3^2 y^2_t - \dfrac{27}{4} y_t^4 + \dfrac{3}{8} \lambda^2+\dfrac{\delta_2^2}{8} \nonumber \\[2mm]
 \gamma_A^{(1)} &=&  \gamma_S^{(1)} = 0 \label{eq:gammaSA}\\
 \gamma_A^{(2)} &=&  \gamma_S^{(2)} = \dfrac{\delta_2^2+d_2^2}{4} \; . \nonumber 
\end{eqnarray}

\section{One loop input relations} \label{app:OneLoopInput}

In this section we analyse the effect of correcting the initial data, used in the RGE running at two-loops, with one-loop input relations, using the dark matter phase to illustrate the effects. Here we have corrected our scalar sector tree level input data (whose scan boxes are detailed in the next paragraph) at one-loop using the effective potential approach (to compute $p^2=0$ contributions) and the variation of the one-loop scalar self-energies (for the $p^2$ dependent terms) as described below.

In our scans, we chose the physical scalar masses, mixing matrix angles, vacuum expectation and a subset of the scalar couplings (which remain independent) as input (seven parameters in total which corresponds to the number of couplings in our potential). All remaining scalar couplings are functions of this input. At one-loop, we use a similar strategy, i.e. we provide these parameters as input and then we correct the dependent couplings/parameters at one-loop to perform the RGE running consistently. This is done by recalling the one-loop definitions of the physical vacuum state, and of the scalar state masses. As mentioned in Sect.~\ref{sec:Veff}, for a translation invariant vacuum, the VEVs, $v_i$, at the minimum are determined by the stationary points of the effective potential (these are the tadpole equations), namely
\begin{equation}
\dfrac{\partial V_{\rm eff}}{\partial v_i}=0\Leftrightarrow \dfrac{\partial V^{(0)}}{\partial v_i}+\varepsilon \dfrac{\partial V^{(1)}}{\partial v_i}+\ldots\equiv T(L,v_k)_i=0\; .
\end{equation}
Again we have denoted the set of couplings collectively by $L$. This provides a set of constraints relating the couplings and the VEVs. Regarding the physical scalar particle states, they are defined through the poles of the (Dyson resummed) inverse scalar propagator which we denote
\begin{eqnarray}
\mathbf{G}^{-1}_{ij}&=&i\left(-p^{2}\delta_{ij}+\mathbf{M}^{2}_{ij}+\Pi\left(p^2\right)_{ij}\right) \; \\
\Leftrightarrow \mathbf{G}^{-1}_{ij}&=&i\left(-p^{2}\delta_{ij}+\partial^{2}_{ij}V_{\rm eff}+\Delta\Pi\left(p^2\right)_{ij}\right) \; .
\end{eqnarray}
Here $\mathbf{M}^{2}_{ij}$ is the mass matrix obtained from the tree level potential, $\Delta\Pi\left(p^2\right)=\Pi\left(p^2\right)-\Pi\left(0\right)$, and we have used the relation (in matrix notation for brevity)
\begin{equation}
-\mathbf{\partial^{2}V}_{\rm eff}=i\left(\mathbf{G}^{-1}\right)_{p^{2}=0}=-\mathbf{M}^{2}-\Pi(0) \; .
\end{equation}
The physical pole masses of the scalar states $a=\left\{h,s_1,s_2\right\}$ are then defined through
\begin{equation}
0=\det\left(-m_{a}^{2}\mathbf{1}+\mathbf{\partial^{2}V}_{\rm eff}+\Delta\Sigma(m_{a}^{2})\right) \; ,
\end{equation}
which means that the scalar eigenstate $a$ is an eigenvector ($E^j_{\phantom{j}a}$) of $\mathbf{G}^{-1}$ with null eigenvalue, i.e.
\begin{equation}
\mathbf{G}^{-1}_{ij}E^j_{\phantom{j}a}\equiv P\left(L,v_i,m_a^2,E^j_{\phantom{j}a}\right)_{ia}=0 \; .
\end{equation}
Our approach consists on solving, perturbatively (i.e. in an expansion in powers of $\varepsilon$), for the one-loop corrections of the subset of dependent couplings in $\{L\}$ given input data for the physical scalar state pole masses and VEVs. This is determined by the system
\begin{equation}\label{eq:1loopLowScale}
\begin{cases}
T(L,v_k)_i=0& \\
P\left(L,v_i,m_a^2,E^j_{\phantom{j}a}\right)_{ia}=0& \; .
\end{cases}
\end{equation}
Since we have used the effective potential computed in the MS-bar scheme in the Landau gauge, we have also used the scalar self energies computed under the same conditions by S. Martin in~\cite{Martin:2005eg}. We have perturbatively expanded the system~\eqref{eq:1loopLowScale}, \cite{FuturePaper}, in the dark matter phase and found explicit expressions for the corrected couplings $\left\{m^2, \lambda,\delta_2,b_2,d_2,b_1\right\}$ and the overlap of the second mixing scalar $H_{\rm new}$ with the Higgs doublet fluctuation, i.e. the correction to $\kappa_{H_{\rm new}}$. If we define the general notation for the loop expansion of some coupling/parameter by $L=L^{(0)}+\varepsilon L^{(1)}+\ldots$ then the final one-loop corrections in the dark phase are
\begin{eqnarray}
{m^{2}}^{(1)}&=&\frac{1}{v}\left[3 V_h - v_S \left(W_{10} - 2 \kappa_{H_{\rm new}}^{(0)2} W_{10} + \kappa_{H_{126}}\kappa_{H_{\rm new}}^{(0)} (W_{00} - W_{11})\right) + \right. \\
&&\left.
 -v \left(\kappa_{H_{126}}^{2} W_{00} -  2\kappa_{H_{126}}\kappa_{H_{\rm new}}^{(0)}W_{10} + \kappa_{H_{\rm new}}^{(0)2} W_{11}\right)\right]\\
\lambda^{(1)}&=&\dfrac{2}{v^3}\left[-V_h+v\left(\kappa_{H_{126}}^{2}W_{00}-2\kappa_{H_{126}}\kappa_{H_{\rm new}}^{(0)}W_{10}+\kappa_{H_{\rm new}}^{(0)2}W_{11}\right)\right] \\
\delta_2^{(1)}&=&\dfrac{2}{v\,v_S}\left[W_{10} - 2 \kappa_{H_{\rm new}}^{(0)2}W_{10} + \kappa_{H_{126}}\kappa_{H_{\rm new}}^{(0)} (W_{00} - W_{11})\right]\\
b_2^{(1)}&=&\frac{1}{v_S}\left[2 V_s - v W_{10} - v_S W_{11} - \kappa_{H_{126}}\kappa_{H_{\rm new}}^{(0)} \left(2 v_S W_{10} + v (W_{00} - W_{11})\right) + \right. \nonumber\\
&&\left.+ \kappa_{H_{\rm new}}^{(0)2} \left(2 v W_{10} + v_S (W_{11} - W_{00})\right) + v_S W_{22}\right] \\
d_2^{(1)}&=&\frac{2}{v_S^3} \left[-V_s + v_S \left(2 \kappa_{H_{126}}\kappa_{H_{\rm new}}^{(0)} W_{10} + \kappa_{H_{\rm new}}^{(0)2} (W_{00} - W_{11}) + W_{11}\right)\right]\\
b_1^{(1)}&=&\frac{V_s}{v_S}-W_{22} \\
\kappa_{H_{\rm new}}&=&\dfrac{\kappa_{H_{\rm new}}^{(0)}+\delta\kappa_{H_{126}}}{\sqrt{1+\delta^2}}\;\;,\;\; \delta\equiv\varepsilon \dfrac{W_{01}-W_{10}}{m_{H_{\rm new}}^2-m_{H_{126}}^2}
\end{eqnarray}
All masses ($m_i^2$), VEVs ($v_i$), the other $\kappa_{H_{126}}$ and $a_1$ were set as weak scale input\footnote{Observe that since we perform a global scan, to some extent, what is used as input is arbitrary because the scan will run over all possible scenarios. Our choice was guided by the principle of providing as input as many physical quantities as possible, while keeping the scan efficient.}. Also note that, on the last line, we have written the corrected $\kappa_{H_{\rm new}}$ in full using the normalisation condition for the corresponding corrected eigenvector, for consistency.
The loop corrections are encoded in the quantities
\begin{equation}
V_{a}=\delta_a^hy_{t}\sqrt{2}m_{t}^{3}\left(\log\tfrac{m_{t}^{2}}{\mu^{2}}-1\right)-\dfrac{1}{2}R_a^{\phantom{a}i}\lambda_{\phantom{k}ki}^{k}m_{k}^{2}\left(\log\tfrac{m_{k}^{2}}{\mu^{2}}-1\right)
\end{equation}
and 
\begin{eqnarray}
W_{ia}&=&R_{\phantom{h}i}^{h}R_{\phantom{h}a}^{h}y_{t}^{2}\left[m_{t}^{2}\left(3\log\tfrac{m_{t}^{2}}{\mu^{2}}-1\right)-\left(m_{a}^{2}-4m_{t}^{2}\right)I_{0}\left(\tfrac{m_{a}^{2}}{m_{t}^{2}}\right)-m_{a}^{2}\log\left(\tfrac{m_{t}^{2}}{\mu^{2}}\right)\right]+\nonumber \\
&&+\lambda_{\phantom{\epsilon_{1}\epsilon_{2}}i}^{\epsilon_{1}\epsilon_{2}}\left[1-\frac{1}{2}\log\left(\tfrac{m_{a}^{2}}{\mu^{2}}\right)\right]\lambda_{\epsilon_{1}\epsilon_{2}a}-\dfrac{1}{2}\left[\lambda_{kia}^{k}m_{k}^{2}\left(\log\tfrac{m_{k}^{2}}{\mu^{2}}-1\right)+\right. \\&&\left.+\lambda_{i}^{kl}\lambda_{kla}\left(\log\tfrac{m_{k}^{2}}{\mu^{2}}-1+f\left(m_{k}^{2},m_{l}^{2}\right)+I_{1}\left(\tfrac{m_{l}^{2}}{m_{k}^{2}},\tfrac{m_{a}^{2}}{m_{k}^{2}}\right)-I_{2}\left(\tfrac{m_{l}^{2}}{m_{k}^{2}}\right)\right)\right]_{k,l\neq {\rm goldstones}}\ ,\nonumber
\end{eqnarray}
where the indices $i,a=\left\{0,1,2,3,4,5\right\}$  in $W_{ia}$ correspond respectively to the eigenstates $\left\{H_{\rm new},H_{126},H_{\rm dark},G_1,G_2,G_3\right\}$  ($G_i$ are Goldstone degrees of freedom) whereas in $V_a$ the index $a$ is a weak basis index. 
Note that on the second line we have explicitly cancelled out the infrared divergence coming from the Goldstone masses between the derivatives of the effective potential and the self-energies. The term which is left over contains a sum over $\epsilon_1$ and $\epsilon_2$ which runs over Goldstone directions only (first term on the second line). We have also checked explicitly that the correct cancelations occur such that the Goldstones are present in the one-loop inverse propagator and that we recover the one loop results in the SM limit. Finally the loop function combinations appearing are defined
\begin{eqnarray}
f(x,y)&\equiv& x\frac{\log x-\log y}{x-y}\\
I_0(x)&\equiv&\Re\left[\int_{0}^{1}dt\log\left[1-t(1-t)\left(x+i\epsilon\right)\right]\right] \\
I_1(x,y)&\equiv&\Re\left[\int_{0}^{1}dt\log\left[t+(1-t)x-t(1-t)y-i\epsilon\right]\right]\\
I_2(x)&\equiv&-1+\dfrac{x\log x}{x-1}\; .
\end{eqnarray}
For brevity here we define $I_0$ and $I_1$ through their integral forms though they can be integrated explicitly.

\begin{figure}
\begin{center}
\includegraphics[width=0.49\linewidth,clip=true,trim=40 35 25 45]{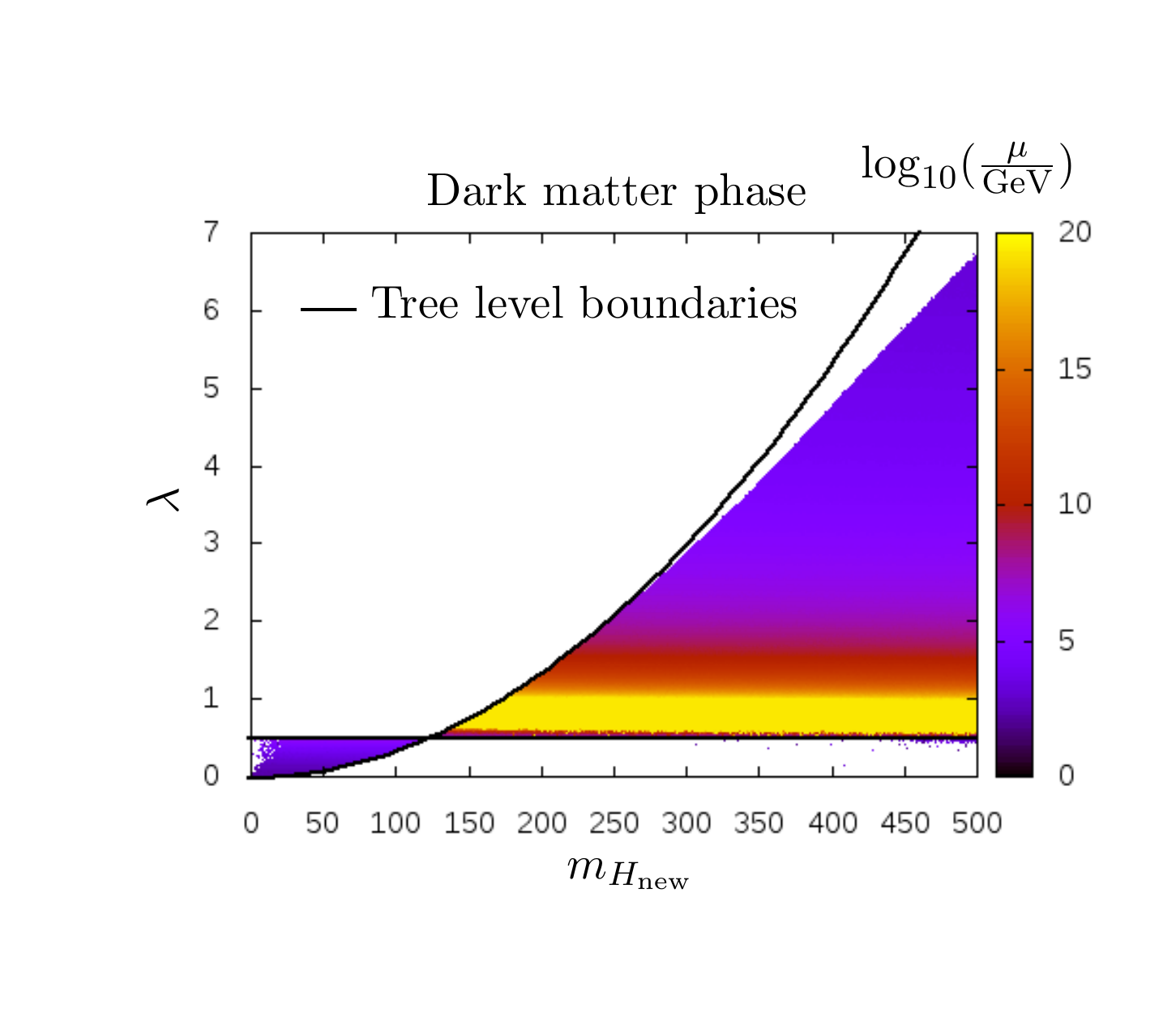}\hspace{0.01\linewidth}\includegraphics[width=0.49\linewidth,clip=true,trim=40 35 25 45]{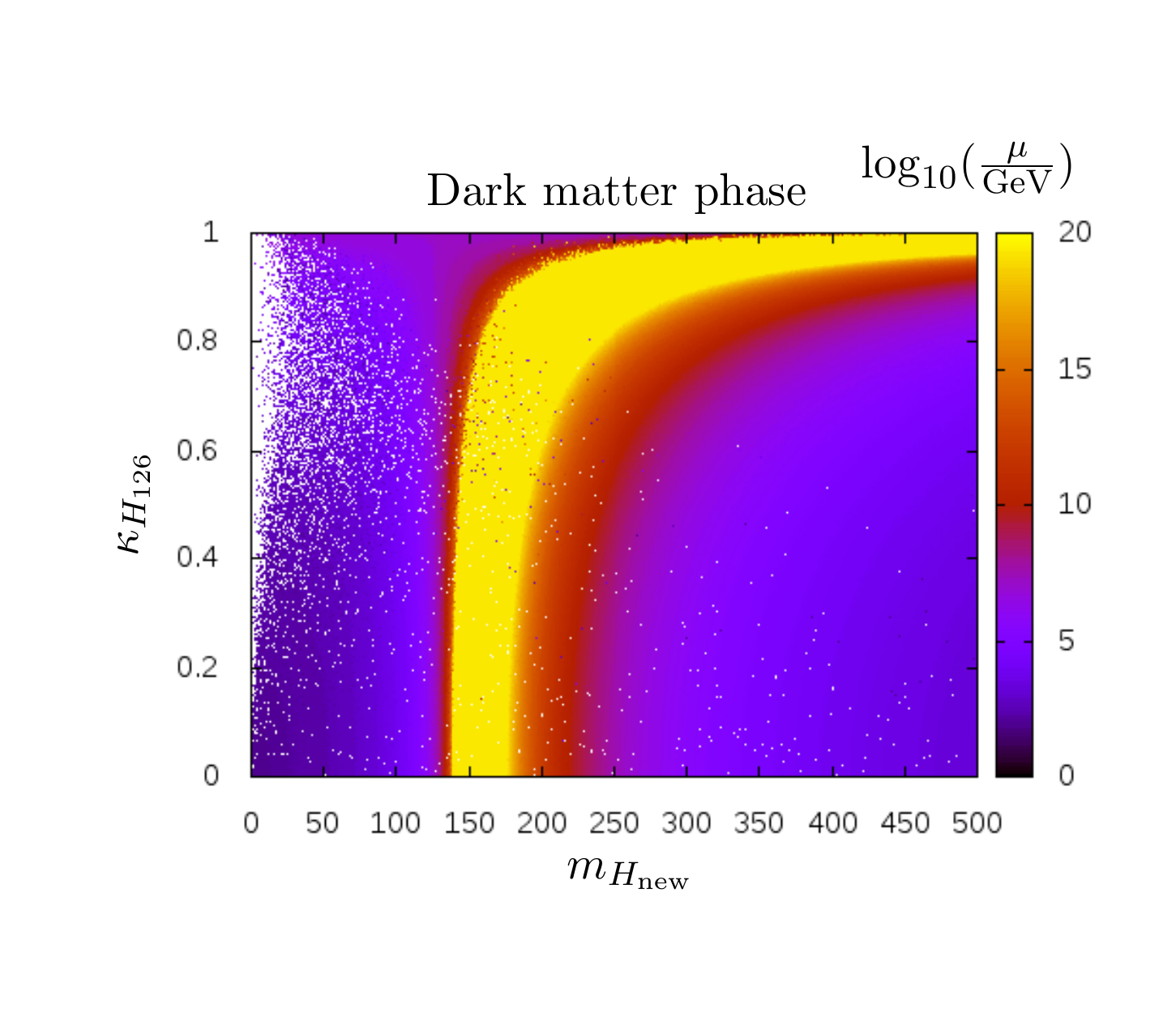}
\end{center}
\caption{{\em Dark matter phase corrected}: Here we display the same quantities as in Fig.~\ref{onlyRGE_mvisible_lambda} with the one-loop corrected initial data. On the left panel, we also indicate the line which defines the tree level upper boundary for comparison.}
\label{onlyRGE_mvisible_lambda_corrected}
\end{figure}
In Fig.~\ref{onlyRGE_mvisible_lambda_corrected} we finally present a comparison of the regions generated for Fig.~\ref{onlyRGE_mvisible_lambda}, but now with the one-loop corrected initial data. It is clear that the main conclusions of our study, which correlate with shape of the stability band, are not affected by these corrections. In particular, the lower bound for the mass of the new visible scalar which is responsible for stabilizing the scalar potential remains the same. The only relevant difference is a small thickening of the stability band for larger masses (see right hand side panel) and a lowering of the upper boundary  on the $(\lambda,m_{H_{\rm new}})$ plane which is related to the corrected minimum conditions that define the vacuum of the theory (see left panel).

\begin{figure}
\begin{center}
\includegraphics[width=0.49\linewidth,clip=true,trim=25 35 25 55]{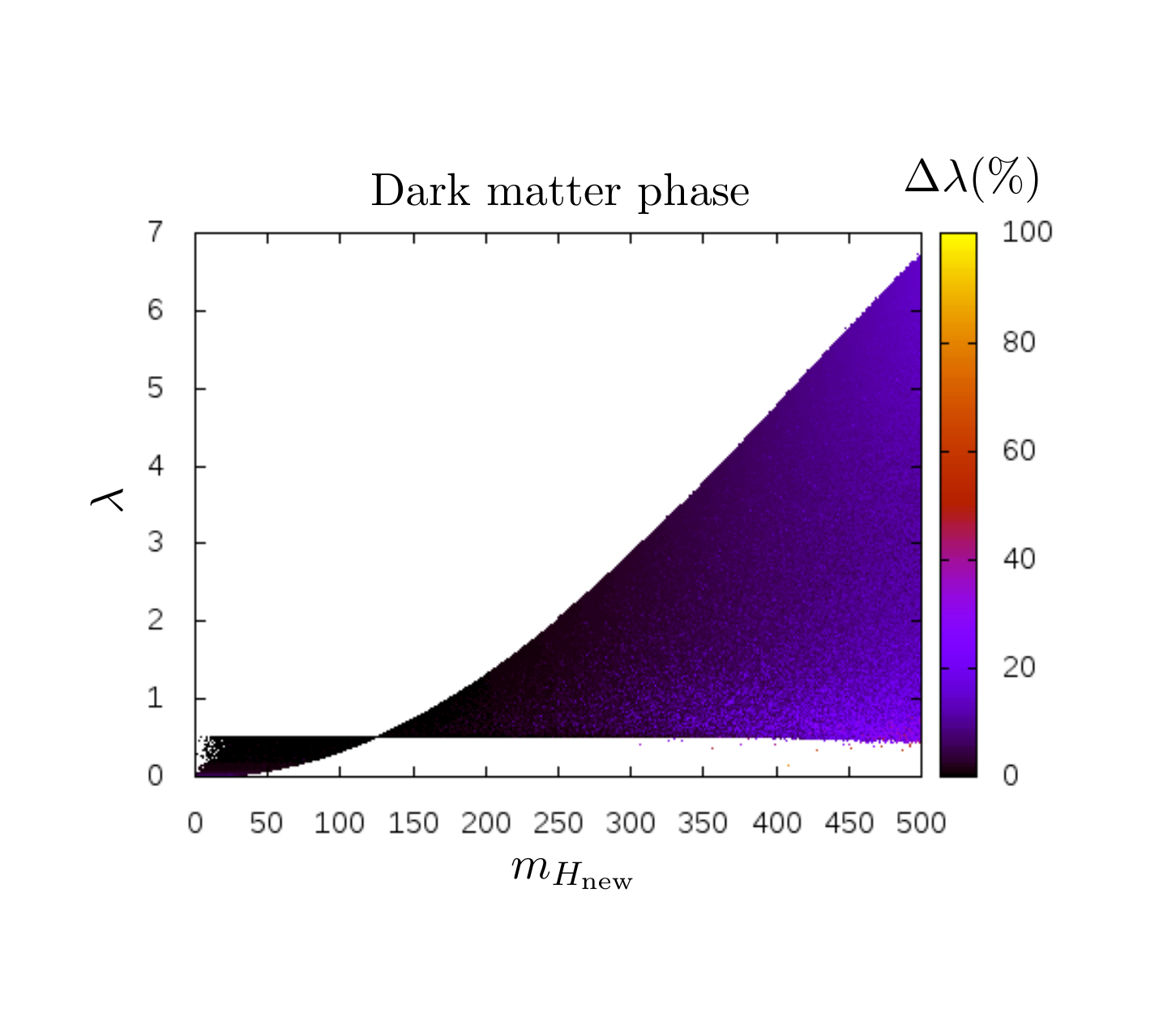}\hspace{0.01\linewidth}\includegraphics[width=0.49\linewidth,clip=true,trim=20 35 25 55]{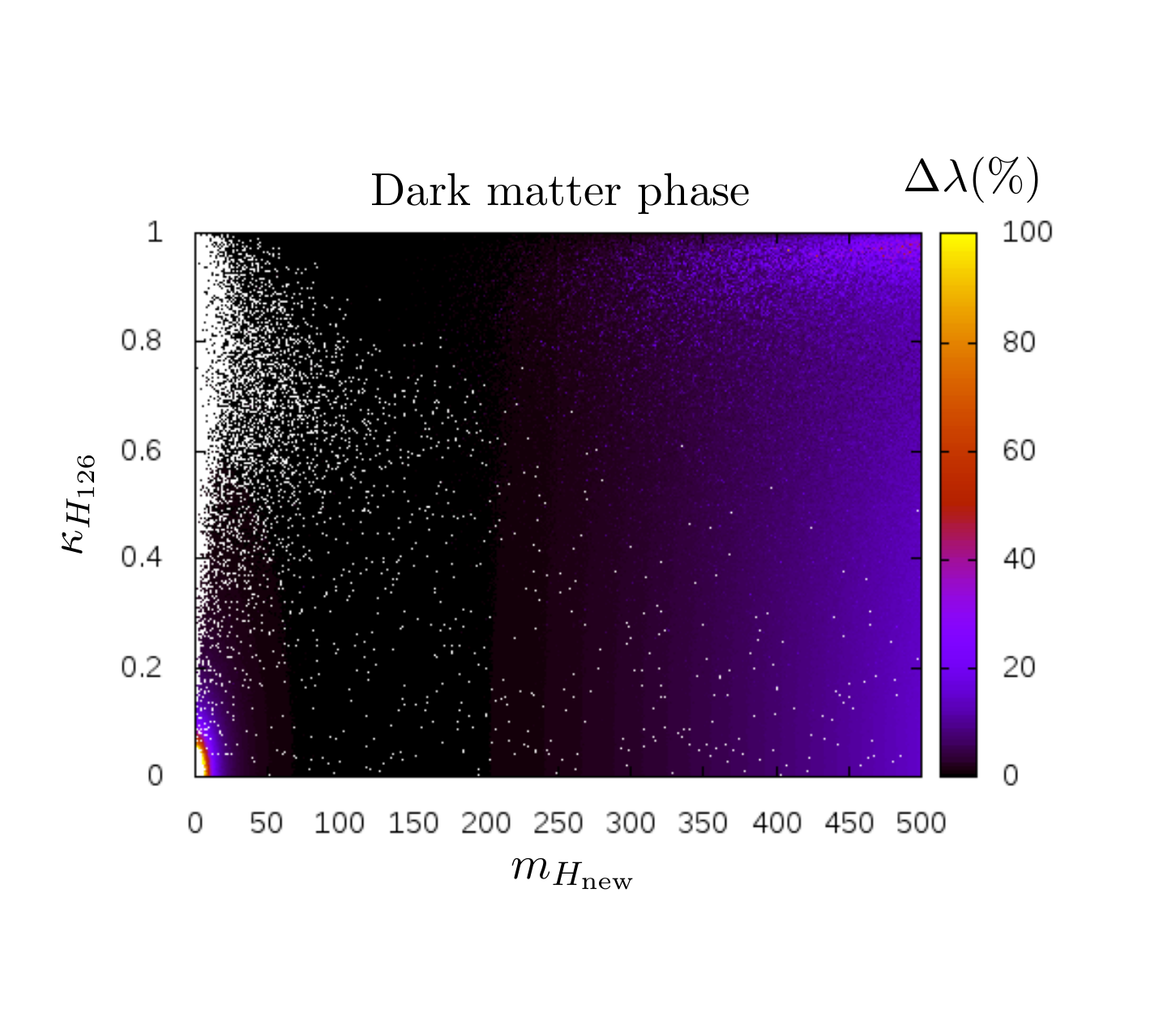}
\end{center}
\caption{{\em Dark matter phase corrected}: Here we display the same quantities as in Fig.~\ref{onlyRGE_mvisible_lambda_corrected}. The color gradient shows the relative diference to the $\lambda$ between tree level and with the one-loop initial data correction, defined $\Delta \lambda = |\varepsilon \lambda^{(1)}|/\sqrt{\lambda^{(0)2}+\left(\lambda^{(0)}+\varepsilon \lambda^{(1)}\right)^2}$.}
\label{onlyRGE_mvisible_lambda_corrected_err}
\end{figure}
 Finally in Fig.~\ref{onlyRGE_mvisible_lambda_corrected_err} we show the relative difference between tree level and one-loop intput (in percentage) for the $\lambda$ coupling in the color scale. Everywhere the error is small especially in the intermediate mass region where the stability lower bound is obtained (The only exception is close to the origin in the right panel which can be checked to be just due to $\lambda$ going to zero, so the relative error definition becomes ill defined).

\bibliographystyle{JHEP}       
\bibliography{references}   

\end{document}